\documentclass[pdflatex,sn-basic,iicol]{sn-jnl} 

\usepackage{graphicx}%
\usepackage{multirow}%
\usepackage{amsmath,amssymb,amsfonts}%
\usepackage{amsthm}%
\usepackage{mathrsfs}%
\usepackage[title]{appendix}%
\usepackage{xcolor}%
\usepackage{textcomp}%
\usepackage{manyfoot}%
\usepackage{booktabs}%
\usepackage{algorithm}%
\usepackage{algorithmicx}%
\usepackage{algpseudocode}%
\usepackage{listings}%
\usepackage{orcidlink}
\usepackage{multicol}
\usepackage{lipsum}
\usepackage{balance}

\theoremstyle{thmstyleone}%
\theoremstyle{thmstyletwo}%

\theoremstyle{thmstylethree}%

\def \lesssim {\mathrel{\vcenter
     {\offinterlineskip \hbox{$<$}\hbox{$\sim$}}}}
\def \gtrsim {\mathrel{\vcenter
     {\offinterlineskip \hbox{$>$}\hbox{$\sim$}}}}

\newcommand{\nue}{\ensuremath{\nu_e}}
\newcommand{\nuebar}{\ensuremath{\bar{\nu}_e}}
\newcommand{\nux}{\ensuremath{\nu_x}}
\renewcommand{\vec}[1]{\ensuremath{\pmb{#1}}}

\begin{document}

\title[Neutron Star and Black Hole Kicks]{Interplay Between Neutrino Kicks and Hydrodynamic
Kicks of Neutron Stars and Black Holes}


\author*[1]{\fnm{Hans-Thomas} \sur{Janka}\orcidlink{0000-0002-0831-3330}} \email{thj@mpa-garching.mpg.de}

\author[1,2]{\fnm{Daniel} \sur{Kresse}\orcidlink{0000-0003-1120-2559}}\email{danielkr@mpa-garching.mpg.de}



\affil*[1]{\orgname{Max Planck Institute for Astrophysics}, \orgaddress{\street{Karl-Schwarzschild-Stra\ss e~1}, \city{Garching}, \postcode{85748}, \country{Germany}}}

\affil[2]{\orgname{Technical University of Munich}, \orgdiv{TUM School of Natural Sciences, Physics Department}, \orgaddress{\street{James-Franck-Stra{\ss}e 1}, \city{Garching}, \postcode{85748}, \country{Germany}}}



\abstract{
Neutron stars (NSs) are observed with high space velocities and elliptical orbits in binaries. The magnitude of these effects points to natal kicks that originate from asymmetries during the supernova (SN) explosions. Using a growing set of long-time 3D SN simulations with the \textsc{Prometheus-Vertex} code, we explore the interplay of NS kicks that are induced by asymmetric neutrino emission and by asymmetric mass ejection. Anisotropic neutrino emission can arise from a large-amplitude dipolar convection asymmetry inside the proto-NS (PNS) termed LESA (Lepton-number Emission Self-sustained Asymmetry) and from aspherical accretion downflows around the PNS, which can lead to anisotropic neutrino emission (absorption/scattering) with a neutrino-induced NS kick roughly opposite to (aligned with) the kick by asymmetric mass ejection. In massive progenitors, hydrodynamic kicks can reach up to more than 1300\,km\,s$^{-1}$, whereas our calculated neutrino kicks reach (55--140)\,km\,s$^{-1}$ (estimated upper bounds of (170--265)\,km\,s$^{-1}$) and only $\sim$(10--50)\,km\,s$^{-1}$, if LESA is the main cause of asymmetric neutrino emission. Therefore, hydrodynamic NS kicks dominate in explosions of high-mass progenitors, whereas LESA-induced neutrino kicks dominate for NSs born in low-energy SNe of the lowest-mass progenitors, when these explode nearly spherically. Our models suggest that the Crab pulsar with its velocity of $\sim$160\,km\,s$^{-1}$, if born in the low-energy explosion of a low-mass, single-star progenitor, should have received a hydrodynamic kick in a considerably asymmetric explosion. Black holes, if formed by the collapse of short-lived PNSs and solely kicked by anisotropic neutrino emission, obtain velocities of only some km\,s$^{-1}$.
}




\maketitle

\section{Introduction}
\label{sec:intro}

Various observational findings provide evidence that neutron stars (NSs) obtain natal kicks during their birth in the supernova (SN) explosions of massive stars. Radio pulsars are found to possess space velocities of up to $\sim$1500\,km\,s$^{-1}$ \citep[e.g.,][]{Arzoumanian+2002,Hobbs+2005,Chatterjee+2005}, NSs associated with core-collapse supernova (CCSN) remnants exhibit displacements from the geometric center of the explosion (e.g., Puppis~A, \citealt{Mayer+2020}; Cas~A, \citealt{Holland-Ashford+2023}), and orbital parameters of NSs in binary systems require kicks of the compact objects \citep[for a summary of relevant observational aspects, see][]{Lai+2001}. These observations cannot be explained by the electromagnetic recoil due to the Harrison-Tademaru effect \citep{Lai+2001} or the Blaauw mechanism \citep{Blaauw1961}, which predicts binary systems to become affected and even unbound by the mass loss connected to the SN explosion of one component. Instead, intrinsic asymmetries during the SN explosions must be the dominant effect for kicking the NSs. Different studies have pointed to the NS kick distribution being unimodal \citep{Hobbs+2005,Faucher-Giguere+2006} or bimodal \citep{Arzoumanian+2002,Verbunt+2017} with an average velocity of 200--500\,km\,s$^{-1}$. In particular, a fraction of the NS population must have received relatively lower kick velocities in order to account for the pulsars found gravitationally bound in globular clusters \citep[e.g.,][]{Pfahl+2002a,Abbate+2023,Heywood+2023,Wu+2023}. Also NSs born in low-mass X-ray binaries \citep{Bahramian+2023,Doherty+2023} and high-mass X-ray binaries \citep{Pfahl+2002b,Kim+2023} as well as those in the binary NSs observed in our Galaxy \citep{Gaspari+2024} and measured by gravitational waves in their latest pre-merger evolution \citep{Abbott+2017a,Abbott+2017b} were not kicked out of their orbits but stayed bound to their companions.

Anisotropic mass-energy loss through asymmetrically ejected gas and neutrinos must be considered as the primary mediators of the recoil momentum transferred to NSs during their birth, causing hydrodynamic NS kicks and neutrino-induced NS kicks \citep[e.g.,][]{Janka+1994,Burrows+1996,Fryer+2006}, since the energy radiated in gravitational waves by collapsing stars is minuscule (typically less than $10^{-7}$\,M$_\odot c^2$; \citealt{Radice+2019,Powell+2019,Andresen+2021,Mezzacappa+2023,Vartanyan+2023}) compared to the gravitational binding energy of the new-born NS radiated in neutrinos (several $10^{53}$\,erg or $\gtrsim$\,0.1\,M$_\odot c^2$; \citealp{Sukhbold+2016,Kresse+2021,VartanyanBurrows2023}) and compared to the kinetic energy of the SN ejecta (exceeding $\sim$10$^{50}$\,erg; \citealp{KasenWoosley2009,PejchaPrieto2015b,Martinez+2022}). Abundant theoretical work has been devoted to understanding of these NS kicks \citep[e.g.,][]{Janka+1994,Scheck+2004,Burrows+1996,Scheck+2006,Nordhaus+2010,Nordhaus+2012,Wongwathanarat+2010,Wongwathanarat+2013,Janka2017,Mueller+2017,Gessner+2018,Mueller+2019,Nagakura+2019,Powell+2020,Bollig+2021,Rahman+2022,Powell+2023,Coleman+2022,Burrows+2023b}, based mainly on axisymmetric (2D) CCSN simulations until about 2010 and later on three-dimensional (3D) simulations.

Interestingly, in their neutrino-driven 3D SN simulations, \citet{Wongwathanarat+2013} found that iron-group nuclei and intermediate-mass elements (IMEs, mainly those heavier than $^{24}$Mg) are ejected preferentially, i.e., with higher abundances, in the hemisphere opposite to the direction of a large hydrodynamic NS kick, because the SN explosion and mass ejection are stronger on this side. Such a geometry is dictated by linear momentum conservation, which implies that the net momentum of the ejected gas is balanced by the opposite momentum received by the kicked NS. Since the SN shock and the gas outflow that is driven by the neutrino energy deposition are stronger where the SN explosion is more powerful, also the explosive nucleosynthesis is more efficient on this side, which enhances the production of IMEs by nuclear burning in shock-heated matter and of iron-group nuclei, trans-iron species, and radioactive isotopes such as $^{44}$Ti in the freeze-out from NSE in neutrino-heated material. 

\citet{Wongwathanarat+2017} showed that this explosion-nucleosynthesis-kick asymmetry shapes the entire large-scale morphology of Cassiopeia~A (Cas~A), which is the remnant of a type-IIb SN of a progenitor star that had stripped almost all of its hydrogen envelope, allowing the evolving ejecta to retain the explosion geometry very clearly even at the stage of the aging SN remnant \citep{Orlando+2021}. The NuSTAR (Nuclear Spectroscopic Telescope Array) mapping of the radioactive $^{44}$Ti distribution in Cas~A by \citet{Grefenstette+2014} and \citet{Grefenstette+2017} provides general support of the picture drawn by the 3D simulations and thus for Cas~A being the relic of a neutrino-driven SN explosion with the NS having received its natal kick mainly by the hydrodynamic recoil of asymmetrically ejected gas. A closer analysis of the gas geometry \citep{Holland-Ashford+2017} and of the  distribution of IMEs \citep{Katsuda+2018} in larger samples of SN remnants also reveals overall consistency with this scenario.

Although neutrino-induced NS kicks had also been evaluated in some of the mentioned simulation-based studies \citep[e.g.,][]{Scheck+2006,Nordhaus+2010,Wongwathanarat+2013,Nagakura+2019,Rahman+2022}, the models considered there were still handicapped by their 2D nature or their simplified treatment of the neutrino physics and transport, or both. A reliable determination of the neutrino kicks requires a sophisticated treatment of the neutrino transport in full 3D simulations conducted over sufficiently long evolution times of, at best, several seconds. Such simulations have become available only recently \citep{Stockinger+2020,Bollig+2021,Coleman+2022,Burrows+2023b}. We supplement this recent work by a detailed and comprehensive analysis of a growing set of 3D stellar core-collapse simulations performed with the \textsc{Prometheus-Vertex} SN code of the Garching group including cases that lead to black-hole (BH) formation. 

The situation is observationally less consolidated for natal BH kicks than for NS kicks \citep[e.g.,][]{Andrews+2022}. Many stellar population synthesis studies assume BH kicks with a momentum distribution similar to that of NS kicks, i.e., with a BH velocity distribution that is reduced by the BH-to-NS mass ratio or the fraction of the stellar envelope that falls back \citep[e.g.,][]{Fryer+2012,Mandel+2020,Mandel+2021,Oh+2023,Stevance+2023}. However, referring to the observation of some low-mass X-ray binaries containing BHs at large distances above the plane of the Milky Way, \citet{Repetto+2012}, \citet{Repetto+2015}, and \citet{Repetto+2017} concluded that the velocity distribution, instead of the momentum of natal kicks, of at least this population of BHs might be more similar to that of NSs \citep[but see][for counter-arguments]{Mandel2016}. Detailed observations of individual low-mass X-ray binaries and BH X-ray binaries indeed confirm that natal kicks of considerable magnitude (up to several 100\,km\,s$^{-1}$ at least) can be imparted to the BHs in such systems \cite[e.g.,][]{Atri+2019,Kimball+2023,Brown+2024}. \cite{Janka2013} explained this possibility by low-mass BHs that form due to asymmetric mass accretion onto the transiently existing NS when initially ejected material falls back, because the SN explosion is not strong enough to unbind the inner layers of the collapsing star. The thus born BH may receive a large hydrodynamic kick provided the SN explosion is still sufficiently powerful to expel a considerable fraction of the asymmetrically distributed innermost ejecta. Recent 3D simulations of such fallback SNe within the framework of the neutrino-driven mechanism lend support to the viability of this predicted scenario \citep{Chan+2020} and even show quantitative agreement \citep{Burrows+2023} with the analytic estimates by \citet{Janka2013}. 

Based on theoretical considerations one also expects a population of BHs that are born in stellar core-collapse events without a SN explosion or where the SN blast is too weak to produce asymmetric ejecta. Under these circumstances either little mass ejection by the wimpy SN occurs or mass loss is absent or effectively spherical if (a part of) the hydrogen envelope of the progenitor is stripped due to the so-called mass-decrement effect \citep{Nadezhin1980,Lovegrove+2013} when the gravitational potential of the NS prior to BH formation is reduced by the energy loss through neutrino emission. In such cases the BH receives a natal kick that is solely caused by anisotropic neutrino emission and should remain very low, namely of the order of a few km\,s$^{-1}$, if the BH forms within fractions of a second \citep{Rahman+2022}, or at most some 10\,km\,s$^{-1}$, if the NS lives for a few seconds before it collapses to a BH \citep{Burrows+2023b}. Very recently, the BH binary VFTS~243 was identified by its well determined system properties as the first case where the BH has received a very low kick velocity predominantly by asymmetric neutrino emission \citep{Vigna-Gomez+2024}.  

The present paper aims at a consistent and comprehensive evaluation of the hydrodynamic kicks together with the corresponding neutrino-induced kicks of new-born NSs and BHs for a set of long-term 3D stellar collapse simulations performed by the Garching group for single-star progenitors with zero-age-main-sequence (ZAMS) masses between 9\,M$_\odot$ and 75\,M$_\odot$. We thus intend to shed light on the physical origin of these kicks by tracking down in detail the processes that play a role for the acceleration of the compact remnants. Our special focus will be on the complex interplay of neutrino and hydrodynamic momentum transfer that is a consequence of their mutual coupling due to neutrino interactions in the stellar matter. In the course of this effort we also intend to relate our results to those of similar analyses reported in recent literature and thus to clarify some misconceptions discussed there. In particular, we will furnish further evidence that the effect that acts longest in accelerating NSs (and some BHs) is the gravitational attraction between the compact remnant and the asymmetric ejecta gas. Since the momentum transfer to the compact remnant by the force of gravity accounts for the dominant contribution to the hydrodynamic kicks, we will argue that the term ``gravitational tug-boat mechanism'' for the hydrodynamic recoil of the compact objects \citep{Wongwathanarat+2013} is well justified. Our work is further motivated by the goal to understand the relevance of neutrino-induced kicks for the total NS kicks, because this might have consequences for the interpretation of the mentioned observations of SN remnants where the NS kick vector points opposite to the direction of
a dipole deformation of the ejecta or enhanced concentrations of IMEs and iron-group elements.

Our paper is structured as follows. In Section~\ref{sec:models}, we briefly describe the simulation code and its inputs, the investigated progenitors, and the results of their long-term 3D core-collapse simulations, and we provide the details of our numerical methods to evaluate these 3D simulations for the hydrodynamic and neutrino-induced NS and BH kicks. In Section~\ref{sec:results}, we report our results for the hydrodynamic and neutrino kicks of NSs in models with successful SN explosions as well as our results for neutrino-induced BH kicks in models that fail to explode. Section~\ref{sec:conclusions} contains a summary, discussion, and conclusions. In Appendices~\ref{secA1}--\ref{secA:bh} we present, for all of our 3D simulations with sophisticated neutrino transport, supplementary plots which show in detail the time evolution of quantities that are especially relevant for understanding the physical effects that contribute to the neutrino-induced NS and BH kicks.

\section{Numerical modeling}
\label{sec:models}

\subsection{Simulation code and inputs}

The 3D CCSN models discussed in this work were computed with the \textsc{Prometheus-Vertex} neutrino-hydrodynamics code \citep{Rampp+2002,Buras+2006}. The \textsc{Prometheus} hydrodynamics module \citep{Fryxell+1989,Keil1997,Kifonidis+2003} is based on a higher-order Godunov scheme for the Newtonian hydrodynamics equations, which are integrated with directional splitting and an exact Riemann solver for ideal gases. The \textsc{Vertex} transport module solves the energy and velocity (${\cal O}(v/c)$) dependent three-flavor neutrino transport by a two-moment scheme with Boltzmann closure \citep{Rampp+2002}. The three-dimensionality of the transport problem is handled by a ray-by-ray-plus (RbR+) approach, which computes only radial neutrino flux components because it applies the assumption that the neutrino phase space distribution is axially symmetric around the radial direction. The good overall agreement of neutrino properties and hydrodynamic evolution in 3D CCSN simulations with the RbR+ approximation and full multi-dimensional transport was demonstrated by \citet{Glas+2019}. The neutrino energy bins are fully coupled for the propagator and interaction terms and the reported simulations employed an energy grid with 12--15 bins up to a maximum energy of 380\,MeV. The code includes general relativistic effects in the monopole term of the gravitational potential \citep[Case~A of][]{Marek+2006} as well as corrections for general relativistic redshift and time dilation in the neutrino transport. Some of the 3D simulations were presented in previous papers (see Table~\ref{tab:models}), where details of the specific numerical setups can be found. A recent summary of the most relevant features of the code, its input physics for the equation of state (EoS) of the stellar plasma and the nuclear composition, and the neutrino reactions taken into account in the transport solver is provided by \citet{Fiorillo+2023}.

Most of the 3D simulations leading to successful SN explosions and NS formation were computed with the nuclear EoS of \citet{Lattimer+1991} (LS220, using an incompressibility modulus of 220\,MeV), but in some of the reported models, in particular for BH formation, we also employed the DD2 EoS of \citet{Typel+2010}, \citet{Hempel+2010}, and \citet{Fischer+2014} and the SFHo EoS of \citet{Hempel+2012} and \citet{Steiner+2013} (see Table~\ref{tab:models}).

\onecolumn
\addtolength{\tabcolsep}{-2pt}
\begin{table}[h]
\caption{Overview of investigated models: progenitor properties and setups of 3D CCSN simulations}
\label{tab:models}
\begin{tabular*}{\textwidth}{@{\extracolsep\fill}lcccccccccc}
        \toprule
        Model & $M_\mathrm{ZAMS}$ & $M_\mathrm{prec}$ & $M_\mathrm{He}$ & $Z$ & $\xi_{2.5}$ & Progenitor & EoS & Grid & $N_r$ & Refs. \\[0.5ex]
        & [M$_\odot$] & [M$_\odot$] & [M$_\odot$] & [Z$_\odot$] && Dim./Rot. && early/late && \\
        \midrule
        s9.0 & 9.00 & 8.75 & 1.57 & 1 & 0.00004 & 1D, non-rot. & LS220 & YY-SMR/2$^\circ$ & 400--626 & [1],[9,17] \\
        z9.6 & 9.60 & 9.60 & 1.70 & 0 & 0.00008 & 1D, non-rot. & LS220 & YY-2$^\circ$ & 400--700 & [2],[10,17] \\
        s12.28 & 12.28 & 11.13 & 3.29 & 1 & 0.03167 & 3D, non-rot. & SFHo & YY-3.5$^\circ$ & 550--729 & [3,4],[11,18] \\
        m15 & 15.00 & 11.23 & 4.77 & 1 & 0.10602 & 1D, rot.$^\ast$ & LS220 & YY-2$^\circ$ & 400--800 & [5],[12,18] \\
        m15e & " & " & " & " & " & " & LS220 & YY-2$^\circ$ & 400--800 & [5],[12,18] \\
        s18.88 & 18.88 & 14.34 & 5.79 & 1 & 0.28335 & 3D, non-rot. & LS220 & YY-2$^\circ$ & 550--730 & [6],[13,18] \\
        s20 & 20.00 & 15.93 & 6.33 & 1 & 0.28462 & 1D, non-rot. & LS220 & SP/YY-2$^\circ$ & 400--750 & [7],[14,18] \\
        s20e & " & " & " & " & " & " & LS220 & SP/YY-2$^\circ$ & 400--740 & [7],[14,18] \\
        \midrule
        s40 & 40.00 & 15.34 & 15.34 & 1 & 0.54399 & 1D, non-rot. & LS220 & YY-5$^\circ$ & 400--667 & [7],[15,16,18] \\
        u75\_DD2 & 75.00 & 74.05 & 54.84 & $10^{-4}$ & 0.88157 & 1D, non-rot. & DD2 & YY-5$^\circ$ & 400--646 & [8],[15,18] \\
        u75\_LS220\_1 & " & " & " & " & " & " & LS220 & YY-5$^\circ$ & 400--562 & [8],[15,18] \\
        u75\_LS220\_2 & " & " & " & " & " & " & LS220 & YY-5$^\circ$ & 400--562 & [8],[15,18] \\
        u75\_LS220\_hr & " & " & " & " & " & " & LS220 & YY-SMR & 400--573 & [8],[15,16,18] \\
        u75\_SFHo & " & " & " & " & " & " & SFHo & YY-5$^\circ$ & 400--706 & [8],[15,18] \\
        \botrule
\end{tabular*}
\footnotetext{\textbf{Note:} $M_\mathrm{ZAMS}$, $M_\mathrm{prec}$, and $M_\mathrm{He}$ are the progenitor's ZAMS mass, pre-collapse mass, and helium-core mass (defined by
50\% hydrogen
mass fraction); $Z$ and $\xi_{2.5}$ are the progenitor's metallicity and (pre-collapse) compactness parameter \citep{OConnor+2011}; 3D/1D and rot./non-rot. indicates whether the pre-collapse evolution has been simulated in 3D/1D and with/without rotation [$^{(\ast)}$model m15 at 10\,ms after the bounce: iron-core mass $M_\mathrm{Fe}=1.42$\,M$_\odot$, radius $R_\mathrm{Fe}=8.2\times10^7$\,cm, moment of inertia $I_\mathrm{Fe}=8.9\times10^{47}$\,g\,cm$^2$, angular momentum $J_\mathrm{Fe}=9.7\times10^{47}$\,g\,cm$^2$s$^{-1}$, average rotation period $\bar{P}_\mathrm{Fe}=2\pi I_\mathrm{Fe}/J_\mathrm{Fe}=5.8$\,s]; nuclear EoS: LS220, SFHo, or DD2; computational grid: Yin-Yang (YY) or spherical polar (SP), with uniform angular resolution $\Delta\theta\in\{2^\circ,3.5^\circ,5^\circ\}$ or with static mesh refinement (SMR) for angular grid ($0.5^\circ\leqslant\Delta\theta\leqslant2^\circ$); number of radial zones ($N_r$; min.--max.). \textbf{References (progenitor models):} [1]\,\citet{WoosleyHeger2015}, [2]\,\citet{Heger2012}, [3]\,\citet{Sukhbold+2018}, [4]\,N.\,Yadav (2023, priv. com.), [5]\,\citet{Heger+2005}, [6]\,\citet{Yadav+2020}, [7]\,\citet{WoosleyHeger2007}, [8]\,\citet{Woosley+2002}. \textbf{References (SN/BH models):} [9]\,\citet{Melson+2020}, [10]\,\citet{Melson+2015a}, [11]\,R.\,Bollig (2023, priv. com.), [12]\,\citet{Summa+2018}, [13]\,\citet{Bollig+2021}, [14]\,\citet{Melson+2015b}, [15]\,A.\,Summa (2018, priv. com.), [16]\,\citet{Walk+2020}, [17]\,\citet{Stockinger+2020}, [18]\,Kresse et al (2024, in preparation).}
\end{table}
\addtolength{\tabcolsep}{2pt}

\begin{multicols}{2}

The refined treatment of the neutrino transport with the \textsc{Vertex} module was applied for several seconds only in two of our most recent 3D simulations, whereas in all of the other models the \textsc{Vertex} transport was used for at most about 600\,ms after core bounce (time $t_\mathrm{f}^\nu$ listed in Tables~\ref{tab:hydro_kicks} and~\ref{tab:neutrino_kicks}). At that time either the NS began to collapse to a BH and the simulation was stopped or the calculation was continued with a more approximate description of the neutrino heating and cooling inside and around the proto-neutron star (PNS), applying neutrino transport results from detailed 1D PNS cooling simulations that took into account PNS convection by a mixing-length treatment. The numerical method for this transport approximation, which we call \textsc{Nemesis} (Neutrino-Extrapolation Method for Efficient SImulations of Supernova explosions), was first introduced in Appendix~E of \citet{Stockinger+2020} and further improved by \citet{Kresse2023} (D.~Kresse et al, 2024, in preparation).

\subsection{Investigated models}
\label{sec:3Dsimulations}

Table~\ref{tab:models} lists the basic properties of the single-star progenitors considered in our study and some elementary information of the computed core-collapse and SN models: the ZAMS mass, pre-collapse mass, helium core mass, stellar metallicity, compactness parameter
\begin{equation}
\xi_M \equiv \frac{M/\mathrm{M}_\odot}{r(M)/1000\,\mathrm{km}} 
\label{eq:compactness}
\end{equation}
\citep{OConnor+2011} for a chosen value of $M = 2.5\,\mathrm{M}_\odot$ of the enclosed mass ($r(M)$ is the corresponding radius inside the star), the dimension of the pre-collapse model and the total angular momentum as well as the average rotation period of the iron core shortly after core bounce for the rotating 15\,M$_\odot$ model, the nuclear EoS used, the computational grids employed in the CCSN simulation (with/without \textsc{Vertex} neutrino transport), and the corresponding angular resolution as well as the (time-dependent) number of radial grid
\end{multicols}
\twocolumn
\noindent
zones. We also provide references for the sources of the progenitor models and for publications in which results from the core-collapse simulations were reported before. 

\begin{figure}[t]
        \centering
        \includegraphics[width=\columnwidth]{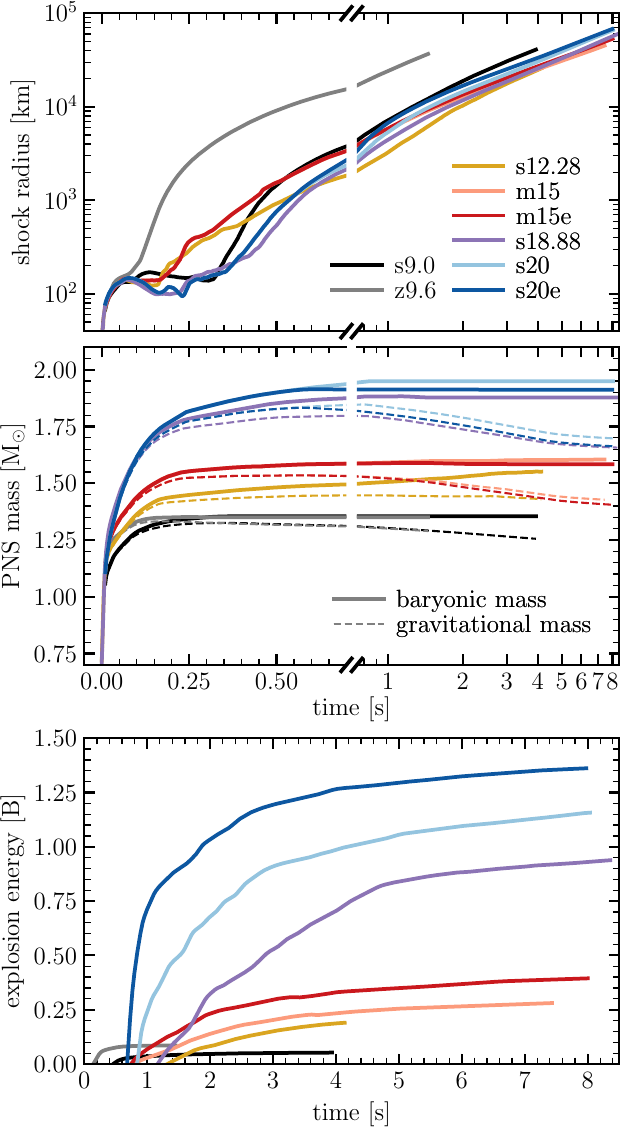}
        \caption{Angle-averaged shock radii ({\em top}), baryonic PNS masses (solid lines) and gravitational PNS masses (dashed lines; {\em middle}), and explosion energies (with overburden energy taken into account; {\em bottom}; $1\,\mathrm{B} = 1\,\mathrm{bethe} = 10^{51}$\,erg) for all successfully exploding models. Note that the time axes in the top and middle panels are different from that of the bottom panel for better visibility}
        \label{fig:rshock_eexpl}
\end{figure}

\begin{figure}[t]
        \centering
        \includegraphics[width=\columnwidth]{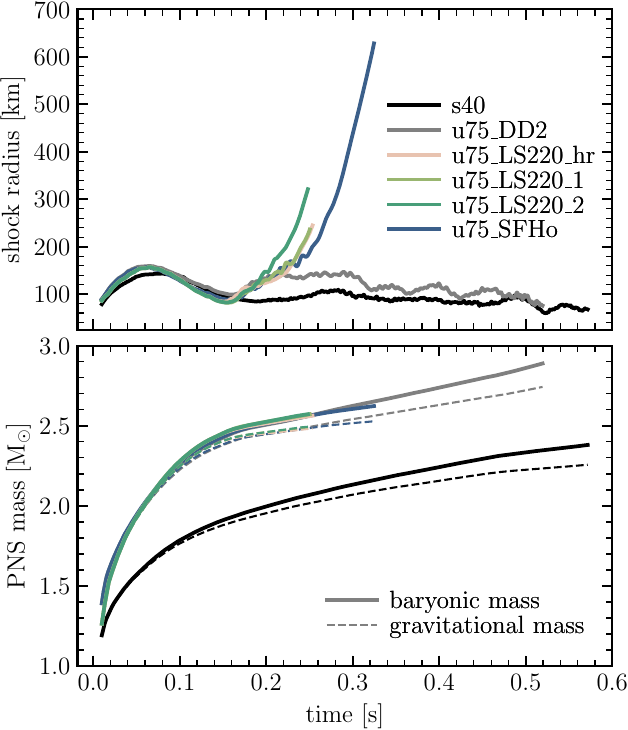}
        \caption{Angle-averaged shock radii ({\em top}) and PNS masses ({\em bottom}) for all BH-forming models. The solid lines show the baryonic PNS masses, the dashed lines the gravitating PNS masses. The lines end when the PNS begins to collapse to a BH}
        \label{fig:rshock_mpns}
\end{figure}

Our compilation includes 3D explosion models with ZAMS masses of 9.0, 9.6, 12.28, 15 (rotating), 18.88, and 20\,M$_\odot$ as well as non-exploding, BH-forming 40\,M$_\odot$ and 75\,M$_\odot$ models. The \textsc{Vertex} neutrino transport was applied for different periods of time $t_\mathrm{f}^\nu$ and the hydrodynamic modeling was subsequently continued with the \textsc{Nemesis} scheme until time $t_\mathrm{f}$ (Table~\ref{tab:hydro_kicks}). The shock radii, PNS masses, and explosion energies $E_\mathrm{exp}^\mathrm{OB-}$ are displayed as functions of time in Figure~\ref{fig:rshock_eexpl}. These energies take into account the overburden of stellar material ahead of the SN shock (i.e., the binding energies of the stellar layers exterior to the SN shock are subtracted). At the end of all of the simulations the explosion energies have nearly reached their final values, because their growth rates have become rather small. Correspondingly, the differences between $E_\mathrm{exp}^\mathrm{OB-}$ and the diagnostic energies $E_\mathrm{exp}^\mathrm{diag}$ (defined as the total internal plus kinetic plus gravitational energy of all postshock matter with a positive value of this sum) are minor when the computational runs were stopped (Table~\ref{tab:hydro_kicks}).

For the 12.28\,M$_\odot$ and 18.88\,M$_\odot$ stars a final evolution period prior to core collapse (seven minutes for the latter and one hour for the former) were simulated in 3D in order to self-consistently generate asymmetric initial conditions of the density, velocity field, and chemical composition for the subsequent SN simulations \citep[see][]{Yadav+2020}. Such pre-collapse asymmetries turned out to be crucial for obtaining explosions in these two cases \citep[see][for the 18.88\,M$_\odot$ model]{Bollig+2021}. The 9.0\,M$_\odot$ \citep{Melson+2020,Stockinger+2020} and 9.6\,M$_\odot$ \citep{Melson+2015a,Stockinger+2020} models explode quite readily (e.g., only $\sim$100\,ms after core bounce in the case of the 9.6\,M$_\odot$ model)
because of their steep temporal decline of the mass infall rate to the stalled shock, whereas the explosion of the 15\,M$_\odot$ progenitor was supported by fairly rapid rotation \citep{Summa+2018} and that of the 20\,M$_\odot$ progenitor by a minor change in the neutral-current neutrino-nucleon scattering cross sections \citep{Melson+2015b}. 

Table~\ref{tab:models} lists two variants of both the 15\,M$_\odot$ and the 20\,M$_\odot$ simulations. Each pair is based on the same 3D neutrino-hydrodynamics simulations with the \textsc{Vertex} transport code applied until $t_\mathrm{f}^\nu$. The main difference between the two cases is a somewhat different treatment of the postshock heating and cooling around the PNS with the \textsc{Nemesis} scheme during the long-time simulations from $t_\mathrm{f}^\nu$ until $t_\mathrm{f}$. These variations were supposed to test, within plausible limits, the sensitivity of the final SN and NS properties on possible uncertainties of the physics of neutrino heating and cooling during this long-term evolution \citep{Kresse2023}. In the \textsc{Nemesis} calculation of the model m15e we assumed constant neutrino luminosities and mean energies instead of using the time-dependent results of a 1D PNS cooling simulation in the m15 model. In the model s20e we adopted the neutrino cooling profiles of the 1D PNS simulation between neutrinosphere and gain radius instead of calculating the neutrino cooling from the local conditions obtained in the 3D simulation of model s20 with its longer-lasting PNS accretion. Both models with the letter ``e'' appended to their names explode a bit more energetically because of higher neutrino energy deposition or less neutrino energy loss outside the PNS, but the differences are moderate in consideration of the fairly radical changes.

Several 3D core-collapse simulations of the two very massive 40 and 75\,M$_\odot$ progenitors were performed by A.~Summa with different nuclear EoSs and varied angular resolution. The evolution was followed up to the moment when the PNS became gravitationally unstable and started to form a BH (times $t_\mathrm{f}^\nu$ in Table~\ref{tab:neutrino_kicks}). Neutrino emission characteristics of two of these simulations were analyzed by \citet{Walk+2020}. In the calculation of the 40\,M$_\odot$ progenitor, which employed the LS220 EoS, the shock recedes after it has reached a maximum radius of about 150\,km between 50\,ms and 100\,ms post bounce (Figure~\ref{fig:rshock_mpns}). A similar behavior is obtained for the 75\,M$_\odot$ progenitor when the DD2 EoS is used. In contrast, with the LS220 and SFHo EoSs the average shock evolves through a minimum radius at $\sim$150\,ms after bounce but afterwards it exhibits a phase of rapid neutrino-driven expansion until the moment when the PNS begins to collapse to a BH and the simulations were stopped. Since the shock radius hardly exceeds 600\,km even in the most extreme case (Figure~\ref{fig:rshock_mpns}), all the matter heated by neutrinos is likely to fall back to the BH and successful 
explosions with SN-like mass ejection cannot be expected to happen in these 75\,M$_\odot$ models \citep[for core-collapse simulations with BH formation and concomitant shock expansion in different kinds of
progenitor stars, see, e.g.,][]{Kuroda+2018,Chan+2018,Ott+2018,Chan+2020,Pan+2021,Powell+2021,Rahman+2022,Burrows+2023}. 
In their 3D calculations of the 40\,M$_\odot$ progenitor, \citet{Burrows+2023} obtained an explosion despite BH formation and significant fallback (fallback SN), in contrast to our failed SN for the same progenitor. They attribute this difference to insufficient angular resolution in our CCSN run. This argument, however, is not likely to be the correct explanation, because our 3D simulations of the 75\,M$_\odot$ models with the LS220 EoS do not reveal any significant resolution dependence when runs with 5$^\circ$ and much higher angular resolution using a static mesh refinement (SMR; model u75\_LS220\_hr) are compared (Table~\ref{tab:models}).\footnote{Our model u75\_LS220\_hr with SMR grid has an angular resolution of 1~degree in the gain layer compared to about 1.4~degrees used by \citet{Burrows+2023} in their 40\,M$_\odot$ simulation. Our simulation of the 40\,M$_\odot$ model employed the LS220 EoS, whereas \citet{Burrows+2023} applied SFHo. Although SFHo permits a slightly longer life time of the PNS, which could favor a stronger explosion \citep[as shown by][]{Powell+2021}, the difference is small and we usually find earlier and easier explosions with LS220 \citep[see][for 18.88\,$M_\odot$ simulations]{Janka+2023}. We therefore hypothesize that the differences of the 40\,M$_\odot$ models might be connected to higher neutrino heating rates in the simulation by \citet{Burrows+2023}. This possibility is supported by the differences between our models u75\_LS220\_1 and u75\_LS220\_2, where the latter model shows stronger shock expansion due to slightly enhanced neutrino heating in the postshock layer. But conclusive results would require a detailed comparison using the same EoS and resolution.} Our set of BH forming models also includes two realizations of the 75\,M$_\odot$ simulations with LS220 and 5$^\circ$ resolution that show slightly different shock expansion due to minor differences in the postshock neutrino heating.


\subsection{Evaluation for PNS kicks}
\label{sec:PNSkickeval}

NSs receive natal kicks connected to their formation in stellar core-collapse events by mass ejection connected to a SN explosion as well as anisotropic neutrino emission. Anisotropic loss of energy in gravitational waves plays a negligible role for kicking the NSs, because the energy carried away by these waves is several orders of magnitude less than the energy lost in neutrinos, which in turn amounts to up to 10\% of the rest-mass energy of the compact remnant. 

\subsubsection{Kick evaluation including the PNS mass evolution}
\label{sec:massdepanalysis}

In order to account for the changes of the PNS's mass associated with the neutrino emission, we evaluate the recoil acceleration of the PNS directly from the forces leading to a change of the PNS momentum:
\begin{equation}\label{eq:pnsdot2}
        \vec{\dot{p}}_\mathrm{NS}
        = \underbrace{-\left(\vec{\dot{p}}_\mathrm{gas} + \vec{\dot{p}}_{\nu}^\mathrm{tot}\right)}_{\substack{\\\text{accelerating forces}
        \\ \text{(hydro and neutrino)}}}
        -\underbrace{c^{-2}L_{\nu}^\mathrm{tot}\,\vec{v}_\mathrm{NS}}_{\substack{\text{momentum change} \\ \text{by neutrino emission} \\ \text{in PNS rest frame}}} \!\!\!\!\!.
\end{equation}
In this equation, $L_{\nu}^\mathrm{tot}$ is the total neutrino luminosity (see Equation~\ref{eq:nulum}), $\vec{v}_\mathrm{NS}$ the PNS velocity, and $c$ the speed of light. $\vec{\dot{p}}_\mathrm{gas}$ is the time derivative of the volume-integrated linear momentum,\footnote{In practice, we numerically differentiate the linear gas momentum of the ejecta for Equation~(\ref{eq:pnsdot2}) with a time-sampling interval of $\sim$1\,ms.} 
\begin{equation}
 \vec{p}_\mathrm{gas}(t) = \int_{r> R_\mathrm{NS}}\mathrm{d}V\,\rho\,\vec{v} \,,  
 \label{eq:pgas}
\end{equation}
of the stellar gas exterior to the PNS surface radius at $R_\mathrm{NS}$ (typically defined at a density of $10^{11}$\,g\,cm$^{-3}$), with $\rho$ and $\vec{v}$ being the gas density and velocity, respectively. The time derivative of the total linear momentum of the escaping neutrinos, $\vec{\dot{p}}_{\nu}^\mathrm{tot}$, is evaluated in the rest frame of the PNS, where the neutrino transport of the CCSN models was computed. It can be written as a function of the energy-flux densities $\vec{F}_{\nu_i}$ of all neutrino species $\nu_i$ ($= \nu_e, \bar\nu_e, \nu_\mu, \bar\nu_\mu, \nu_\tau, \bar\nu_\tau$) as 
\begin{eqnarray}
\vec{\dot{p}}_{\nu}^\mathrm{tot}(R,t)
&=& \sum_{\nu_i}\, \vec{\dot{p}}_{\nu_i}(R,t) \label{eq:pdotnui} \\
&=& \frac{R^2}{c} \oint_{r = R}\mathrm{d}\Omega\,\sum_{\nu_i} (\vec{F}_{\nu_i}\otimes \vec{\hat{r}})\cdot \vec{\hat{r}} \nonumber \\ 
&=& \frac{R^2}{c} \oint_{r = R}\mathrm{d}\Omega\,\sum_{\nu_i} \vec{F}_{\nu_i}(\vec{r},t) \nonumber \\
&\cong& \frac{R^2}{c} \oint_{r = R}\mathrm{d}\Omega\,\sum_{\nu_i} F_{\nu_i}^r(\vec{r},t) \,\vec{\hat{r}} \,, 
\label{eq:pdotnu}
\end{eqnarray}
where $\mathrm{d}\Omega$ is a solid angle element on the sphere of radius $R$, $\vec{\hat{r}}$ the unit vector locally perpendicular to the sphere, and the last line holds as an equality in our analysis, because our RbR+ transport treatment provides only radial components $F_{\nu_i}^r$ of the neutrino energy fluxes. Since the surface integral of Equation~(\ref{eq:pdotnu}) has to be evaluated in the free-streaming limit, typically at $R = 400$\,km, i.e., far outside the neutrinospheres \citep{Stockinger+2020}, any nonradial flux components are negligibly small anyway.\footnote{Here the neutrino fluxes are those measured in the laboratory frame of a distant observer, not those in the comoving frame of the stellar fluid. Note that the PNS is fixed at the coordinate center of the computational grid for numerical reasons, and therefore, the rest frame of the PNS is identical with the observer frame.} Accordingly, the energy loss rate of the PNS by its neutrino emission (i.e., the direction-averaged total neutrino luminosity) is given by
\begin{eqnarray}
    L_\nu^\mathrm{tot}(R,t) &=& \sum_{\nu_i} \, L_{\nu_i}(R,t) \label{eq:nuilum} \\
    &=& R^2 \oint_{r = R}\mathrm{d}\Omega\,\sum_{\nu_i} F_{\nu_i}^r(\vec{r},t) \,.
    \label{eq:nulum}
\end{eqnarray}

We stress that our evaluation of the NS kick is based on a consideration of the total linear momentum carried away by the escaping neutrinos and by gas exterior to the PNS. Due to momentum conservation, the sum of these two momentum terms must be compensated by the negative momentum received by the NS. For this evaluation it is crucial to compute the neutrino-induced kick {\em in the free-streaming limit} in order to include the effects of neutrino emission, absorption, and scattering in material outside the neutrinosphere \citep{Stockinger+2020}. If, for example, the neutrino kick is evaluated at the neutrinosphere, one misses the contributions from anisotropic neutrino emission, absorption, and scattering that occur in massive accretion flows around the PNS. Therefore, placing the evaluation radius at different locations can, on the one hand, significantly alter the results for the neutrino-induced PNS kick (as visible by comparing numbers in \citealt{Coleman+2022} and \citealt{Burrows+2023b}). On the other hand, the momentum deposited by neutrino interactions in the gas exterior to the neutrinosphere (which for all neutrino species is close to our PNS radius $R_\mathrm{NS}$) changes the gas momentum integrated by Equation~(\ref{eq:pgas}) in the volume $r > R_\mathrm{NS}$. This implies a corresponding change of the NS momentum that is inferred from the total gas momentum (according to Equation~\ref{eq:pnsdot2}). However, the momentum transferred to the gas in the vicinity of the PNS is subtracted from the momentum carried by the neutrinos, or in other words, the corresponding change of the NS's hydrodynamic kick is compensated by a negative change of the neutrino-induced kick of the PNS, as correctly concluded when the total neutrino momentum is measured at a very large distance from the NS. In contrast, shifting the evaluation radius for the neutrino emission asymmetry to the neutrinosphere instead of placing it in the free streaming regime can imply a double counting of momentum contributions in the neutrino and hydro sectors and can thus lead to an overestimation of the total (neutrino plus hydro) NS kick. We will come back to this point in our later discussion.

The last term on the rhs of Equation~(\ref{eq:pnsdot2}) accounts for the change of the PNS's momentum due to the fact that neutrinos leaving the moving PNS carry away linear momentum in the observer frame, because the term $-\,c^{-2}L_{\nu}^\mathrm{tot}\vec{v}_\mathrm{NS}$ denotes the rate of momentum loss connected to the mass equivalent of the rate of neutrino energy loss by the PNS. As we will see in the following, this term will ensure that the PNS's velocity does not change in the case external forces on the PNS are absent, the baryonic mass of the PNS, $M_\mathrm{NS,b}$, is constant, but the PNS radiates neutrinos isotropically in its rest frame and accordingly its gravitational mass, $M_\mathrm{NS,g}$, changes with time. 

Generally, the temporal variation of the PNS's momentum can be caused by changes of the PNS's velocity and gravitational mass:   
\begin{equation}\label{eq:pnsdot1}
\begin{split}
        \vec{\dot{p}}_\mathrm{NS} =&~ M_\mathrm{NS,g} \vec{\dot{v}}_\mathrm{NS} + \dot{M}_\mathrm{NS,g} \vec{v}_\mathrm{NS}\\ 
        =&~ M_\mathrm{NS,g} \vec{\dot{v}}_\mathrm{NS} + \left(\dot{M}_\mathrm{NS,b} - c^{-2}L_{\nu}^\mathrm{tot}\right) \vec{v}_\mathrm{NS} \,,
\end{split}
\end{equation}
where, in the second line, we have introduced the reduction of the gravitational mass by neutrino emission, and the time derivative of the baryonic mass indicates possible mass gain or loss by the PNS through accretion or neutrino-driven outflows, respectively. The baryonic mass of the PNS, $M_\mathrm{NS,b}(t)$, is obtained by integrating the baryon number density, multiplied with the average nucleon rest mass, over the PNS volume interior to the PNS radius $R_\mathrm{NS}(t)$ at a density of $10^{11}$\,g\,cm$^{-3}$. The time-dependent gravitational PNS mass, $M_\mathrm{NS,g}(t)$, is related to this mass by subtracting the mass equivalent corresponding to the energy loss via neutrino emission:
\begin{equation}
 M_\mathrm{NS,g}(t) = M_\mathrm{NS,b}(t) - \frac{1}{c^2} \int_{t_\mathrm{i}}^t \mathrm{d}t'\,L_\nu^\mathrm{tot}(t') \,,
 \label{eq:mpnsg}
\end{equation}
taking as starting point $t_\mathrm{i}$ of the time integration the beginning of the \textsc{Prometheus-Vertex} simulations to include the (small) neutrino energy loss during the collapse of the degenerate stellar core before core bounce (i.e., $t = 0$). Now combining Equations~(\ref{eq:pnsdot2}) and (\ref{eq:pnsdot1}), we get for the PNS acceleration:
\begin{equation}
   \vec{\dot{v}}_\mathrm{NS}(t) = -\,\frac{\vec{\dot{p}}_\mathrm{gas}(t) + \vec{\dot{p}}_{\nu}^\mathrm{tot}(t)}{M_\mathrm{NS,g}(t)} - \frac{\dot{M}_\mathrm{NS,b}(t)\,\vec{v}_\mathrm{NS}(t)}{M_\mathrm{NS,g}(t)} \,.
   \label{eq:vnsdot}
\end{equation}
From this equation we can verify what we mentioned above: If no hydrodynamic forces act on the PNS, i.e., $\vec{\dot{p}}_\mathrm{gas} = 0$, and the PNS radiates neutrinos isotropically in its rest frame, i.e., $\vec{\dot{p}}_{\nu}^\mathrm{tot} = 0$, and the baryonic PNS mass is constant, i.e., $\dot{M}_\mathrm{NS,b} = 0$, the PNS is not accelerated, $\vec{\dot{v}}_\mathrm{NS} = 0$, and the PNS's velocity $\vec{v}_\mathrm{NS}$ is therefore constant in time, too.

In practice, we separate this equation into expressions for the accelerations connected to the hydrodynamic kick and the neutrino-induced kick. The recoil acceleration due to the hydrodynamic PNS kick is given by
\begin{equation}\label{eq:vnsdot_hyd}
\begin{split}
        \vec{\dot{v}}_\mathrm{NS}^\mathrm{hyd}(t) =&~ -\frac{\vec{\dot{p}}_\mathrm{gas}(t)}{M_\mathrm{NS,g}(t)} - \frac{\dot{M}_\mathrm{NS,b}(t)\,\vec{v}_\mathrm{NS}^\mathrm{hyd}(t)}{M_\mathrm{NS,g}(t)}\\
        \simeq&~ -\frac{\vec{\dot{p}}_\mathrm{gas}(t)}{M_\mathrm{NS,g}(t)} \,,
\end{split}
\end{equation}
and, in analogy, the recoil acceleration leading to the neutrino-induced PNS kick velocity is given by
\begin{equation}\label{eq:vnsdot_nu}
\begin{split}
        \vec{\dot{v}}_\mathrm{NS}^{\nu}(t) =&~ -\frac{\vec{\dot{p}}_{\nu}^\mathrm{tot}(t)}{M_\mathrm{NS,g}(t)} - \frac{\dot{M}_\mathrm{NS,b}(t)\,\vec{v}_\mathrm{NS}^\nu(t)}{M_\mathrm{NS,g}(t)}\\
        \simeq&~ -\frac{\vec{\dot{p}}_{\nu}^\mathrm{tot}(t)}{M_\mathrm{NS,g}(t)} \,.
\end{split}
\end{equation}
The approximate expressions in the second lines of Equations~\eqref{eq:vnsdot_hyd} and \eqref{eq:vnsdot_nu} neglect the term $\propto$$\dot{M}_\mathrm{NS,b}\vec{v}_\mathrm{NS}$. This is well justified most of the time, because during the early postbounce evolution, during which the baryonic mass of the PNS grows rapidly by accretion, $\vec{v}_\mathrm{NS}$ is still small, and at late postbounce times, i.e., later than $\sim$1\,s after bounce, the baryonic PNS mass is effectively constant. In our standard procedure for the numerical post-processing of the core-collapse simulations, the hydrodynamic kicks and neutrino-induced kicks of the PNSs are determined by time integration of the accelerations in Equations~\eqref{eq:vnsdot_hyd} and \eqref{eq:vnsdot_nu}, neglecting the terms $\propto$$\dot{M}_\mathrm{NS,b}\vec{v}_\mathrm{NS}$: 
\begin{eqnarray}
   \vec{v}_\mathrm{NS}^\mathrm{hyd}(t) &=& \int_0^t\mathrm{d}t'\,\vec{\dot{v}}_\mathrm{NS}^\mathrm{hyd}(t') \,, 
   \label{eq:vhyd} \\
   \vec{v}_\mathrm{NS}^{\nu}(t) &=& \int_0^t\mathrm{d}t'\,\vec{\dot{v}}_\mathrm{NS}^{\nu}(t') \,, \label{eq:vneut}
\end{eqnarray}
with time zero being typically at core bounce. Correspondingly, we obtain the total PNS kick velocity as function of
time, $\vec{v}_\mathrm{NS}^\mathrm{tot}(t)$ as the vector sum of $\vec{v}_\mathrm{NS}^\mathrm{hyd}(t)$ and $\vec{v}_\mathrm{NS}^{\nu}(t)$:
\begin{equation}
    \vec{v}_\mathrm{NS}^\mathrm{tot}(t) = \vec{v}_\mathrm{NS}^\mathrm{hyd}(t) + \vec{v}_\mathrm{NS}^{\nu}(t) \,.
    \label{eq:vtot}
\end{equation}

Solving Equations~\eqref{eq:vnsdot_hyd} and \eqref{eq:vnsdot_nu} including the terms $\propto$$\dot{M}_\mathrm{NS,b}\vec{v}_\mathrm{NS}$ for the hydrodynamic and neutrino-induced PNS kick velocities $\vec{v}_\mathrm{NS}^\mathrm{hyd}(t)$ and $\vec{v}_\mathrm{NS}^{\nu}(t)$, e.g., with a Runge-Kutta integrator, yields results that differ from the ones obtained with our standard, slightly more approximate evaluation by less than $\sim$(2--3)\% in all models and at any time during their evolution. 
Such small deviations are on the same level as the numerical inaccuracies connected to the discretized computations of the time derivatives and integrals of the strongly fluctuating force vectors $\vec{\dot{p}}_\mathrm{gas}(t)$ and $\vec{\dot{p}}_{\nu}^\mathrm{tot}(t)$. In the figures and tables of this paper we will therefore present results where the terms $\propto$$\dot{M}_\mathrm{NS,b}\vec{v}_\mathrm{NS}$ are neglected in Equations~\eqref{eq:vnsdot_hyd} and \eqref{eq:vnsdot_nu}. In the following, we will hint at this minor limitation of accuracy by denoting the total NS kick velocities with $\vec{v}_\mathrm{NS}^\mathrm{tot}$ instead of $\vec{v}_\mathrm{NS}$.

\subsubsection{Approximate estimates}
\label{sec:approxestimates}

As said before, the time integrals of Equations~(\ref{eq:vhyd}) and~(\ref{eq:vneut}) permit us to include the temporal variation of the inertial/gravitational PNS mass in the kick calculations. Alternatively, hydrodynamic and neutrino kick velocities can be estimated by considering global momentum conservation in the observer frame of reference, balancing the NS momentum, $\vec{p}_\mathrm{NS}$, and the gas plus neutrino momentum, 
$\vec{p}_\mathrm{gas} + \vec{p}_\nu^\mathrm{tot}$, as done by \citet{Scheck+2006,Wongwathanarat+2013,Gessner+2018,Stockinger+2020}, and \citet{Bollig+2021}. This approach is suitable to obtain estimates of the kick velocities in very good agreement with those from the analysis of Section~\ref{sec:massdepanalysis} at early postbounce times when the baryonic and the gravitational PNS mass are still very similar. Minor differences between the results from Equation~(\ref{eq:vtot}) and Equation~(\ref{eq:vnsmc2}) below during this phase are mostly connected to inaccuracies of the numerical integration, because the direction and absolute value of the PNS acceleration still fluctuate wildly due to the violently turbulent gas flows in the postshock layer. At later times, when the PNS acceleration has developed a preferred direction, both calculations disagree up to about 10\%, corresponding to the typical relative difference of gravitational and baryonic NS mass.

In Figures~\ref{fig:models9}--\ref{fig:models20} and \ref{fig:models40}--\ref{fig:modelsu75deg5sfho} in Appendices~\ref{secA1} and \ref{secA:bh}, which provide detailed information for most of our core-collapse simulations with explosions or BH formation, we also show for comparison the evolution of the hydrodynamic kick and total (i.e., hydrodynamic plus neutrino) kick computed approximately from the total gas momentum according to momentum conservation:
\begin{eqnarray}
    \vec{v}_\mathrm{NS}^\mathrm{hyd,mc}(t) &=& -\,\frac{\vec{p}_\mathrm{gas}(t)}{M_\mathrm{NS,b}(t)} \,,\label{eq:vnsmc1}  \\
    \vec{v}_\mathrm{NS}^\mathrm{tot,mc}(t) &=& -\,\frac{\vec{p}_\mathrm{gas}(t) + \vec{p}_{\nu}^\mathrm{tot}(t)}{M_\mathrm{NS,b}(t)} \,. \label{eq:vnsmc2}
\end{eqnarray}
Dividing by the baryonic PNS mass in Equations~(\ref{eq:vnsmc1}) and~(\ref{eq:vnsmc2}) ensures that the velocity of the PNS does not change in the absence of accelerating forces when its gravitational mass decreases due to isotropic neutrino emission in the PNS's rest frame. However, these estimates usually lead to lower bounds on the actual kicks and they work best in the extreme case that the accelerating forces act very fast and early when the mass decrement of the PNS by neutrino emission is still small. In contrast, dividing the total momentum by the gravitational mass, $\vec{v}_\mathrm{NS}^\mathrm{upper}(t) = - (\vec{p}_\mathrm{gas}(t) + \vec{p}_{\nu}^\mathrm{tot}(t))/M_\mathrm{NS,g}(t)$, tends to yield an upper bound of the NS kick; this approximation would work well if the NS acceleration happened mainly after the PNS has radiated most of its gravitational binding energy in neutrinos and therefore the inertial (gravitational) mass of the NS has converged to its final value. The true kick velocities evaluated via time integration of Equations~(\ref{eq:vnsdot}), (\ref{eq:vnsdot_hyd}), and (\ref{eq:vnsdot_nu}) lie between these lower and upper limits.
 
\subsubsection{Forces of the hydro kicks}
\label{sec:forces}

According to \citet{Scheck+2006}, the hydrodynamic force kicking the PNS can be separated into different contributions corresponding to the momentum transfer to the PNS by (a) the pressure of surrounding gas, (b) momentum carried by the accretion flows or outflows, and (c) the action of gravitational forces between the PNS and the matter of the collapsing and exploding star. Integrating the Euler equation over the relevant volume exterior to a radius $R_0$ around the PNS, one obtains:
\begin{eqnarray}
    \vec{\dot{p}}_\mathrm{hyd} &=& -\,\vec{\dot{p}}_\mathrm{gas} = \vec{\dot{p}}_\mathrm{pres} + \vec{\dot{p}}_\mathrm{mom} + \vec{\dot{p}}_\mathrm{grav} \nonumber \\
    &=& -\oint_{r = R_0}\mathrm{d}S\,P\,\vec{\hat{r}} - \oint_{r = R_0}\mathrm{d}S\, \rho\,(\vec{v}\otimes\vec{v})\cdot \vec{\hat{r}} \nonumber \\
    &\phantom{=}& + \int_{r>R_0}\mathrm{d}m\, GM_\mathrm{NS}\,\frac{\vec{\hat{r}}}{r^2}\,,
\label{eq:hydroforces}
\end{eqnarray}
where $\mathrm{d}S = R_0^2\mathrm{d}\Omega$ is the surface element with solid angle $\mathrm{d}\Omega$ on the sphere of radius $R_0$, $P = P(|\vec{r}| = R_0,t)$ is the gas pressure on the sphere, and the term $\rho(\vec{v}\otimes\vec{v})\cdot \vec{\hat{r}} = \rho\vec{v}v_r$ denotes the momentum flux of gas flowing radially inward or outward through this sphere. By default, we use $R_0 = 2\,R_\mathrm{NS}$,
where $R_\mathrm{NS}$ is the PNS radius corresponding to an angle-averaged density of $10^{11}$\,g\,cm$^{-3}$.
This choice is justified by the fact that in all cases this radius of evaluation yields excellent agreement of the time-integrated vector sum of the three force contributions, $\int_0^t \vec{\dot{p}}_\mathrm{hyd} \mathrm{d}t'$, and the negative of the total gas momentum, $-\,\vec{p}_\mathrm{gas}$, of the ejecta according to Equation~(\ref{eq:pgas}). Minor discrepancies occur only in phases when the absolute value of the NS kick velocity is very small and thus sensitive to low-amplitude variations and minor numerical discretization errors. The agreement becomes considerably worse in some cases when $R_0$ is moved to other locations between $\sim$$R_\mathrm{NS}$ and $\sim$$3\,R_\mathrm{NS}$. Moreover, inappropriate choices of $R_0$ also cause quantitative changes of the three terms added up in Equation~(\ref{eq:hydroforces}). This may lead to incorrect conclusions drawn from the entire analysis. In particular, the magnitudes of the pressure and momentum flux terms are quite sensitive to the radius where they are evaluated. We will come back to the critical role of the force-evaluation radius in the context of neutrino-induced kicks in Section~\ref{sec:noteeval}.

Any significant acceleration of the PNS by pressure forces and momentum transfer through gas flows requires sufficiently dense gas in the close vicinity of the PNS. Therefore, these forces are relevant only during the early postbounce phase and shortly after the onset of the SN explosion. During this period, the hydrodynamic PNS kick just starts to develop and all of the three forces contributing in Equation~(\ref{eq:hydroforces}) have similar magnitudes and partly cancel each other due to opposite signs \citep{Scheck+2006}. Gravitational forces, in contrast, are important during all of the evolution. In particular, because of its long-range and non-saturating nature, gravity can play a role still at late time after the beginning of the explosion \citep[see the long-time CCSN simulations discussed by][]{Janka+2022}. Most of the hydrodynamic PNS kick is therefore transferred to the PNS once the explosion has attained its final asymmetry and a preferred kick direction has developed, and a major fraction of the asymptotic value of the PNS velocity due to the hydrodynamic kick can be attributed to the action of gravitational forces between the PNS and the matter ejected by the SN. We will again verify this fact in the simulations discussed in the next section. For this reason, the hydrodynamic PNS kick has been termed ``gravitational tugboat'' effect \citep{Wongwathanarat+2013}.\footnote{This naming has been criticized as inappropriate in some recent papers \citep{Coleman+2022,Burrows+2023b}. We will demonstrate and argue in Section~\ref{sec:hydrokicks} that the term ``gravitational tugboat mechanism/effect'' is in fact well justified.}

\subsubsection{Asymmetry parameters}
\label{sec:asymmetryparameters}

The hydrodynamic and neutrino kicks of the PNSs can be related to the asymmetry of the SN ejecta and neutrino emission by dimensionless asymmetry parameters. 

For the hydrodynamic kicks one sets the total linear momentum of the gas exterior to the PNS, $\vec{p}_\mathrm{gas}(t)$ (Equation~\ref{eq:pgas}), in relation to the scalar quantity
\begin{equation}
  p_\mathrm{ej}(t) = \int_{R_\mathrm{NS}}^{R_\mathrm{s}}\mathrm{d}V\,\rho\,|\vec{v}|  \,,
  \label{eq:pej}
\end{equation}
which represents the total momentum of the gas between the PNS surface and the direction-dependent shock radius $R_\mathrm{s}$. With both of these quantities one can define the momentum-asymmetry parameter of the ejecta by
\begin{equation}
    \bar{\alpha}_\mathrm{ej}(t) = \frac{|\vec{p}_\mathrm{gas}(t)|}{p_\mathrm{ej}(t)} 
    \label{eq:alphaej}
\end{equation}
(\citealt{Janka+1994}; \citealt{Scheck+2006}; \citealt{Janka2017}; \citealt{Gessner+2018}; \citealt{Janka+2022}). The values of $\bar{\alpha}_\mathrm{ej}(t)$ for our 3D CCSN simulations at the end of the computational runs at times $t = t_\mathrm{f}$ are listed in Table~\ref{tab:hydro_kicks}.

Using this parameter and Equation~(\ref{eq:vnsmc1}), the kick velocity deduced from momentum conservation can be recovered: $v_\mathrm{NS}^\mathrm{hyd,mc}(t)= |\vec{v}_\mathrm{NS}^\mathrm{hyd,mc}(t)| = \bar{\alpha}_\mathrm{ej}(t)p_\mathrm{ej}(t)M_\mathrm{NS,b}^{-1}(t)$. One should note that the numerical value of $\bar{\alpha}_\mathrm{ej}(t)$ exhibits considerable variation in time with the general tendency to decrease during the long-time evolution of the SN explosions, because $|\vec{p}_\mathrm{gas}|$ converges to a constant value when the ejecta decouple from the PNS hydrodynamically and gravitationally, whereas $p_\mathrm{ej}(t)$ continues to grow until the SN shock breaks out from the stellar surface and the ejecta reach homologous expansion. \citet{Janka2017} derived a simple scaling relation $v_\mathrm{NS}^\mathrm{hyd,mc} \propto \bar{\alpha}_\mathrm{ej} E_\mathrm{exp} M_\mathrm{NS}^{-1}$, which connects the hydrodynamic kick with the explosion energy $E_\mathrm{exp}$ and the ejecta asymmetry $\bar{\alpha}_\mathrm{ej}$. The scaling factor in this relation together with the normalization of $\bar{\alpha}_\mathrm{ej}$ depend on the phase of the SN evolution, which has to be specified in order to compare explosion asymmetries of different SN models.

Also for the neutrino-induced NS kicks, time-dependent parameters $\alpha_\nu(t)$ and $\bar{\alpha}_\nu(r)$ can be introduced, following \citet{Janka+1994}, \citet{Gessner+2018}, and \citet{Stockinger+2020}. These parameters characterize the neutrino emission asymmetry that leads to a net linear momentum associated with the energy carried away by the neutrinos escaping the NS. For the instantaneous value of a quantity defined in this way we can write
\begin{eqnarray}
    \alpha_{\nu_i}(R,t) &=& c\,\frac{|\vec{\dot{p}}_{\nu_i}(R,t)|}{L_{\nu_i}(R,t)} \,, \label{eq:alpnui} \\
    \alpha_{\nu}^\mathrm{tot}(R,t) &=& c\,\frac{|\vec{\dot{p}}_\nu^\mathrm{tot}(R,t)|}{L_\nu^\mathrm{tot}(R,t)} \,,\label{eq:alpnu}
\end{eqnarray}
using the linear momenta associated with the neutrino emission as defined in Equations~(\ref{eq:pdotnui}) and~(\ref{eq:pdotnu}) and the neutrino luminosities as given by Equations~(\ref{eq:nuilum}) and~(\ref{eq:nulum}).
Time averages of these anisotropy parameters of the neutrino emission can be obtained by integrating the terms in the numerators and denominators over the time intervals of interest, for example $[0,t]$ for the evolution between core bounce and moment $t$:
\begin{eqnarray}
    \bar{\alpha}_{\nu_i}(R,t) &=& c\,\frac{|\int_0^t\mathrm{d}t'\,\vec{\dot{p}}_{\nu_i}(R,t')|}{\int_0^t\mathrm{d}t'\,L_{\nu_i}(R,t')} \,, \label{eq:baralpnui} \\
    \bar{\alpha}_{\nu}^\mathrm{tot}(R,t) &=& c\,\frac{|\int_0^t\mathrm{d}t'\,\vec{\dot{p}}_\nu^\mathrm{tot}(R,t')|}{\int_0^t\mathrm{d}t'\,L_\nu^\mathrm{tot}(R,t')} \,.\label{eq:baralpnu}
\end{eqnarray}
The values of these parameters at the end of the 3D neutrino transport simulations with the \textsc{Vertex} code at times $t = t_\mathrm{f}^\nu$ are listed in Table~\ref{tab:neutrino_kicks}.

For the neutrino kicks at times $t > t_\mathrm{f}^\nu$, we can only estimate the possible effects of anisotropic neutrino emission, and for that we employ assumptions about the average values of the asymmetry parameter $\bar{\alpha}_{\nu,\infty}^\mathrm{tot}$ at these late times. Thus extrapolating, we get for the neutrino-induced kick:
\begin{equation}
    \tilde{v}_\mathrm{NS}^{\nu,\infty} = v_\mathrm{NS}^\nu(t_\mathrm{f}^\nu) + 
    \frac{\bar{\alpha}_{\nu,\infty}^\mathrm{tot}}{c} \int_{t_\mathrm{f}^\nu}^\infty \mathrm{d}t'\,
    \frac{L_\nu^\mathrm{tot}(t')}{M_\mathrm{NS,g}(t')} \,.
    \label{eq:vnuinfty}
\end{equation}
The time integration is performed until the PNS has effectively radiated its entire gravitational binding energy in neutrinos, which for all of our models is beyond the time $t_\mathrm{f}$ when the 3D CCSN simulations were stopped. However, data from 1D PNS cooling simulations with \textsc{Prometheus-Vertex} including a mixing-length treatment of PNS convection were available for much longer periods \citep[see][]{Fiorillo+2023} and allowed us to carry out the time integration of Equation~(\ref{eq:vnuinfty}) until the hot PNS has become a cold, neutrino-transparent NS.

For low-mass models, where the anisotropy of the neutrino emission is dominated by the Lepton-number Emission Self-sustained Asymmetry effect~\citep[LESA;][]{Tamborra+2014}, we will find that the values of the instantaneous asymmetry parameter $\alpha_{\nu}^\mathrm{tot}(t)$ decrease towards the end of our \textsc{Vertex} simulations, for which reason we use Equation~(\ref{eq:vnuinfty}) with a value of $0.1\,\bar{\alpha}_\nu^\mathrm{tot}(t_\mathrm{f}^\nu) \lesssim \bar{\alpha}_{\nu,\infty}^\mathrm{tot} \lesssim 0.3\,\bar{\alpha}_\nu^\mathrm{tot}(t_\mathrm{f}^\nu)$. This choice is guided by the time evolution of $\alpha_\nu^\mathrm{tot}(t)$ and/or $\bar{\alpha}_\nu^\mathrm{tot}(t)$ obtained in models s9.0, z9.6, and models s12.28 (in the last model evaluated at $10^{12}$\,g\,cm$^{-3}$), where the emission asymmetry is barely influenced by accretion fluctuations and therefore stable in its direction (see Figures~\ref{fig:vnu_z96}, \ref{fig:vnu_s1228}, \ref{fig:models9}, \ref{fig:modelz96}, and~\ref{fig:models1228}).

For high-mass models, where the anisotropy of the neutrino emission is strongly affected by accretion effects and corresponding directional fluctuations, we assume that the value of the time-averaged asymmetry parameter remains roughly constant during the evolution that follows after the \textsc{Vertex} neutrino transport calculations have ended, i.e., we use $\bar{\alpha}_{\nu,\infty}^\mathrm{tot} \approx \bar{\alpha}_\nu^\mathrm{tot}(t_\mathrm{f}^\nu)$ with $\bar{\alpha}_\nu^\mathrm{tot}(t_\mathrm{f}^\nu)$ being the neutrino emission asymmetry averaged from core bounce until time $t_\mathrm{f}^\nu$ (see Figures~\ref{fig:vnu_s1228}, \ref{fig:vnu_s1888}, \ref{fig:models1228}, and \ref{fig:models1888}).

The neutrino anisotropy parameters can be directly used to express the momentum transfer to the PNS by its asymmetric neutrino emission in terms of the corresponding energy loss in neutrinos:
\begin{eqnarray}
|\vec{\dot{p}}_\nu^\mathrm{tot}(t)| &=& \alpha_{\nu}^\mathrm{tot}(t)\,\frac{L_\nu^\mathrm{tot}(t)}{c}\,,
\label{eq:pdotalpha} \\
|\vec{p}_\nu^\mathrm{tot}(t)| &=& \bar{\alpha}_{\nu}^\mathrm{tot}(t)\,\frac{E_\nu^\mathrm{tot}(t)}{c}\,,
\label{eq:palpha}
\end{eqnarray}
with $E_\nu^\mathrm{tot}(t) = \int_0^t\mathrm{d}t'\,L_\nu^\mathrm{tot}(t')$.\footnote{The small difference of the total neutrino energy loss when starting the time integration at core bounce ($t = 0$) or at the beginning of the CCSN simulation (i.e., at time $t_\mathrm{i}$) does not matter in this context, because the neutrino emission prior to core bounce is essentially isotropic in all cases.} Moreover, ignoring the (relatively small) variation of the gravitational PNS mass in the integrand of Equation~(\ref{eq:vneut}) (with Equation~\ref{eq:vnsdot_nu}) relative to the baryonic PNS mass,\footnote{In general, the effect of the variation of the gravitational mass of the PNS compared to its baryonic mass is a subdominant effect, which affects the neutrino kicks only on the level of about 10 percent.} we can approximately write for the NS's neutrino kick velocity 
\begin{equation}
v_\mathrm{NS}^\nu(t) = |\vec{v}_\mathrm{NS}^\nu(t)| \approx \frac{\bar{\alpha}_{\nu}^\mathrm{tot}(t)}{c}\,\frac{E_\nu^\mathrm{tot}(t)}{M_\mathrm{NS,b}(t)} \,.
\label{eq:vnuapprox}
\end{equation}
Using this, one obtains a simple estimate of the neutrino kick \citep[e.g.,][]{Gessner+2018}:
\begin{equation}
v_\mathrm{NS}^\nu \approx 167\,\mathrm{\frac{km}{s}}\,\,\frac{\bar{\alpha}_{\nu}^\mathrm{tot}}{0.005}\,\frac{E_\nu^\mathrm{tot}}{3\times 10^{53}\,\mathrm{erg}}\left(\frac{M_\mathrm{NS,b}}{1.5\,\mathrm{M}_\odot}\right)^{\!\! -1} \!\!,
\label{eq:vnuestimate}
\end{equation}
which implies that an 0.5\% long-time (net) asymmetry of the neutrino emission, e.g., due to PNS convection and LESA or asymmetric accretion, can lead to a NS kick of typically around 150\,km\,s$^{-1}$.

If the total neutrino luminosity has a dipole asymmetry, i.e.,
\begin{equation}
    \frac{\mathrm{d}L_\nu^\mathrm{tot}}{\mathrm{d}\Omega} = \frac{L_\nu^\mathrm{tot}}{4\pi}\,(1+ a_\mathrm{d}\cos\theta)\,,
    \label{eq:ldipole}
\end{equation}
where $a_\mathrm{d}$ is the normalized amplitude of the dipole component, then the surface integral of 
Equation~(\ref{eq:pdotnu}) yields 
\begin{equation}
 \dot{p}_\nu^\mathrm{tot} = |\vec{\dot{p}}_\nu^\mathrm{tot}| = \alpha_\nu^\mathrm{tot}\,\frac{L_\nu^\mathrm{tot}}{c} 
                          = \frac{a_\mathrm{d}}{3}\,\frac{L_\nu^\mathrm{tot}}{c} \,,
\label{eq:dipolekick}
\end{equation}
which means that $\alpha_\nu^\mathrm{tot} = a_\mathrm{d}/3$. A value of 0.5\% for $\bar{\alpha}_\nu^\mathrm{tot}$ as considered in Equation~(\ref{eq:vnuestimate}) thus implies an emission dipole of 1.5\% of the monopole.

\onecolumn
\addtolength{\tabcolsep}{-2pt}
\begin{table}[tb]
\caption{Hydrodynamic PNS kicks and explosion energies of 3D SN models}
\label{tab:hydro_kicks}
\begin{tabular*}{\textwidth}{@{\extracolsep\fill}lcccccccccccccc}
        \toprule
        Model & $t_\mathrm{f}^{\nu}$ & $t_\mathrm{f}$ & $M_\mathrm{NS,b}^\mathrm{f}$ & $M_\mathrm{NS,g}^\mathrm{f}$ & $M_\mathrm{NS,g}^\infty$ & $v_\mathrm{NS}^\mathrm{hyd}$ & $v_\mathrm{NS}^\nu$ & $v_\mathrm{NS}^\mathrm{tot}$ & $\bar{\alpha}_\mathrm{ej}$ & $\theta_\mathrm{hyd}^{\nu,\mathrm{f}}$ & $\theta_\mathrm{spin}^\mathrm{kick}$ & $P_\mathrm{NS}$ & $E_\mathrm{exp}^\mathrm{diag}$ & $E_\mathrm{exp}^\mathrm{OB-}$ \\[0.5ex]
        & [s] & [s] & [M$_\odot$] & [M$_\odot$] & [M$_\odot$] & [km/s] & [km/s] & [km/s] && [$^\circ$] & [$^\circ$] & [ms] & [B] & [B]\\
        \midrule
        s9.0   & 0.488 & 3.936 & 1.355 & 1.255 & 1.240 &    30.4 &   44.4 &    57.0 & 0.095 &   82.4 &  135.4 & 1,518 & 0.054 & 0.052 \\
        z9.6   & 0.450 & 1.450 & 1.350 & 1.290 & 1.237 &    10.4 &   49.4 &    59.6 & 0.039 &   12.1 &  140.7 & 1,715 & 0.086 & 0.086 \\
        s12.28 & 4.139 & 4.139 & 1.551 & 1.429 & 1.402 &   243.1 &  138.6 &   301.5 & 0.173 &   79.3 &   70.8 & 7.5 & 0.228 & 0.190 \\
        m15    & 0.457 & 7.433 & 1.605 & 1.427 & 1.416 &   627.0 &   54.4 &   621.1 & 0.196 &   98.8 &   80.4 & 2.1 & 0.332 & 0.281 \\
        m15e   & 0.457 & 8.002 & 1.583 & 1.405 & 1.396 &   561.6 &   54.4 &   549.0 & 0.147 &  106.1 &   60.9 & 2.1 & 0.436 & 0.394 \\
        s18.88 & 1.675 & 8.360 & 1.878 & 1.657 & 1.652 &   462.3 &   95.7 &   506.5 & 0.074 &   67.7 &  145.1 & 7.8 & 1.000 & 0.938 \\
        s20    & 0.506 & 8.038 & 1.949 & 1.698 & 1.690 &  1291.5 &   55.0 &  1305.5 & 0.167 &   76.4 &   89.9 & 36.3 & 1.229 & 1.157 \\
        s20e   & 0.506 & 7.980 & 1.912 & 1.663 & 1.655 &   878.5 &   55.0 &   918.8 & 0.100 &   44.2 &   48.7 & 316.1 & 1.431 & 1.361 \\
        \botrule
\end{tabular*}
\footnotetext{\textbf{Note:} $t_\mathrm{f}^{\nu}$ and $t_\mathrm{f}$ are the postbounce times at the end of the full-transport simulation with \textsc{Prometheus-Vertex} and at the end of the \textsc{Nemesis} extension run, respectively; 
$M_\mathrm{NS,b}^\mathrm{f}$ and $M_\mathrm{NS,g}^\mathrm{f}$ are the final baryonic mass and gravitational mass (Equation~\ref{eq:mpnsg}) of the NS at time $t_\mathrm{f}$; 
$M_\mathrm{NS,g}^\infty$ is the corresponding final gravitational NS mass for $t\to\infty$
(assuming $M_\mathrm{NS,b}(t) = \mathrm{const}$ for $t>t_\mathrm{f}$);
$v_\mathrm{NS}^\mathrm{hyd}$ is the magnitude of the hydrodynamic NS kick velocity at $t=t_\mathrm{f}$ (Equation~\ref{eq:vhyd}),
$v_\mathrm{NS}^\nu$ that of the neutrino-induced NS kick velocity at $t=t_\mathrm{f}^{\nu}$ (Equation~\ref{eq:vneut}; see Table~\ref{tab:neutrino_kicks} for estimated upper bounds of the neutrino-induced kick velocity for $t\to\infty$), and 
$v_\mathrm{NS}^\mathrm{tot}$ is the total computed NS kick velocity as the absolute value of the vector sum of $\vec{v}_\mathrm{NS}^\mathrm{hyd}(t_\mathrm{f})$ and $\vec{v}_\mathrm{NS}^\nu(t_\mathrm{f}^{\nu})$; 
$\bar{\alpha}_\mathrm{ej}$ is the anisotropy parameter of the SN ejecta (Equation~\ref{eq:alphaej}) at time $t_\mathrm{f}$;
$\theta_\mathrm{hyd}^{\nu,\mathrm{f}} = \theta(\vec{v}_\mathrm{NS}^\mathrm{hyd}(t_\mathrm{f}),\vec{v}_\mathrm{NS}^{\nu}(t_\mathrm{f}^{\nu}))$ is the angle between the hydrodynamic NS kick vector (at $t=t_\mathrm{f}$) and the neutrino-induced NS kick vector (at $t=t_\mathrm{f}^{\nu}$);
$\theta_\mathrm{spin}^\mathrm{kick}=\theta(\vec{J},\vec{v})$ is the angle between the NS spin vector and its kick vector at $t=t_\mathrm{f}$;
$P_\mathrm{NS}$ is the NS spin period at $t=t_\mathrm{f}$ (evaluated as described in Appendix~D of \citealt{Stockinger+2020} with an assumed final radius of the cold NS of 12\,km; see also \citealt{Kresse2023});
$E_\mathrm{exp}^\mathrm{diag}$ and $E_\mathrm{exp}^\mathrm{OB-}$ are the final diagnostic explosion energy (i.e., the integrated gravitational plus kinetic plus internal energy of all gas between PNS radius and shock radius with a positive value of this total energy) and the final explosion energy with the overburden energy (i.e., the binding energy of stellar matter outside of the SN shock) subtracted, both computed at $t=t_\mathrm{f}$. $1\,\mathrm{B} = 1\,\mathrm{bethe} = 10^{51}$\,erg.}
\end{table}
\addtolength{\tabcolsep}{2pt}
\begin{multicols}{2}

\section{Results}
\label{sec:results}

In the following, we will first discuss our results for hydrodynamic NS kicks due to asymmetric mass ejection and will compare them to the neutrino-induced NS kicks caused by asymmetric neutrino emission. Subsequently, we will elaborate on the neutrino kicks of NSs and BHs in somewhat more detail, based on a time-dependent analysis of all of the discussed models.

\subsection{Hydrodynamic kicks}
\label{sec:hydrokicks}

Table~\ref{tab:hydro_kicks} provides an overview of parameters that characterize the explosions and NS properties (masses, kicks, and spins) of our set of 3D SN models. The explosion energies have nearly reached saturation at the end of our long-term simulations. The values of $E_\mathrm{exp}^{\mathrm{OB}-}$ include the remaining (negative) binding energy of stellar shells exterior to the shock. In all of the listed cases this overburden energy makes less than $\sim$15\% of the ``diagnostic'' explosion energy $E_\mathrm{exp}^\mathrm{diag}$ (for a definition, see the notes of Table~\ref{tab:hydro_kicks}) of the postshock gas. The values of $E_\mathrm{exp}^{\mathrm{OB}-}$ span a wide range from about $5\times 10^{49}$\,erg for the lowest-mass progenitor to close to $1.4\times 10^{51}$\,erg for the 20\,$M_\odot$ models (see also Figure~\ref{fig:rshock_eexpl}). The explosion energy grows monotonically with the progenitor's ZAMS mass and compactness $\xi_{2.5}$ (Equation~\ref{eq:compactness} and Table~\ref{tab:models}) in our limited set of models.

The final gravitational NS masses are between $\sim$1.24\,M$_\odot$ and $\sim$1.70\,M$_\odot$, depending on the progenitor and, as a secondary effect, on the explosion energy. The middle panel of Figure~\ref{fig:rshock_eexpl} shows the time evolution of both the NS baryonic masses and the corresponding gravitational masses. In our set of models, lower-mass progenitors give birth to lower-mass NSs, but this dependence can also be non-monotonic because of a non-monotonic relation between the core structure of pre-collapse stars and the ZAMS mass \citep[][]{Sukhbold+2014,Sukhbold+2018}. More energetic neutrino-driven explosions typically correlate with higher NS masses because of the higher neutrino luminosities and more matter being heated by neutrinos \citep{Nakamura+2015,Burrows+2021}, but more energetic explosions for the same progenitor lead to slightly lower NS masses (Table~\ref{tab:hydro_kicks}).

\begin{figure*}[ht!]
        \centering
        \includegraphics[width=0.8\textwidth]{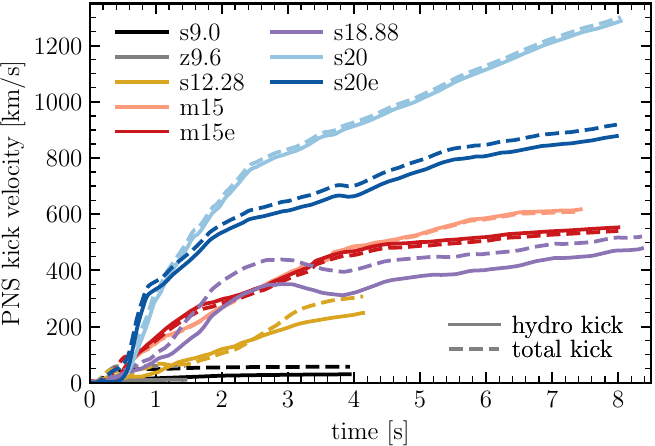}
        \caption{Evolution of the PNS kick velocities in all successfully exploding 3D models. The solid lines show the hydrodynamic PNS kick velocities, the dashed lines display the total PNS kick velocities as the vector sums of hydrodynamic plus neutrino-induced kick velocities. Note that the neutrino kick velocity vector is fixed here to the results obtained at the end of the \textsc{Vertex} neutrino transport calculations (at $t_\mathrm{f}^\nu$; Tables~\ref{tab:hydro_kicks} and~\ref{tab:neutrino_kicks})}
        \label{fig:vhyd_vtot}
\end{figure*}

\begin{figure*}[t!]
        \centering
        \includegraphics[width=0.9\textwidth]{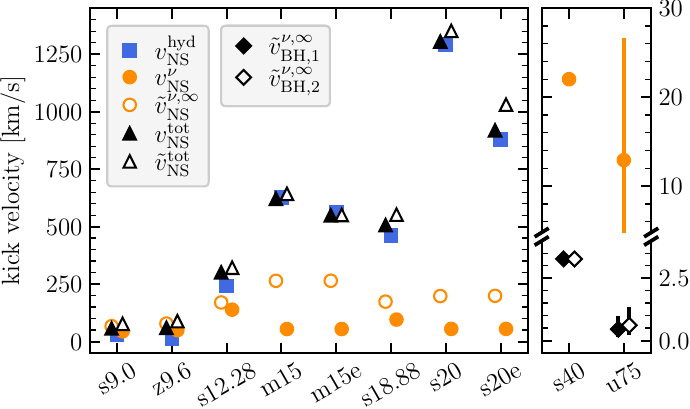}
        \caption{Overview of the main results for NS kicks ({\em left panel}) and BH kicks ({\em right panel}). The filled orange circles mark the neutrino-induced kick velocities of the PNSs at times $t_\mathrm{f}^\nu$ (see Tables~\ref{tab:hydro_kicks} and~\ref{tab:neutrino_kicks}), i.e., when the 3D simulations with \textsc{Vertex} neutrino transport ended or when the PNS collapsed to a BH. {\em Left panel:} The blue squares denote the hydrodynamic NS kicks at later times $t_\mathrm{f}\ge t_\mathrm{f}^\nu$ (Table~\ref{tab:hydro_kicks}) when the 3D \textsc{Prometheus-Vertex/Nemesis} simulations were stopped. The filled black triangles represent the total kick velocities, i.e., $|\vec{v}_\mathrm{NS}^\mathrm{tot}| = |\vec{v}_\mathrm{NS}^\mathrm{hyd}(t_\mathrm{f}) + \vec{v}_\mathrm{NS}^\nu(t_\mathrm{f}^\nu)|$, at time $t_\mathrm{f}$ when the 3D SN simulations were stopped. Note that in some models the velocities still grow with significant rates at the end of our simulations (see Figure~\ref{fig:vhyd_vtot}).
        The open orange circles are extrapolated values of the neutrino-induced kick velocities according to Equation~(\ref{eq:vnuinfty}) with $\bar{\alpha}_{\nu,\infty}^\mathrm{tot} \approx 0.2\,\bar{\alpha}_\nu^\mathrm{tot}(t_\mathrm{f}^\nu)$ for the LESA-dominated low-mass (9.0 and 9.6\,M$_\odot$) models and $\bar{\alpha}_{\nu,\infty}^\mathrm{tot} \approx \bar{\alpha}_\nu^\mathrm{tot}(t_\mathrm{f}^\nu)$ for the explosion models of more massive (12.28, 15, 18.88, and 20\,M$_\odot$) progenitors. The open black triangles show extrapolated values of the total (neutrino plus hydrodynamic) kicks, $|\vec{\tilde{v}}_\mathrm{NS}^\mathrm{tot}| = |\vec{v}_\mathrm{NS}^\mathrm{hyd}(t_\mathrm{f}) + \vec{\tilde{v}}_\mathrm{NS}^{\nu,\infty}|$, i.e., assuming that the neutrino kick can be projected from $t_\mathrm{f}^\nu$ into the future using its direction as given at $t_\mathrm{f}^\nu$.
        {\em Right panel:} The filled black diamonds in the right panel are the final BH kicks when the whole pre-collapse mass of the progenitor is swallowed by the BH, whereas the open back diamonds assume that only the He-core mass falls into the BH and the hydrogen envelope of the pre-collapse progenitor gets stripped. Note the different vertical scale of the right panel with a scale break at 4\,km\,s$^{-1}$. The vertical bars in the case of model u75 indicate the stochastic variations of our different 3D collapse simulations of the 75\,M$_\odot$ progenitor (see Table~\ref{tab:neutrino_kicks}). The plot shows that for SN explosions of low-mass progenitors (here the 9.0 and 9.6\,M$_\odot$ models) neutrino-induced NS kicks are expected to dominate the hydrodynamic kicks, for intermediate-mass progenitors (here the 12.28\,M$_\odot$ model) both contributions to the NS kick are of similar magnitude, whereas for SN explosions of massive progenitors (here the 15, 18.88, and 20\,M$_\odot$ models) the hydrodynamic NS kicks are dominant. Note that in the 15\,M$_\odot$ simulations, rapid rotation has an effect on the results (see text). In the case of BH formation without concomitant explosion and asymmetric mass ejection, the neutrino-induced BH kicks are only on the order of some km\,s$^{-1}$}
        \label{fig:scatter_plot}
\end{figure*}

The hydrodynamic NS kicks are expected to scale with the explosion energy and the explosion asymmetry:
\begin{equation}
v_\mathrm{NS}^\mathrm{hyd} \propto \bar{\alpha}_\mathrm{ej}\, E_\mathrm{exp}\,M_\mathrm{NS}^{-1} 
\label{eq:vhydscaling}
\end{equation}
\citep[Equation~(11) in][]{Janka2017}. This expectation is compatible with our results of Table~\ref{tab:hydro_kicks} and those of \citet{Burrows+2023b}. The low-mass progenitors, which explode with the lowest energies, also develop small ejecta asymmetries (measured by $\bar{\alpha}_\mathrm{ej}$ of Equation~\ref{eq:alphaej}) because of their explosion early after core bounce and their rapid ejecta expansion \citep[see the discussion in, e.g.,][]{Gessner+2018,Mueller+2019,Stockinger+2020,Burrows+2023b}. Therefore, their hydrodynamic kicks converge to their final values within only 1--2\,s (Figure~\ref{fig:vhyd_vtot}) and at only a few 10\,km\,s$^{-1}$. Our explosion models for the 12.28, 15, and 18.88\,M$_\odot$ progenitors reach hydro kicks of 240--630\,km\,s$^{-1}$ until the simulations were stopped, but there is a trend of monotonic increase continuing beyond this time (Figure~\ref{fig:vhyd_vtot}). The two explosion models of the 20\,M$_\odot$ progenitor develop the highest explosion energies and therefore also the biggest hydro kicks of nearly 900\,km\,s$^{-1}$ and 1300\,km\,s$^{-1}$ with considerable growth rates even after 8\,s, in particular in the model with the most extreme NS kick. This model is likely to reach a NS kick of around 1500\,km\,s$^{-1}$ in the end.

It is interesting that in the two rotating models, m15 and m15e, the explosion is strongest close to the equatorial plane due to a powerful spiral mode of the standing accretion shock instability \citep[spiral SASI;][]{Blondin+2007} driving the onset of shock expansion in this simulation \citep[see][]{Summa+2018}. For this reason, an explosion with a pronounced hemispheric asymmetry {\em perpendicular} to the rotation axis develops, i.e., the ejecta expand most strongly around the equator and possess a sizable dipolar deformation. Therefore, in contrast to naive expectations based on 2D simulations of rotation-supported explosions, the hydrodynamic NS kicks in these 3D simulations have big inclination angles relative to
the rotation axis of the 15\,M$_\odot$ progenitor. The relative angles between the progenitor's spin vector and the vector direction of the NS's final hydro kicks are 81.1$^\circ$ for model m15 and 61.5$^\circ$ for model m15e. These angles are close to those between the NS spin and kick vectors of models m15 and m15e (see Table~\ref{tab:hydro_kicks}), because the NS spin directions are basically identical with the spin axis of the progenitor, and the total (hydro plus neutrino) kicks of the NSs are far dominated by the hydrodynamic kicks. In model m15 (and m15e), also the neutrino-induced kick is near the equatorial plane at time $t_\mathrm{f}^{\nu}$ when the \textsc{Vertex} neutrino transport was switched off; the angle between the progenitor's spin vector and the neutrino-kick vector is 92.3$^\circ$.

Table~\ref{tab:hydro_kicks} also provides the values of the neutrino-induced NS kicks at times $t_\mathrm{f}^\nu$. These values and the corresponding kick directions are fixed when plotting the subsequent evolution of the total kicks in Figure~\ref{fig:vhyd_vtot}. The neutrino kicks at times $t_\mathrm{f}^\nu$ are typically lower estimates of the final values. We also list our estimates of the possible final neutrino-kick velocities, $\tilde{v}_\mathrm{NS}^{\nu,\infty}$, in Table~\ref{tab:neutrino_kicks} and plot them in Figure~\ref{fig:scatter_plot}. They are obtained by using Equation~(\ref{eq:vnuinfty}) and by making the assumptions that the neutrino emission asymmetry $\bar{\alpha}_{\nu,\infty}^\mathrm{tot}$ that holds for the entire PNS neutrino cooling phase at $t > t_\mathrm{f}^\nu$ is (10--30)\% of the value averaged from core bounce until $t_\mathrm{f}^\nu$ in the LESA-dominated low-mass models, and that $\bar{\alpha}_{\nu,\infty}^\mathrm{tot}$ is equal to the value time-

\end{multicols}
\twocolumn

\noindent
averaged until $t_\mathrm{f}^\nu$ in the high-mass models (see Section~\ref{sec:asymmetryparameters}). The results thus derived can be considered as crude extrapolations for the long-time evolution of the neutrino kicks.\footnote{Most extreme upper limits may be obtained by replacing $\bar{\alpha}_{\nu,\infty}^\mathrm{tot} = \bar{\alpha}_\nu^\mathrm{tot}(t_\mathrm{f}^\nu)$ in Equation~(\ref{eq:vnuinfty}) by $\alpha_\nu^\mathrm{tot}(t_\mathrm{f}^\nu)$ (averaged over a time interval of several 10\,ms near $t_\mathrm{f}^\nu$, using Equation~\ref{eq:baralpnu}). The thus computed values for the neutrino kicks extrapolated to very late times can be up to about a factor of two greater than those listed for $\tilde{v}_\mathrm{NS}^{\nu,\infty}$ in Table~\ref{tab:neutrino_kicks} in models where dense, neutrino absorbing, asymmetric gas flows around the NS exist until late after bounce (i.e., m15, s18.88, s20). However, we consider these values as unrealistically high upper bounds, because the directions of such flow asymmetries are usually strongly time dependent. Therefore, the neutrino emission asymmetry in the free-streaming limit exhibits considerable directional fluctuations in time, and the time-averaged neutrino emission asymmetry $\bar{\alpha}_\nu^\mathrm{tot}(t)$ is usually much lower than the instantaneous values $\alpha_\nu^\mathrm{tot}(t)$ (see Section~\ref{sec:NSneutrinokicks} and Appendix~\ref{secA1}).} In Figure~\ref{fig:scatter_plot}, we also estimate the final total NS kicks by adding the hydro kicks at the end points of our simulations, $t_\mathrm{f}$, and our extrapolated values of the neutrino kicks, assuming that the vector directions of the neutrino kicks at times $t_\mathrm{f}^\nu$ remain unchanged during the later neutrino emission.

Figure~\ref{fig:scatter_plot} visualizes our progenitor dependent results for hydrodynamic, neutrino, and total kicks graphically; successfully exploding models are shown in the left panel. In the cases of the low-mass models s9.0 and z9.6 the total kicks ($|\vec{v}_\mathrm{NS}^\mathrm{tot}|$ and $|\vec{\tilde{v}}_\mathrm{NS}^\mathrm{tot}|$) are clearly dominated by the neutrino-induced kicks, in the intermediate-mass model s12.28 both neutrino and hydro kicks yield similarly important contributions to the total kick, and in the high-mass models m15, m15e, s18.88, s20, and s20e the hydrodynamic kicks account for the far dominant contributions to the total kicks. These results are in line with the findings reported by \citet{Stockinger+2020}, \citet{Bollig+2021}, and \citet{Burrows+2023b}.

It is important to note that the hydrodynamic kicks in most models start growing initially more slowly than the neutrino kicks (see plots in Appendix~\ref{secA1}), and they rise rapidly only after the SN shock has expanded to several thousand kilometers and the final ejecta asymmetry has developed. Before this time, the rapid directional variability of the gas flows around the PNS does not permit the formation of a clear direction for the hydrodynamic kicks. The absolute values of the hydro kicks remain small during this early phase because they are reduced by the averaging of acceleration forces in different directions. The faster emergence of neutrino kicks of significant amplitudes and their complex interplay with the hydro kicks will be discussed in detail in Section~\ref{sec:neutrinokicks}.

Despite the fact that the neutrino kicks, in particular the estimated final ones, can reach sizable magnitudes of up to 200--260\,km\,s$^{-1}$ (Table~\ref{tab:neutrino_kicks}), their relative shares in the absolute values of the total NS kicks of our high-mass models turn out to be more moderate (Figure~\ref{fig:scatter_plot}). This finding is explained by the large angles $\theta_\mathrm{hyd}^\nu$ between the directions of the hydro and neutrino kicks (see Table~\ref{tab:neutrino_kicks} for the values at time $t_\mathrm{f}^\nu$ and Table~\ref{tab:hydro_kicks} for those at $t_\mathrm{f}$), especially in our rapidly rotating models m15 and m15e, where the relative angles are around 100$^\circ$ at $t_\mathrm{f}$. In all other models, the directions of hydrodynamic and neutrino kick fall into the same hemisphere. In contrast, the spiral-SASI assisted explosion of the 15\,M$_\odot$ star does not only force both the hydro kick and the neutrino kick to directions nearly perpendicular to the spin axis of progenitor and NS, but it also leads to a very large angle between the two kick directions.

\begin{figure*}[t]
        \centering
        \includegraphics[width=\textwidth]{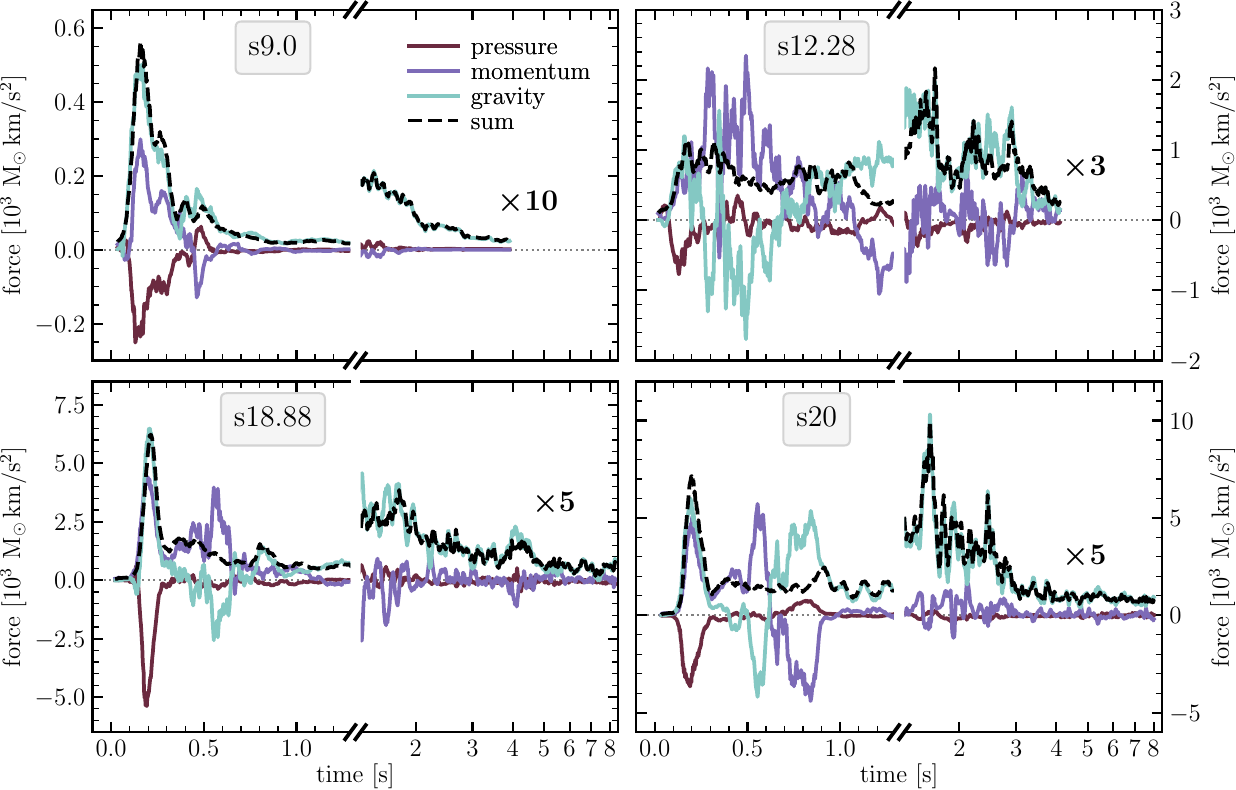}
        \caption{Hydrodynamic forces on the PNS as functions of postbounce time for our 3D CCSN models s9.0 ({\em top left}), s12.28 ({\em top right}), s18.88 ({\em bottom left}), and s20 ({\em bottom right}). Colored solid lines as labeled in the upper left panel show the contributions to the total momentum transfer by gas pressure, momentum carried by mass gained or lost in accretion streams and outflows, and by the gravitational attraction between PNS and ejecta gas, as given in Equation~(\ref{eq:hydroforces}) and evaluated with $R_0 = 2\,R_\mathrm{NS}$. The black dashed lines display the sum of these three components. In order to visualize the values that effectively create the PNS acceleration, we display the projections of the different forces onto the total accelerating force: $\dot{\vec{p}}_i\cdot\dot{\vec{p}}_\mathrm{hyd}/|\dot{\vec{p}}_\mathrm{hyd}|$ with $i\in\{\text{pres},\text{mom},\text{grav}\}$. Note the change of the time axis from linear to logarithmic for the evolution after 1.3\,s. For better visibility the force values during this long-time evolution are scaled by the factors indicated in each panel. It is obvious that the gravitational force accounts for the dominant effect during most phases and is the far dominant contribution during the long-time evolution. Model~s12.28 shows the largest fluctuations in all three force contributions until late times ($\gtrsim$1\,s) because of the ongoing, considerable mass accretion onto the PNS even during this late evolution (see middle panel of Figure~\ref{fig:rshock_eexpl})}
        \label{fig:hydro_forces}
\end{figure*}

Finally, in Table~\ref{tab:hydro_kicks} we also list the NS spin periods $P_\mathrm{NS}$ at $t_\mathrm{f}$ \citep[evaluated as described in Appendix~D of][using $r_0=R_\mathrm{NS}$]{Stockinger+2020} and the angles $\theta_\mathrm{spin}^\mathrm{kick}$ between the NS spin and kick directions at this time. The PNSs are spun up by the angular momentum carried to the compact remnant by anisotropic accretion flows \citep[e.g.,][]{Blondin+2007,Wongwathanarat+2010,Rantsiou+2011,Wongwathanarat+2013,Kazeroni+2016,Kazeroni+2017,Mueller+2017,Mueller+2018,Mueller+2019,Powell+2020,Stockinger+2020,Coleman+2022,Burrows+2023b}, but their values and directions are only transient and can be massively influenced by later fallback accretion \citep[see the detailed discussions in][]{Janka+2022,Mueller2023}. In the 15\,M$_\odot$ models, the NS spin period is determined by the rapid rotation of the collapsing stellar core.

We witness no tendency of spin-kick alignment at $t_\mathrm{f}$, in agreement with results discussed in \citet{Mueller+2019} and \citet{Burrows+2023b}, and also in agreement with the long-term 3D CCSN simulations reported by \citet{Janka+2022}, which include late-time fallback. However, our sample of models is small. Spin-kick alignment is, in particular, also absent in the two rapidly spinning m15 models, because the directions of the NS kicks (hydro and neutrino contributions) are close to the equatorial plane due to the geometry of the explosion, which is shaped by the spiral-SASI that initiated the shock expansion. The situation might be different, however, in 3D explosions of rapidly rotating stars that explode either with a more axially symmetric, oblate deformation of the ejecta or a prolate deformation along the rotation axis with a large hemispheric asymmetry \citep{Janka2017,Burrows+2023b}. In such cases, one can envision a NS kick that is correlated with the spin axis of the NS, whose rotation could be determined by the angular momentum inherited from the progenitor, induced by spiral-SASI activity, or imposed by equatorial accretion continuing beyond the onset of the explosion. The amount of rotation needed to influence the 3D explosion and kick geometry seems to be sizable, however \citep{Summa+2018}, and may not be common to the majority of core-collapse events \citep[see, e.g.,][for the pulsar PSR~J0538+2817, which is estimated to have rotated slowly at birth]{Yao+2021}. The role of magnetic fields in this context is still unclear. \citet{Janka+2022} considered asymmetric, long-lasting or late-time fallback accretion onto kick-displaced NSs as a possible origin of spin-kick alignment. Such extended evolution phases and their requirements of relevant physics are presently not reached by the available, self-consistent 3D CCSN simulations.

A detailed discussion of the development of the NS spins in connection to the explosion behavior and 3D flow dynamics in our SN simulations will be provided in a separate paper (Kresse et al, in preparation).

\begin{figure*}[t]
        \centering
        \includegraphics[width=\textwidth]{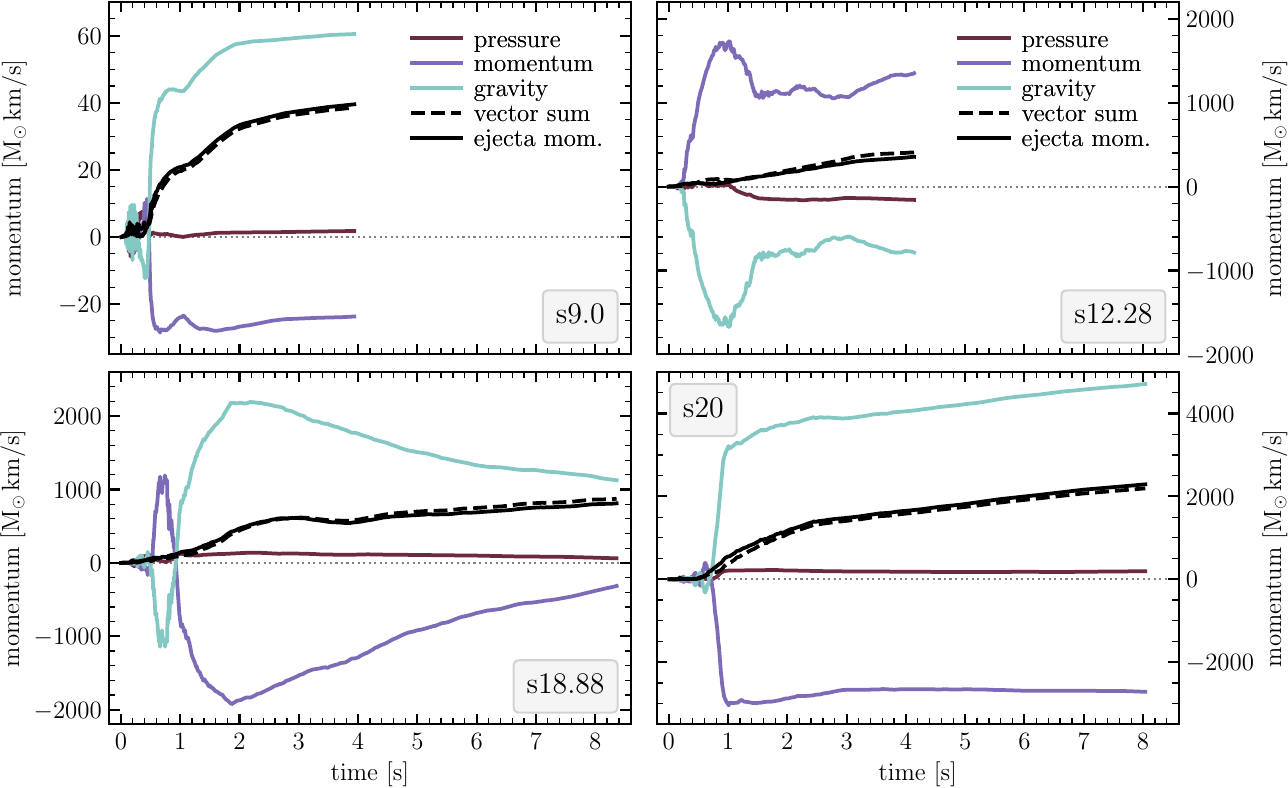}
        \caption{Contributions to the total PNS momentum caused by hydrodynamic forces as functions of postbounce time for our 3D CCSN models s9.0 ({\em top left}), s12.28 ({\em top right}), s18.88 ({\em bottom left}), and s20 ({\em bottom right}). Colored solid lines as labeled in the upper panels show the projections $\vec{p}_i\cdot\vec{p}_\mathrm{hyd}/|\vec{p}_\mathrm{hyd}|$ for $i\in\{\text{pres},\text{mom},\text{grav}\}$, with $\vec{p}_i(t) = \int_0^t \vec{\dot{p}}_i \mathrm{d}t'$ and $\vec{p}_\mathrm{hyd}(t) = \vec{p}_\mathrm{pres} + \vec{p}_\mathrm{mom} + \vec{p}_\mathrm{grav}$.
        This means that the force terms corresponding to
        the momentum transfer by the gas pressure, momentum carried by accretion streams and outflows, and the action of gravity as given in Equation~(\ref{eq:hydroforces}), evaluated with $R_0 = 2\,R_\mathrm{NS}$, are time-integrated until time $t$ and projected onto the
        total momentum transfer to the PNS caused by all hydrodynamic forces until this time. The magnitude of this total hydrodynamic momentum transfer as the vector sum of the three components, $|\vec{p}_\mathrm{hyd}|$, is plotted by black dashed lines. For comparison, the black solid lines give the total momentum of the gas exterior to the PNS, $|\vec{p}_\mathrm{gas}|$, according to Equation~(\ref{eq:pgas}). The agreement is very good in all displayed cases. The gravitational effect dominates also on the absolute momentum scale and not only with its component projected onto the total accelerating force acting on the PNS as shown in Figure~\ref{fig:hydro_forces}. The concordance of the signs of total momentum and gravitational momentum transfer signals that the latter accounts for the decisive effect. Here, as in Figure~\ref{fig:hydro_forces}, model~s12.28 makes an exception due to ongoing, significant mass accretion even at late times $t>1$\,s; in this case, momentum transfer to the PNS is still important and can govern the ongoing acceleration, which is visible by the momentum term dominating and pushing in the direction of the total PNS momentum. However, we expect that the gravitational term will eventually take over once accretion onto the PNS has decreased to very low values as in the other models}
        \label{fig:hydro_forces_integrated}
\end{figure*}

In Figures~\ref{fig:hydro_forces} and~\ref{fig:hydro_forces_integrated}, we show the forces that contribute to the acceleration of the PNS by asymmetric matter ejection and the corresponding integrated momentum transfer caused by hydrodynamic pressure, gas inflows and outflows, and the gravitational attraction of the gas surrounding the PNS. This analysis is based on the description in Section~\ref{sec:forces}. 

Of course, the NS's hydrodynamic kick is a consequence of momentum conservation, since the NS receives a momentum of the same magnitude as the linear momentum of the asymmetric SN ejecta, but in the opposite direction. This fact is obvious from Equation~(\ref{eq:pnsdot2}), which we routinely use to compute the total hydrodynamic NS kick. Our analysis of Figures~\ref{fig:hydro_forces} and~\ref{fig:hydro_forces_integrated} is an alternative way to calculate the kick by adding up the forces acting at a certain radius $R_0$ (doing so, we follow previous analyses by, e.g., \citealt{Scheck+2006,Nordhaus+2010,Wongwathanarat+2013,Mueller+2017,Coleman+2022,Burrows+2023b}). In our study, this radius is chosen
such that the agreement of the NS momentum vectors obtained by both methods is optimal, i.e., the negative of the ejecta momentum and the sum of the time integrals of the three contributing forces show best agreement (Figure~\ref{fig:hydro_forces_integrated}). Thus, we ensure that the evaluation radius $R_0$ for the force calculation (Equation~\ref{eq:hydroforces}) is properly chosen. We empirically determined the optimal radius to be $R_0 \approx 2R_\mathrm{NS}$.
This location is sufficiently far outside the PNS that the momentum transfer to the surrounding gas by neutrino interactions exterior to the neutrinospheres is accounted for already.

From Figures~\ref{fig:hydro_forces} and~\ref{fig:hydro_forces_integrated} it is clear that a major part of the NS acceleration takes place in most models after the gas accretion by the PNS has become unimportant, i.e., later than about 1\,s after bounce, but before fallback of initial ejecta sets in many seconds later when the neutrino heating abates. The term for the gravitational force plays the dominant role on the long time scales, because gravity continuously acts as a long-range force, even after PNS accretion has diminished. This is obvious from the forces plotted in Figure~\ref{fig:hydro_forces}. And Figure~\ref{fig:hydro_forces_integrated} also shows that during the late, post-accretion phase, the final NS momentum is aligned with the direction of the integrated contribution by the gravitational effect (see the signs of the different components and that of the summed effect). Therefore, \citet{Wongwathanarat+2013} introduced the name ``gravitational tugboat mechanism'' for the acceleration process, thus expressing the great importance of the gravitational term for the final NS kick. In contrast, as long as PNS accretion delivers significant amounts of matter onto the PNS (in model s12.28 this is the case during all of the displayed evolution), the momentum transfer by gas flows is the leading effect. But once the accretion subsides to an irrelevant level, any long-lasting NS acceleration will be governed by the gravitational interaction between the NS and the asymmetric ejecta, independent of the question whether some fraction of this gravitationally pulling matter ultimately falls back and is accreted by the NS or not.\footnote{In some recent papers~\citep{Coleman+2022,Burrows+2023b} it was argued that the name ``gravitational tugboat effect/mechanism'' is inappropriately chosen and misleading, because (1) the importance of gravity was misjudged when the simulations were halted before the final kicks were reached, and (2) because the gravitational tug is in the wrong direction, and (3) because some of the pulling matter may get accreted onto the NS in the end. We contradict this reasoning, because our analysis demonstrates that the first two arguments are not valid: (1) Our long-term simulations showed that gravity as a long-range force indeed accounts for the biggest contribution to the NS acceleration on long time scales, and (2) the slower, denser ejecta tow the NS away from the faster ejecta expelled in the opposite direction (thus ensuring overall conservation of the total linear momentum). Concerning point (3) we stress that in the case that some of the gravitationally pulling matter falls back onto the NS, this leads to an increase of the {\em net} ejecta momentum in the opposite direction \citep[see the right panel of figure 7 in][]{Scheck+2006}. Thus, the NS momentum balancing this ejecta momentum increases
towards the side where the gravitationally attracted matter fell inward to the NS. The dominant force mediating the NS acceleration is therefore gravity, even if some involved matter ultimately merges with the NS (due to gravitational interaction). These facts are the reason why the accelerating process was called gravitational tugboat mechanism/effect, although in the end the NS's motion is, of course, just a consequence of global momentum conservation. The analogue of the tugboat refers to the fact that a tugboat pulls a boat in the direction of its own motion, while momentum conservation is ensured by water being propelled faster in the opposite direction. In the SN case the denser, usually more massive and slower ejecta are the tugboat, because they pull the NS by their gravitational attraction in the direction of their own motion, and momentum conservation holds because the faster ejecta move towards the other side. (Note that the tugboat in this picture is {\em not} the faster “bulk” of the ejecta, but the slower, more massive ejecta.)} We expect that ultimately also in model s12.28 the gravitational tug adopts the leading role once ejecta downdrafts do no longer entail gas accretion by the PNS.

\subsection{Neutrino kicks}
\label{sec:neutrinokicks}
 
Anisotropic neutrino loss and the corresponding neutrino-induced NS kicks have two major sources. Either they originate from neutrino transport asymmetries inside the PNS (Figure~\ref{fig:lesakick}), or they are a consequence of asymmetric neutrino emission and absorption/scattering asymmetries in regions outside the PNS but still close to its surface (Figure~\ref{fig:accretionkick}). 

\subsubsection{PNS convection and LESA effects}
\label{sec:conv+LESA}

\begin{figure}[t]
\centering
\includegraphics[width=\columnwidth]{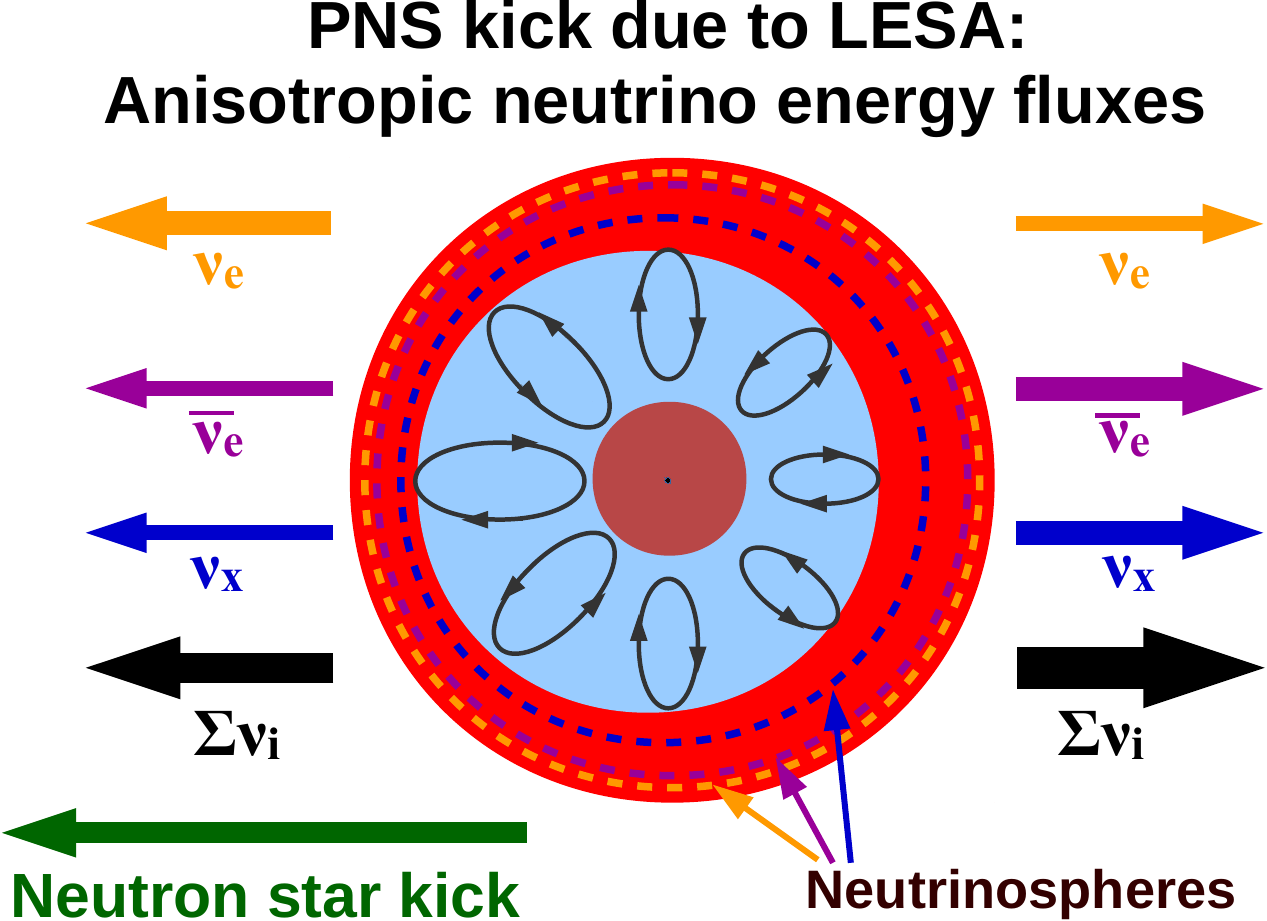}
\caption{PNS kick by anisotropic neutrino emission due to LESA. Stronger convective transport of electron-lepton number out of the PNS core in one hemisphere (in the image on the left side) increases the electron-neutrino ($\nu_e$) emission, whereas the electron antineutrino ($\bar\nu_e$) and heavy-lepton neutrino ($\nu_x$) fluxes are relatively 
stronger in the opposite hemisphere, where also the total neutrino flux as sum over all
neutrino species is highest. Correspondingly, the PNS receives a total neutrino-induced kick (roughly) aligned with the orientation of the LESA dipole, which is defined by the direction of the strongest electron-neutrino lepton-number ($\nu_e$ minus $\bar\nu_e$) flux. The neutrino-induced kick by $\nu_e$ individually is directed opposite to the LESA vector and opposite to the kicks by $\bar\nu_e$ and $\nu_x$ individually}
\label{fig:lesakick}
\end{figure}

\begin{figure*}[t]
\centering
\includegraphics[width=\textwidth]{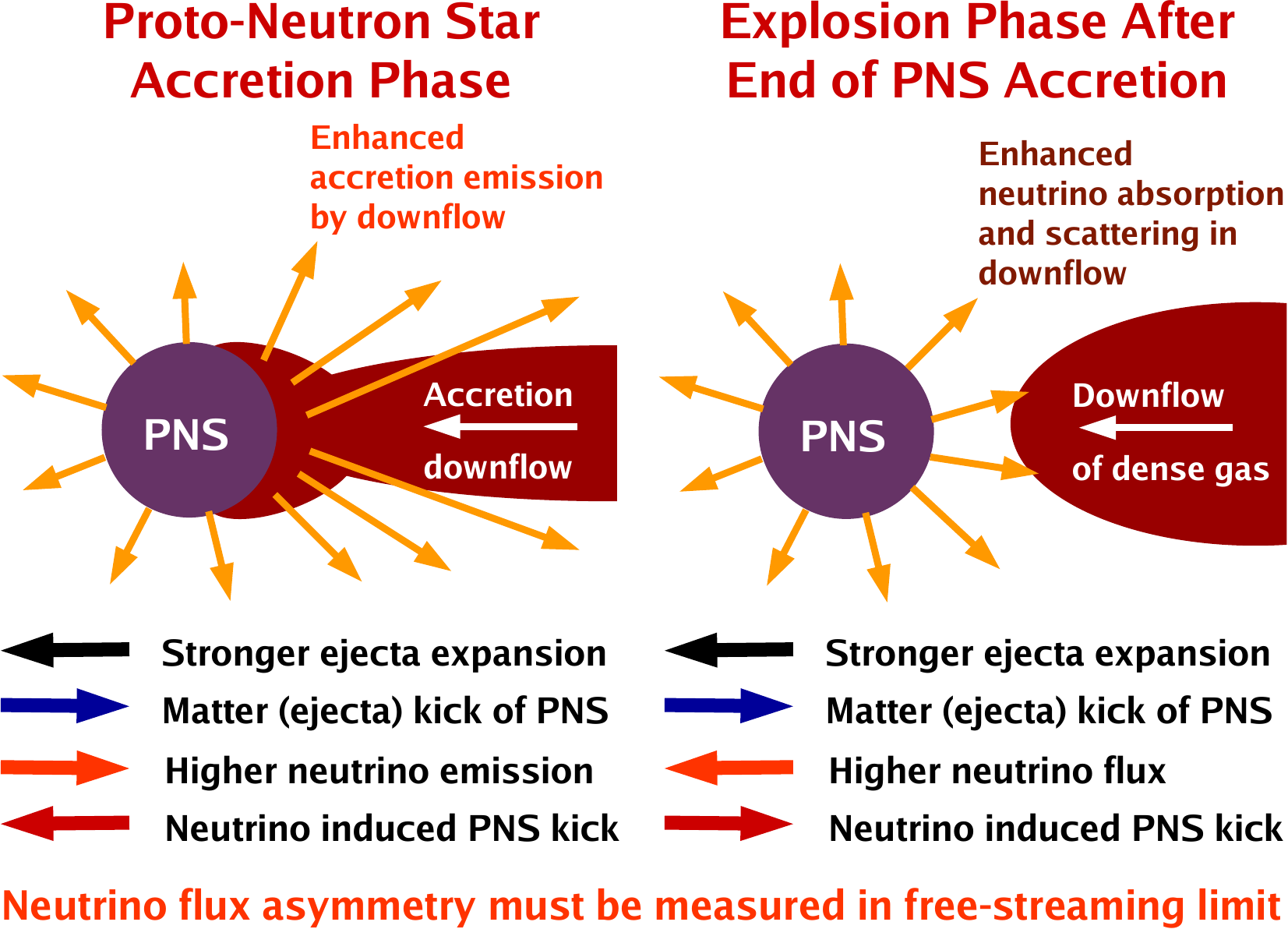}
\caption{Interplay of neutrino-induced recoil acceleration and ejecta (matter) induced hydrodynamic recoil acceleration connected to asymmetric PNS accretion and ejecta distribution in exploding models. Typically, PNS accretion flows ({\em left}) and dense, slow ejecta in the vicinity of the PNS ({\em right}) preferentially occur in the hemisphere where the explosion and mass
ejection are weaker, i.e., on the side the matter-induced PNS kick points to (blue arrows). 
{\em Left:} Gas accreted by the PNS can produce
anisotropic neutrino emission before it settles down onto the PNS surface. The associated
neutrino kick of the PNS, which is opposite to the direction of the enhanced neutrino emission, 
is thus towards the side of the stronger explosion and therefore opposite to the matter kick.
{\em Right:} In the case of enhanced neutrino absorption and scattering by asymmetrically distributed 
dense ejecta near the PNS the neutrino kick and the matter kick are aligned. However, the momentum 
transfer by neutrino absorption and scattering in the dense ejecta reduces the net momentum of the
anisotropically ejected gas in our considered situation, thus reducing the matter kick. 
This reduction of the hydrodynamic NS kick compensates the 
apparent neutrino kick that is associated with the neutrino flux anisotropy caused by the
neutrino absorption and scattering in matter exterior to
the neutrinosphere. Such a compensation is, for example, consistent with the fact that a PNS, 
whose neutrino fluxes leave the neutrinosphere isotropically in all directions in the PNS's rest frame, 
should not receive any net acceleration if a fraction of the neutrinos is absorbed or scattered
at some distance away from the PNS in matter that is detached from the compact remnant. Note that
alignment or anti-alignment of neutrino kicks and hydrodynamic kicks is meant here as a hemispheric
coincidence rather than an exactly parallel orientation of the vectors
}
\label{fig:accretionkick}
\end{figure*}

Convection in the hot mantle layer of the PNS below the neutrinospheres accelerates the transport of electron lepton number by carrying electrons and $\nu_e$ out of the dense PNS core (Figure~\ref{fig:lesakick}). The corresponding convective cells with rising flows of lepton-rich gas and downdrafts of deleptonized plasma create a patchy pattern with areas of enhanced neutrino emission distributed over the neutrinospheres. This pattern of hot-spot emission can lead to NS kicks. Since there is a large number of convective cells that change rapidly on time scales of only ${\cal O}(10\,\mathrm{ms})$, the hot-spot emission varies stochastically in random directions and with fluctuating amplitudes. Therefore, the associated neutrino-induced kicks remain small because of statistical averaging effects. 

However, in addition to these small-scale and rapidly variable asymmetries, also a large hemispheric difference in the strength of PNS convection shows up in 3D simulations \citep{Tamborra+2014,OConnor+2018,Glas+2019,Powell+2019,Vartanyan+2019}, which establishes a hemispheric contrast of the electron fraction in the near-surface layers of the PNS between its convection shell and the neutrinospheres. As a consequence, a dipolar asymmetry of the electron-neutrino lepton-number ($\nu_e$ minus $\bar\nu_e$) flux develops, which was termed Lepton-number Emission Self-sustained Asymmetry \citep[LESA;][]{Tamborra+2014}. This dipole can be the dominant multipole component defining the global asymmetry of the neutrino-lepton number emission of the PNS for longer-lasting (hundreds of milliseconds to seconds) periods of postbounce evolution, and its vector direction was found to be quite stable or to drift only slowly with time.

The LESA phenomenon is also associated with dipolar anisotropies of the number and energy fluxes of all individual neutrino species ($\nu_e$, $\bar\nu_e$, and heavy-lepton neutrinos $\nu_x$) and therefore of the total neutrino luminosity, too. The vector direction of the dipole of the total neutrino luminosity is opposite to the dipole vector of the lepton-number flux. A schematic visualization can be seen in Figure~\ref{fig:lesakick}, where the PNS receives a kick to the left, aligned with the LESA dipole vector and compensating the momentum carried away by the stronger total neutrino emission in the hemisphere on the right side. 

The exact causal origin of LESA is not well understood as yet \citep[see the discussion by][]{Glas+2019}, but its physical properties reported by \citet{Tamborra+2014} received support by a detailed analysis of 3D models with full multi-D neutrino transport including velocity dependence by \citet{OConnor+2018}. The LESA phenomenon was also witnessed in the 3D models of the Princeton group \citep{Vartanyan+2019} and by \citet{Powell+2019} with simpler neutrino transport. Although there is a feedback between internal PNS convective asymmetry, asymmetric neutrino emission, and external neutrino heating and asymmetric (“funneled”) PNS accretion during the shock stagnation phase as described by \citet{Tamborra+2014} and confirmed by \citet{OConnor+2018}, the presence of LESA also in our lowest-mass model z9.6 may disfavor a feedback cycle that is initiated by an external accretion asymmetry \citep[as hypothesized by][]{Tamborra+2014}; this low-mass progenitor explodes like an electron-capture SN (ECSN) without any strong and long-lasting postshock accretion flows. Instead, it seems more likely that LESA originates from a generic trend for the development of a strong dipole component of the PNS convection when the convective layer in the PNS becomes radially sufficiently extended. This possibility was discussed in quite some detail in the paper by \citet{Glas+2019}.

The dipolar asymmetry of the $\nu_x$ emission in addition to the $\nu_e$ and $\bar\nu_e$ dipoles, which was witnessed by \citet{Tamborra+2014}, was also confirmed by \citet{OConnor+2018} with properties in good agreement between their independent models. The $\nu_x$ dipole points in the direction of the stronger $\bar\nu_e$ flux and thus opposite to the LESA dipole, but it has a significantly smaller amplitude than those. Its presence can be understood by the correlation of the LESA emission dipole with a hemispheric asymmetry of the electron fraction in and above the PNS convection layer. This asymmetry leads to enhanced radiation of $\nu_\mu\bar\nu_\mu$ and $\nu_\tau\bar\nu_\tau$ pairs produced by $e^+e^-$ and $\nu_e\bar\nu_e$ annihilation on the side of the PNS with the lower electron fraction and suppressed $\nu_x\bar\nu_x$-pair production on the opposite side \citep{Tamborra+2014}.

A few remarks on the LESA-associated neutrino kicks seem to be indicated. First, since most of our simulations with \textsc{Vertex} neutrino transport cover only periods of $t_\mathrm{f}^\nu < 1$\,s, we estimate the final LESA-induced kicks by extrapolation. For that we use values of the asymmetry parameter $\bar{\alpha}_{\nu,\infty}^\mathrm{tot}$ in Equation~(\ref{eq:vnuinfty}) that are lower than the values obtained at $t_\mathrm{f}^\nu$ (see Sect.~\ref{sec:asymmetryparameters}). There are two reasons for this choice: (i) We find the instantaneous emission asymmetries $\alpha_\nu^\mathrm{tot}$ decreasing towards the end of our simulations, and (ii) consistent with that, the neutrino-induced kicks due to the LESA effect (evaluated at an angle-averaged density of $10^{12}$\,g\,cm$^{-3}$ exterior to the convective PNS layer but inside the PNS) saturate after several 100\,ms post bounce in the two models (s12.28 and s18.88) that we carried to later times with \textsc{Vertex} transport (Figures~\ref{fig:vnu_s1228} and \ref{fig:vnu_s1888}).

Second, our values for the LESA-related kicks have to be taken with a grain of salt because of our RbR+ transport approximation in \textsc{Vertex}. A comparison of RbR+ transport with a full multi-dimensional transport treatment by \citet{Glas+2019} (see figures 1 and 5 there) showed that LESA is present in both cases, but may develop later and with a reduced dipole amplitude in the multi-D transport case. This would lower the expected values for the neutrino-induced kicks when these are dominated by LESA effects.

Third, the development of the LESA emission asymmetry and its strength also depend on the nuclear EoS in the hot PNS \citep{Janka+2016}, because the evolution of PNS convection hinges closely on the variation of the nuclear symmetry energy as a function of density \citep{Roberts+2012}. The consequences of different EoSs are not explored in detail yet. Therefore there is enough uncertainty such that the NS kicks due to LESA might be smaller than the values of $\sim$(30--60)\,km\,s$^{-1}$ that we obtained for slowly rotating NSs at $t_\mathrm{f}^\nu$, and in particular also smaller than those of our crude extrapolation for the low-mass models s9.0 and z9.6 in Table~\ref{tab:neutrino_kicks}. We cannot exclude that they might be as low as about (10--20)\,km\,s$^{-1}$ instead \citep[as suggested by model z9.6 in][]{Wang+2024}.

\subsubsection{Neutrino reactions in PNS accretion flows}
\label{sec:neutrino+accretion}

The second region that contributes in sculpting anisotropic neutrino loss extends between the PNS and the SN shock (Figure~\ref{fig:accretionkick}). In this region, dense downdrafts of cool gas and buoyant plumes of neutrino-heated, high-entropy matter cause asymmetric accretion onto the PNS and concomitant neutrino production as well as anisotropic absorption and scattering of neutrinos leaving the neutrinospheres. These non-spherical gas flows around the PNS are a consequence of convective overturn in the neutrino-heating layer behind the shock on the one hand, and of SASI mass motions as a global, nonradial instability of the stalled accretion shock on the other hand \citep[e.g.,][]{Blondin+2003,Ohnishi+2006,Foglizzo+2007,Blondin+2007,Scheck+2008,Iwakami+2009,Fernandez+2010}. The accretion downdrafts and rising outflows can be highly time dependent with variability of the mass accretion rate on time scales of milliseconds to tens of milliseconds, and their spatial pattern changes accordingly rapidly. However, a strong LESA dipole can have an impact on the gas motions around the PNS such that accretion downdrafts occur, in a time-averaged sense, more frequently in the hemisphere the LESA dipole vector points to, because neutrino heating and thus the explosion is stronger in the opposite hemisphere (see \citealt{Tamborra+2014} and \citealt{OConnor+2018}). This can lead to a near-alignment of the neutrino-induced NS kick produced by the LESA dipole and the hydrodynamic NS kick caused by the asymmetric explosion as witnessed in the low-mass 9.6\,M$_\odot$ explosion model of \citet{Stockinger+2020}, because in this model the LESA dipole determines the neutrino emission asymmetry as well as the ejecta asymmetry.

The general situation, however, is more complex when accretion flows develop randomly without any strong feedback with the LESA anisotropy that is created inside the PNS as visualized in Figure~\ref{fig:lesakick}. Downflows from random directions, as expected for conditions with a weak LESA dipole or massive accretion flows with strong penetration power in high-mass progenitors, can kick the PNS hydrodynamically to arbitrary directions and uncorrelated with the LESA asymmetry, but in a complex interplay with the neutrino emission and absorption occurring in the dense downdrafts. Figure~\ref{fig:accretionkick} provides a schematic and simplified graphical representation of how emission and absorption/scattering effects in the downdrafts in

\onecolumn
\addtolength{\tabcolsep}{-2pt}
\begin{table}[tb]
\caption{Neutrino-induced PNS kicks and characteristic parameters at time $t_\mathrm{f}^{\nu}$ for exploding and BH forming 3D models}
\label{tab:neutrino_kicks}
\begin{tabular*}{\textwidth}{@{\extracolsep\fill}lcccccccccccccc}
        \toprule
        Model & $t_\mathrm{f}^{\nu}$ & $M_\mathrm{NS,b}$ & $M_\mathrm{NS,g}$ & $E_{\nu}^\mathrm{tot}$ & $\bar{\alpha}_{\nue}$ & $\bar{\alpha}_{\nuebar}$ & $\bar{\alpha}_{\nux}$ & $\bar{\alpha}_{\nu}^\mathrm{tot}$ & $v_\mathrm{NS}^{\nu}$ & $v_\mathrm{NS}^\mathrm{LESA}$ & $\theta_\mathrm{hyd}^{\nu}$ & $\tilde{v}_\mathrm{NS}^{\nu,\infty}$ & $\tilde{v}_\mathrm{BH,1}^{\nu,\infty}$ & $\tilde{v}_\mathrm{BH,2}^{\nu,\infty}$ \\[0.5ex]
        & [s] & [M$_\odot$] & [M$_\odot$] & [$10^2$\,B] & [\%] & [\%] & [\%] & [\%] & [km/s] & [km/s] & [$^\circ$] & [km/s] & [km/s] & [km/s]\\
        \midrule
        s9.0           & 0.488 & 1.355 & 1.320 & 0.617 & 1.73 & 3.12 & 0.66 & 0.60 &  44.4 &  47.0 &  76.7 & 67$\pm$11 & --- & --- \\
        z9.6           & 0.450 & 1.350 & 1.319 & 0.556 & 2.06 & 3.81 & 0.84 & 0.76 &  49.4 &  42.0 &   9.8 & 78$\pm$14 & --- & --- \\
        s12.28         & 4.139 & 1.551 & 1.429 & 2.184 & 2.09 & 0.22 & 0.43 & 0.55 & 138.6 &  50.1 &  79.2 & 170.0 & --- & --- \\
        m15            & 0.457 & 1.575 & 1.533 & 0.744 & 0.11 & 0.25 & 1.10 & 0.70 &  54.4 &  11.4 &  48.9 & 264.3 & --- & --- \\
        s18.88         & 1.675 & 1.878 & 1.751 & 2.271 & 0.56 & 0.57 & 0.47 & 0.45 &  95.7 &  35.4 &  10.1 & 173.6 & --- & --- \\
        s20            & 0.506 & 1.901 & 1.828 & 1.310 & 0.82 & 1.53 & 0.57 & 0.46 &  55.0 &  58.7 &  60.3 & 197.8 & --- & --- \\
        \midrule
        s40            & 0.572 & 2.381 & 2.257 & 2.210 & 0.14 & 0.30 & 0.16 & 0.13 &  22.0 &  19.2 & 150.0 &   --- & 3.27 & 3.27 \\
        u75\_DD2       & 0.519 & 2.889 & 2.743 & 2.616 & 0.21 & 0.39 & 0.18 & 0.15 &  26.7 &  18.0 & 118.6 &   --- & 0.99 & 1.34 \\
        u75\_LS220\_1  & 0.250 & 2.572 & 2.493 & 1.418 & 0.15 & 0.14 & 0.07 & 0.05 &   4.8 &   6.9 & 115.3 &   --- & 0.16 & 0.22 \\
        u75\_LS220\_2  & 0.248 & 2.573 & 2.494 & 1.397 & 0.13 & 0.23 & 0.12 & 0.13 &  11.6 &   5.4 & 102.1 &   --- & 0.39 & 0.53 \\
        u75\_LS220\_hr & 0.254 & 2.563 & 2.484 & 1.425 & 0.14 & 0.31 & 0.12 & 0.12 &  11.4 &   7.1 &  87.2 &   --- & 0.38 & 0.52 \\
        u75\_SFHo      & 0.325 & 2.623 & 2.529 & 1.696 & 0.36 & 0.25 & 0.20 & 0.09 &  10.2 &  16.9 & 113.7 &   --- & 0.35 & 0.47 \\
        \botrule
\end{tabular*}
\footnotetext{\textbf{Note:} $t_\mathrm{f}^{\nu}$ is the postbounce time at the end of the evolution simulated in 3D with full neutrino transport, which is identical with the moment when the PNS begins to collapse to a BH in the non-exploding models; 
$M_\mathrm{NS,b}$, $M_\mathrm{NS,g}$, and $E_{\nu}^\mathrm{tot}$ are the baryonic PNS mass, the gravitational PNS mass, and the total radiated neutrino energy, respectively, at time $t_\mathrm{f}^{\nu}$ ($1\,\mathrm{B} = 1\,\mathrm{bethe} = 10^{51}$\,erg);
$\bar{\alpha}_{\nue}$, $\bar{\alpha}_{\nuebar}$, $\bar{\alpha}_{\nux}$, and $\bar{\alpha}_{\nu}^\mathrm{tot}$ are the time-averaged anisotropy parameters of the energy emission in \nue, \nuebar, \nux, and of the total (according to Equations~\ref{eq:baralpnui} and~\ref{eq:baralpnu}) at $t_\mathrm{f}^{\nu}$;
$v_\mathrm{NS}^{\nu} = |\vec{v}_\mathrm{NS}^{\nu}|$ is the magnitude of the neutrino-induced PNS kick velocity at $t_\mathrm{f}^{\nu}$; $v_\mathrm{NS}^\mathrm{LESA}$ is the magnitude of the LESA-induced contribution to the
PNS kick velocity, evaluated at $t_\mathrm{f}^{\nu}$ at an angle-averaged density of $10^{12}$\,g\,cm$^{-3}$;
$\theta_\mathrm{hyd}^{\nu} = \theta(\vec{v}_\mathrm{NS}^\mathrm{hyd}(t_\mathrm{f}^{\nu}),\vec{v}_\mathrm{NS}^{\nu}(t_\mathrm{f}^{\nu}))$ is the angle between the hydro and the neutrino kick vectors at $t=t_\mathrm{f}^{\nu}$;
$\tilde{v}_\mathrm{NS}^{\nu,\infty}$ is the estimated upper bound of the final, neutrino-induced kick velocity (for $t\to\infty$) according to Equation~(\ref{eq:vnuinfty}) with $\bar{\alpha}_{\nu,\infty}^\mathrm{tot} \approx (0.2\pm0.1)\,\bar{\alpha}_\nu^\mathrm{tot}(t_\mathrm{f}^\nu)$ for the LESA-dominated low-mass (9.0 and 9.6\,M$_\odot$) models and $\bar{\alpha}_{\nu,\infty}^\mathrm{tot} \approx \bar{\alpha}_\nu^\mathrm{tot}(t_\mathrm{f}^\nu)$ for the explosion models of more massive (12.28, 15, 18.88, and 20\,M$_\odot$) progenitors; $\tilde{v}_\mathrm{BH}^{\nu,\infty} = v_\mathrm{NS}^{\nu} \cdot M_\mathrm{NS,g} / M_\mathrm{BH}$ is the final BH kick velocity, assuming a final BH mass of $M_\mathrm{BH} = M_\mathrm{prec} - E_{\nu}^\mathrm{tot}/c^2$ (i.e., the whole pre-collapse progenitor collapses into the BH) in the case of $\tilde{v}_\mathrm{BH,1}^{\nu,\infty}$ and $M_\mathrm{BH} = M_\mathrm{He} - E_{\nu}^\mathrm{tot}/c^2$ (i.e., ejection of the entire hydrogen envelope instead of its accretion by the BH) in the case of $\tilde{v}_\mathrm{BH,2}^{\nu,\infty}$; the values of $M_\mathrm{prec}$ and $M_\mathrm{He}$ are given in Table~\ref{tab:models}.}
\end{table}
\addtolength{\tabcolsep}{2pt}

\begin{multicols}{2}

\noindent
the close vicinity of the neutrino radiating PNS can correlate or anti-correlate with the hydrodynamic (matter-induced) kick. The directions of both kicks should not be considered here as perfectly aligned or anti-aligned, but as preferentially coincident in their hemispheres, because the neutrino emission by accretion downstreams can be spread over a wide angle and downflows can also be channeled around the PNS due to their associated angular momentum. 

When strong and frequent downflows cause accretion emission of neutrinos dominantly on one side of the PNS, neutrinos will kick the PNS against the main direction of the hydrodynamic kick. This may sound counter-intuitive, but it can be understood by the fact that the ejecta-induced kick usually points towards the hemisphere with the higher downflow activity, because the explosion is stronger and thus the momentum of the outward ejecta expansion is higher in the opposite hemisphere. In contrast, when downflows do not reach as close to the PNS surface and are not accreted by the PNS, they mostly absorb and scatter neutrinos from the PNS instead of producing accretion emission. The absorbed and scattered neutrinos will transfer momentum to the downflows, thus increasing the gas momentum on this side of the PNS. Consequently, the hydrodynamic kick of the PNS associated with asymmetric gas ejection will be reduced, because this kick is directed towards the downflows and therefore it is collinear with the extra gas momentum caused by neutrino absorption and scattering. The reduction of the effective hydrodynamic kick, however, is compensated by a neutrino kick aligned with the hydrodynamic kick, since the neutrino absorption and scattering in the downflows lead to a relative enhancement of the neutrino radiation (as present in the free-streaming limit) in the opposite hemisphere.

As a special case, one can consider a PNS that radiates neutrinos in its rest frame equally to all directions. Such a PNS does not receive a kick by its neutrino emission. Similarly, if some of these neutrinos interact in downdrafts that are detached from the PNS, this does not directly transfer recoil momentum to the NS. Instead, the neutrinos that

\end{multicols}

\twocolumn

\noindent
get absorbed or scatter in the downdrafts transfer momentum to the gas in the downdrafts. This changes the linear momentum in the ejecta, which is an effect that can potentially be diagnosed by observable asymmetries in SNe and their gaseous remnants. The asymmetry of the neutrino radiation at large distances that is caused by the neutrino interactions in the downdrafts cannot be measured experimentally, but its proper calculation is crucial for computing the correct momentum budget of the NS.

\begin{figure*}[h!]
        \centering
        \includegraphics[width=\textwidth]{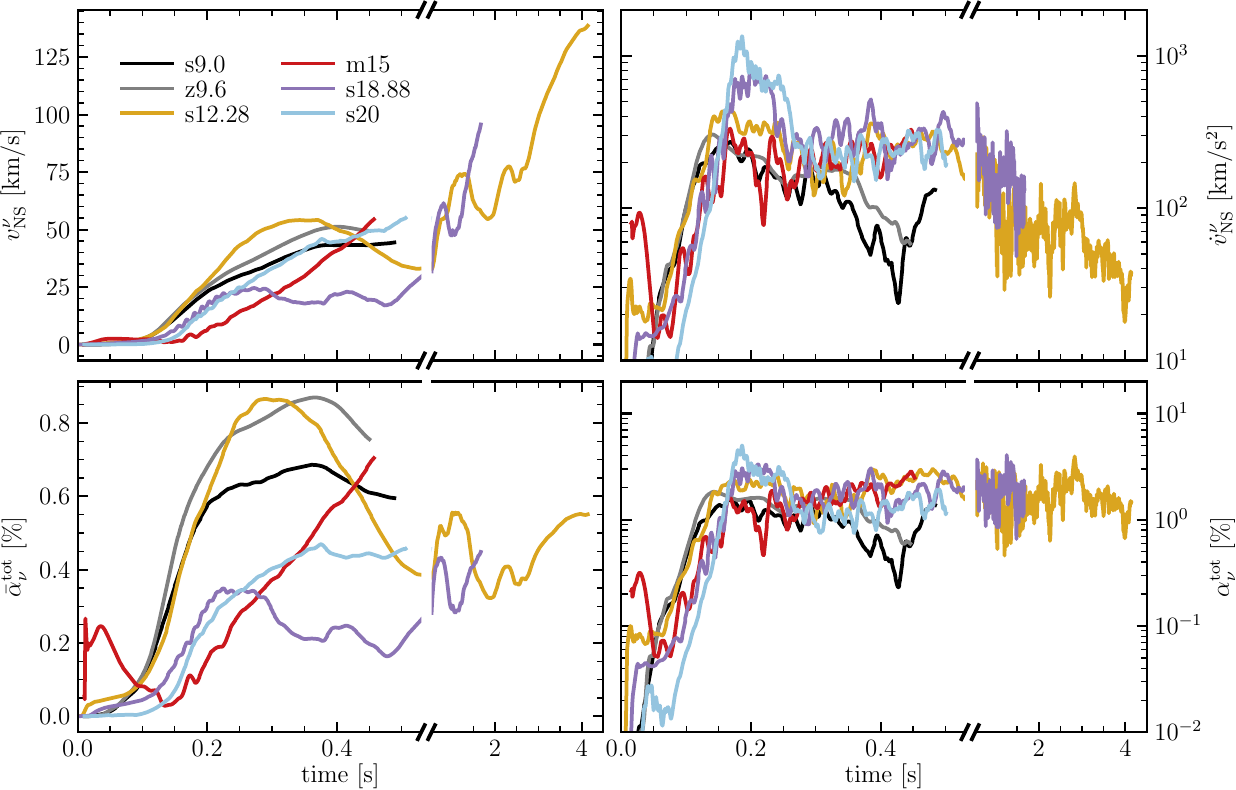}
        \caption{Neutrino-induced PNS kick velocity ({\em top left}), the corresponding PNS acceleration by anisotropic neutrino emission ({\em top right}), the time-integrated anisotropy parameter of the total neutrino emission (Equation~\ref{eq:baralpnu}; {\em bottom left}), and the corresponding instantaneous neutrino anisotropy parameter (Equation~\ref{eq:alpnu}; {\em bottom right}) for all successfully exploding models}
        \label{fig:vnu_anu_alpha}
\end{figure*}

A proper balancing of these counter-working effects of hydrodynamic and neutrino kicks therefore requires the measurement of the momentum asymmetry of the escaping neutrinos in the free-streaming regime at large distances on the one hand and the integration of the entire ejecta momentum on the other hand, as already discussed in Sect.~\ref{sec:PNSkickeval}.\footnote{Alternatively, one can calculate neutrino and hydrodynamic forces acting on the PNS surface at a suitably chosen radius. We will further comment on this issue in Section~\ref{sec:noteeval}.}

One can verify these elementary facts on the basis of our detailed analysis of the neutrino kicks for NSs and BHs in our set of 3D simulations.

\subsubsection{Neutrino kicks of neutron stars}
\label{sec:NSneutrinokicks}

Table~\ref{tab:neutrino_kicks} provides an overview of our results for the neutrino-induced PNS kicks in all of our 3D simulations at time $t_\mathrm{f}^\nu$ until which the \textsc{Vertex} neutrino transport was applied, and Figure~\ref{fig:vnu_anu_alpha} displays the time evolution of characteristic quantities. The neutrino kicks start rising within $\sim$100\,ms after bounce and typically reach higher values more quickly than the hydrodynamic kicks (see Figures~\ref{fig:models9}--\ref{fig:models20}). The former grow as soon as neutrino emission asymmetries develop inside the PNS or in its near-surface accretion layer, whereas the hydrodynamic kicks find a stable direction only after the explosion has set in and then reach higher values in a long-time trend of continuous growth in explosions with pronounced ejecta asymmetries (Figure~\ref{fig:vhyd_vtot}).

The magnitudes of the neutrino kicks reached within about 0.5\,s after bounce are on the order of 50\,km\,s$^{-1}$. While the instantaneous neutrino emission asymmetry parameters $\alpha_\nu^\mathrm{tot}$ fluctuate around a few percent (typically between roughly $\sim$1\% and $\sim$3\%), their time-averaged values are considerably lower (around $\sim$0.5--0.8\%; Table~\ref{tab:neutrino_kicks}) because of the rapid, short-time changes of the directions of the emission asymmetries (Figure~\ref{fig:vnu_anu_alpha}). It is remarkable that in our lower-mass models s9.0 and z9.6, the time-averaged asymmetry parameters of $\nu_e$ and $\bar\nu_e$, and in model s12.28 that of $\nu_e$ stick out by significantly higher values of 2--4\%. This fact is connected to the main origin of the neutrino emission anisotropy from a LESA dipole in these models. We will come back to this issue below.

From Figure~\ref{fig:vnu_anu_alpha}, we can also conclude that the time-averaged neutrino emission asymmetry converges to fairly stable values between about 0.3\% and 0.5\% after several seconds, despite continuous large-amplitude variations of the instantaneous values (see our models s12.28 and s18.88 with long-term neutrino transport). This persistent low-level emission asymmetry leads to a long-lasting neutrino-induced acceleration of the PNS (Figure~\ref{fig:vnu_anu_alpha}, upper right panel), which decreases with time as the neutrino luminosities of the cooling PNS gradually decline. These findings justify the assumptions that we made in our extrapolation formula for the neutrino kicks in Equation~(\ref{eq:vnuinfty}) for SN explosion of higher-mass progenitors. As described before (Section~\ref{sec:conv+LESA}), the two low-mass models s9.0 and z9.6 are special cases, where the LESA-dominated kicks exhibit a trend of saturation after a few 100\,ms post bounce (consistent with the LESA kicks in our long-term \textsc{Vertex} simulations of s12.28 and s18.88).

Inspecting Figures~\ref{fig:vnu_z96}, \ref{fig:vnu_s1228}, and~\ref{fig:vnu_s1888}, as well as the supplementary plots given in Appendices~\ref{secA1} and~\ref{secA:s12}, we can verify the different kinds of neutrino-induced kicks discussed before and visualized graphically in Figures~\ref{fig:lesakick} and \ref{fig:accretionkick}. 

\medskip\noindent
{\bf Model z9.6:}

Model z9.6 is the explosion of a low-mass progenitor with a particularly steep density gradient outside of its iron core, for which reason its bounce shock expands continuously and very rapidly. Therefore, the explosion is nearly spherical (see the value of $\bar\alpha_\mathrm{ej}$ in Table~\ref{tab:hydro_kicks}), and long-lasting accretion downflows towards the PNS do not occur \citep{Melson+2015a,Stockinger+2020}. Accordingly, the hydrodynamic NS kick remains very small (Figures~\ref{fig:vhyd_vtot} and~\ref{fig:scatter_plot}) and the neutrino kick is dominated by the LESA asymmetry of the neutrino emission.

\begin{figure*}[t]
        \centering
        \includegraphics[width=\textwidth]{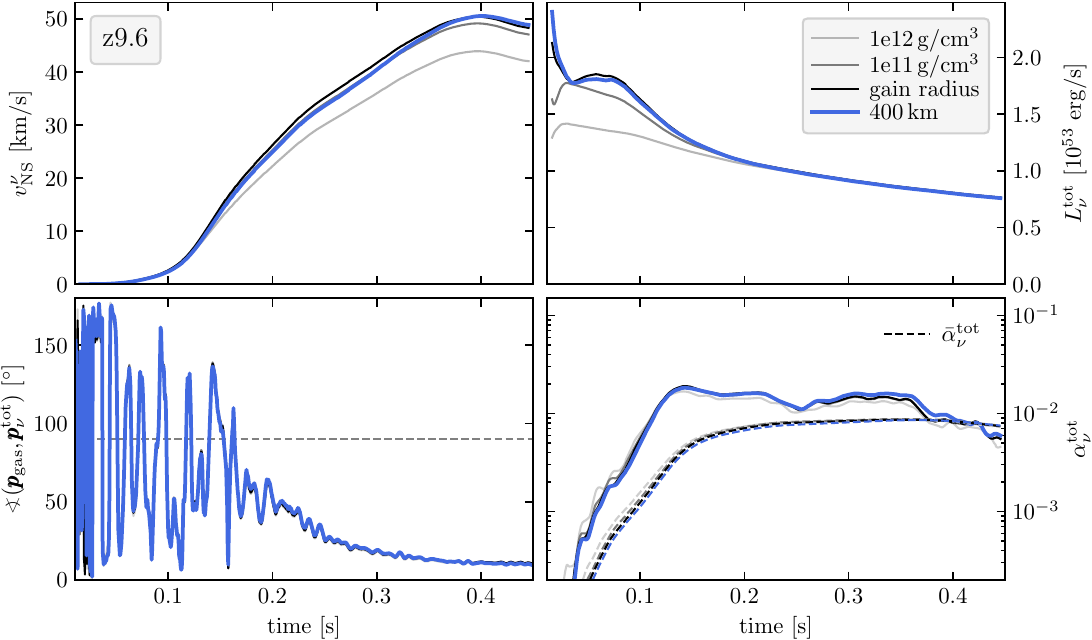}
        \caption{Neutrino-induced NS kick velocity ({\em top left}), total neutrino luminosity ({\em top right}), angle between vector directions of the linear gas momentum and the linear momentum of the neutrino emission ({\em bottom left}),
        instantaneous neutrino anisotropy parameter (Equation~\ref{eq:alpnu}; {\em bottom right}, solid lines), and time-integrated neutrino anisotropy parameter (Equation~\ref{eq:baralpnu}; dashed lines)
        as functions of time for model z9.6. Thick blue lines show the results for our standard evaluation at a fixed radius of 400\,km (i.e., in the free-streaming regime of the neutrinos). Light gray, dark gray, and black thin lines indicate the corresponding results for evaluation at the (angle-averaged) densities of $10^{12}$\,g\,cm$^{-3}$ and $10^{11}$\,g\,cm$^{-3}$, and at the (angle-averaged) gain radius, respectively. Model z9.6 exhibits the typical properties of a LESA-dominated neutrino kick, e.g., aligned neutrino and hydrodynamic kicks, because the low-mass progenitor explodes quickly and long-existent, massive accretion downflows around the PNS are absent. The analysis of the NS kick at all chosen locations yields similar results}
        \label{fig:vnu_z96}
\end{figure*}

The lower left panel of Figure~\ref{fig:vnu_z96} shows the trend of aligned hydro and neutrino kicks that is expected when the LESA dipole asymmetry of the neutrino emission determines also the explosion asymmetry. The mass ejection will then be most powerful in the direction of the highest summed luminosity of $\nu_e$ and $\bar\nu_e$, because these two neutrino species dominate the postshock heating by their absorption on free nucleons. The neutrino-driven explosion is, therefore, strongest in the direction opposite to the LESA vector (i.e., opposite to the dipole vector of the $\nu_e$ minus $\bar\nu_e$ number flux) and thus opposite to both hydro and neutrino kicks \citep[see also][]{Stockinger+2020}. 

In the bottom right panel of Figure~\ref{fig:modelz96}, we witness the expected, LESA-typical alignment of the PNS kicks caused by $\bar\nu_e$ and $\nu_x$ emission asymmetries, which are anti-aligned with the $\nu_e$-induced NS kick, because the direction of the $\nu_e$-luminosity dipole vector is opposite to those of the $\bar\nu_e$ and $\nu_x$ luminosity dipoles (Figure~\ref{fig:lesakick}). The panel also shows that the hydrodynamic acceleration of the NS (represented by $\vec{\dot{p}}_\mathrm{gas}$) fluctuates mostly in directions perpendicular to the acceleration vector of the total neutrino kick (represented by $\vec{\dot{p}}_\nu^\mathrm{tot}$), because of considerable stochastic variations of the directions of the accelerating hydrodynamic forces (weakened in Figure~\ref{fig:modelz96} because the curves are smoothed by applying a running average). For random relative directions of hydro and neutrino accelerations, the angles between both would be symmetrically distributed around a maximum at 90$^\circ$. However, because of the LESA influence on the explosion asymmetry, there is a statistical preference to angles of less than 90$^\circ$, in particular between $\sim$150\,ms and $\sim$400\,ms, so that the net effect leads to aligned directions of the (time-integrated) hydro and neutrino kicks (Figure~\ref{fig:vnu_z96} and third panel from top in the right column of Figure~\ref{fig:modelz96}).

\medskip\noindent
{\bf Model s9.0:}

Some of the phenomena found in model z9.6 can similarly be noticed in model s9.0 (Figure~\ref{fig:models9}), which also has a small hydrodynamic NS kick, mainly because of its low explosion energy and despite a bigger ejecta asymmetry $\bar\alpha_\mathrm{ej}$ than in model z9.6 (see Table~\ref{tab:hydro_kicks} and Equation~\ref{eq:vhydscaling} for the scaling relations). Although model s9.0 exhibits the described characteristic features of a LESA-dominated neutrino kick most of the time, e.g., aligned acceleration directions by $\bar\nu_e$ and $\nu_x$ emission dipoles, anti-aligned to the $\nu_e$-induced kick, there are times when the clear LESA situation in this model is perturbed by accretion downflows. This happens, for example, between $\sim$350\,ms and $\sim$450\,ms after bounce, when massive downflows mainly in the hemisphere opposite to the LESA vector absorb and scatter neutrinos (accretion emission of $\nu_e$ and $\bar\nu_e$ is only a secondary effect). This has several consequences, all of which are visible in Figure~\ref{fig:models9}. First, the NS's hydro kick, which in model s9.0 is not determined by LESA and has a large angle relative to the LESA direction \citep{Stockinger+2020}, begins to grow during this phase shortly after the onset of the explosion (top and second panels in right column); second, the angle between neutrino-induced and hydrodynamic kick velocities finds a stable, slowly evolving direction (third panel from top in right column); and third, the angle between $\nu_e$-induced PNS acceleration (in the anti-LESA direction) and $\bar\nu_e$ and $\nu_x$-induced accelerations shrinks due to the neutrino interactions in the downflows (bottom right panel). However, the direction of the neutrino kick is quite stable and largely unaffected by the downflows. Overall, in model s9.0, in contrast to the situation in model z9.6, the final kick geometry is not affected by the LESA emission dipole. This means that the hydrodynamic kick and the LESA-dominated neutrino kick are basically uncorrelated, implying that the relative angle between hydro kick and neutrino kick at time $t_\mathrm{f}^\nu$ is large, nearly 80$^\circ$ (Table~\ref{tab:neutrino_kicks}).

Interestingly, in model s9.0 as well as in model z9.6, the time-averaged and instantaneous asymmetry parameters of each neutrino species, $\bar{\alpha}_{\nu_i}$ and $\alpha_{\nu_i}$, approach each other towards the end of our \textsc{Vertex} transport calculations (see Figures~\ref{fig:vnu_z96}, \ref{fig:models9}, and \ref{fig:modelz96}). The underlying reason is that in these two simulations, the postbounce phases of high mass accretion rates by the PNS last only for $\sim$140\,ms (in z9.6) and for $\sim$320\,ms (in s9.0) (upper left panels of Figures~\ref{fig:models9} and~\ref{fig:modelz96}), although in model s9.0 occasional downflows at later times can still cause perturbations of the general trends. The values of $\bar{\alpha}_{\nue}$ and $\bar{\alpha}_{\nuebar}$ are particularly large in these simulations where the LESA dipole dominates the neutrino emission anisotropy, namely several percent instead of typically less than about 1\% (see Table~\ref{tab:neutrino_kicks}).

\begin{figure*}[t]
        \centering
        \includegraphics[width=\textwidth]{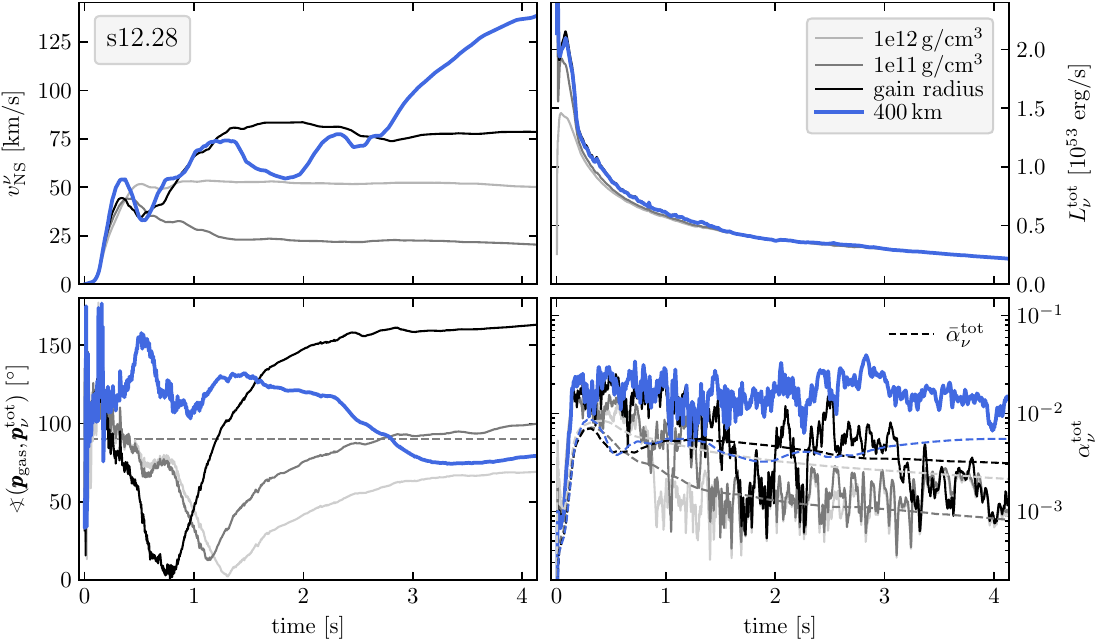}
        \caption{Same as Figure~\ref{fig:vnu_z96}, but for model s12.28. The model is an intermediate case that shows characteristic features of neutrino-induced NS kicks due to the LESA asymmetry in the first $\sim$0.5\,s after bounce, due to asymmetric accretion emission between $\sim$0.5\,s and $\sim$1.5\,s, and due to neutrino absorption and scattering in anisotropic downflows at later times. The results at $t \gtrsim 0.5$\,s after bounce depend strongly on the chosen location of the kick analysis}
        \label{fig:vnu_s1228}
\end{figure*}

\medskip\noindent
{\bf Model s12.28:}

In model~s12.28, neutrino and hydrodynamic NS kicks are of comparable magnitudes (Figure~\ref{fig:scatter_plot}) with a big angle of around 80$^\circ$ between the two vector directions at late times (similar to model s9.0; Tables~\ref{tab:hydro_kicks} and~\ref{tab:neutrino_kicks}). This model is an intermediate case where LESA as well as long-lasting downflows and accretion around the PNS play similarly important roles for the neutrino kicks. Accretion occurs at a considerable strength for roughly the first 1.5\,s after bounce and on a lower level even later. The corresponding neutrino emission is visible by enhanced $\nu_e$ and $\bar\nu_e$ luminosities that show variations over time (Figure~\ref{fig:models1228}). This can also be concluded from Figure~\ref{fig:vnu_s1228}, which demonstrates the important influence of emission and absorption/\-scattering of neutrinos exterior to the neutrinospheres by the fact that the evaluation of the neutrino-induced kick yields drastically different results for magnitude and direction, depending on whether the analysis is performed around and interior to the neutrinospheres (i.e., at densities of $10^{11}$\,g\,cm$^{-3}$ and $10^{12}$\,g\,cm$^{-3}$), or at the gain radius, or in the free-streaming regime at 400\,km. Moreover, since the final angle between hydro and neutrino kicks is large, the ejecta geometry does not seem to be strongly influenced by the LESA asymmetry of the neutrino emission of the PNS in this model. 

These aspects manifest themselves in a variety of specific properties of the total neutrino kick in Figures~\ref{fig:vnu_s1228} and~\ref{fig:models1228}. During the first $\sim$0.5\,s after bounce, the LESA asymmetry of the PNS's neutrino emission mainly determines the neutrino kick, because the analysis for $v_\mathrm{NS}^\nu$, $\alpha_\nu^\mathrm{tot}$, and the angle between hydro kick and neutrino kick yields very similar results at all of the different locations chosen in Figure~\ref{fig:vnu_s1228}. In the following period from 0.5\,s until $\sim$1.5\,s after bounce, the total neutrino kick is accretion-emission dominated and thus anti-aligned to the hydro kick, signaled by a large angle relative to the latter, as expected from the left panel of Figure~\ref{fig:accretionkick}. This is consistent with the fact that the neutrino kick evaluated at the gain radius differs from the kicks obtained at $10^{11}$\,g\,cm$^{-3}$ and $10^{12}$\,g\,cm$^{-3}$ but shows good agreement with the result obtained in the free-streaming limit during this period (Figure~\ref{fig:vnu_s1228}). At later times, the angle between hydro and neutrino kicks shrinks, suggesting that the neutrino kick is now strongly influenced by neutrino scattering and absorption in downflows that do not reach the PNS surface (as sketched in Figure~\ref{fig:accretionkick}, right panel). This picture is confirmed by the fact that the neutrino kick evaluated at 400\,km significantly differs from the result obtained at the gain radius (Figure~\ref{fig:vnu_s1228}).

Further support of this picture is provided by an evaluation of the neutrino kicks of the different neutrino species individually in Figures~\ref{fig:s1228nue}, \ref{fig:s1228barnue}, and~\ref{fig:s1228nux}. In the first $\sim$0.5\,s after bounce, one can witness location-independent kick results as well as the LESA-characteristic opposing directions of the $\nu_e$ kick on the one hand and the $\bar\nu_e$ and $\nu_x$ kicks on the other hand, which thus compensate each other partially. This is visible from the angles between $\vec{p}_{\nu_i}$ and $\vec{p}_\mathrm{gas}$ and from the direction angles $(\vartheta,\varphi)$ of the different neutrino kicks in a global polar coordinate system of the 3D simulation. During the earliest phase, $t\lesssim 0.2$\,s after bounce, one can also notice a LESA-atypical tendency that the (still small) linear gas momentum is aligned with the $\nu_e$ momentum vector and anti-aligned with the $\bar\nu_e$ and $\nu_x$ momentum vectors, which confirms the previous statement that the LESA neutrino emission asymmetry of the PNS has no strong influence on the gas flows between PNS and SN shock. At $t\gtrsim 0.2$\,s, the gas momentum direction moves to the hemisphere opposite to the $\nu_e$ momentum vector (relative angle around 100$^\circ$), before at $t\gtrsim 0.4$\,s, massive accretion downflows develop in the $-z$ hemisphere. This leads to a migration of the vector direction of the gas momentum, $\vec{p}_\mathrm{gas}$, towards the $+z$ axis ($\vartheta \to 0$) of the global coordinate system and thus, incidentally more LESA-compatible, it is now directed nearly against the $\nu_e$ dipole \citep[see][]{Tamborra+2014}. Since the corresponding $\nu_e$ momentum vector of the LESA emission points towards the south ($\vartheta \approx 150^\circ$), the additional $\nu_e$ emission by accretion in the same hemisphere strengthens the $\nu_e$-induced NS kick until about 1.5\,s (visible by the large difference between the kicks evaluated at $10^{11}$\,g\,cm$^{-3}$ and $10^{12}$\,g\,cm$^{-3}$ on the one hand and at the gain radius and in the free streaming regime on the other hand; Figure~\ref{fig:s1228nue}). 

Since the momentum direction of the $\nu_e$ emission remains nearly stable during all of this evolution and at any of the chosen locations for evaluation in Figure~\ref{fig:s1228nue}, the time-averaged asymmetry parameter $\bar{\alpha}_{\nu_e}$ is close to the magnitude of the fluctuating values of $\alpha_{\nu_e}$, similar to our findings for $\bar{\alpha}_{\nu_e}$ and $\bar{\alpha}_{\bar{\nu}_e}$ in the LESA dominated models z9.6 and s9.0. At times $t \gtrsim 1.5$\,s, accretion emission becomes negligibly small and the $\nu_e$-induced NS kick remains nearly constant. Since the gas momentum direction drifts, there is a continuous, slow change of the angle between $\vec{p}_{\nu_e}$ and $\vec{p}_\mathrm{gas}$. 
 
The massive accretion downflows that develop in the $-z$ hemisphere between $\sim$0.4\,s and $\sim$1.5\,s after bounce reduce the $\bar\nu_e$-induced NS kick, in contrast to their effect on the $\nu_e$-induced kick. Since the LESA dipole direction of the $\bar\nu_e$ emission is opposite to the one of $\nu_e$, the additional accretion emission of $\bar\nu_e$ points against the $\bar\nu_e$ LESA dipole and thus leads to a decrease of the NS kick due to the anisotropically radiated $\bar\nu_e$ not only as a function of time, but also when the kick magnitudes are evaluated at locations outside the PNS rather than in the neutrinospheric region (see Figure~\ref{fig:s1228barnue}). Moreover, the accretion emission of $\bar\nu_e$ in the hemisphere opposite to the $\bar\nu_e$ LESA vector causes a considerable migration of the $\vec{p}_{\bar\nu_e}$ vector when computed at locations exterior to the PNS. 

For the NS kick by the $\nu_x$ emission, similar effects as for $\bar\nu_e$ can be observed (Figure~\ref{fig:s1228nux}). But here the accretion in the southern hemisphere between $\sim$0.4\,s and $\sim$1.5\,s tends to suppress the $\nu_x$ diffusion out of the PNS on this side. It therefore enhances the $\nu_x$ LESA dipole pointing northward and thus the $\nu_x$-induced kick evaluated at the gain radius. This effect, however, is nearly compensated by asymmetric $\nu_x$ scattering in the gain layer at larger distances, for which reason the NS kick at 400\,km increases only little until $\sim$1.5\,s. Only at later times, new downdrafts (at new directions, as indicated by the shift of $\vec{p}_\mathrm{gas}$) cause considerable $\nu_x$ scattering between the gain radius and 400\,km (but hardly affect the $\nu_e$ and $\bar\nu_e$ emission). After $\sim$2.4\,s, this increases the $\nu_x$-induced NS kick by nearly a factor of two, connected with a considerable change of the $\nu_x$ momentum direction (Figure~\ref{fig:s1228nux}, bottom left panel). An effect of the same kind can be witnessed for $\bar\nu_e$ at $t \gtrsim 3$\,s and at a radius of 400\,km (Figure~\ref{fig:s1228barnue}). The drifting momentum directions of the radiated $\bar\nu_e$ and $\nu_x$ in the free streaming limit lead to time-averaged values of their $\bar{\alpha}_{\nu_i}$ that are considerably lower than the (fluctuating) magnitudes of the $\alpha_{\nu_i}$ because of the compensating effects of acceleration phases in different directions (Figure~\ref{fig:models1228}). Only the analysis of $\bar{\alpha}_{\bar\nu_e}$ and $\bar{\alpha}_{\nu_x}$ at densities of $10^{11}$\,g\,cm$^{-3}$ and $10^{12}$\,g\,cm$^{-3}$ reveals directional stability, because at these radii the LESA dipole determines the $\bar\nu_e$ and $\nu_x$ emission asymmetries (Figures~\ref{fig:s1228barnue} and~\ref{fig:s1228nux}).

\begin{figure*}[t]
        \centering
        \includegraphics[width=\textwidth]{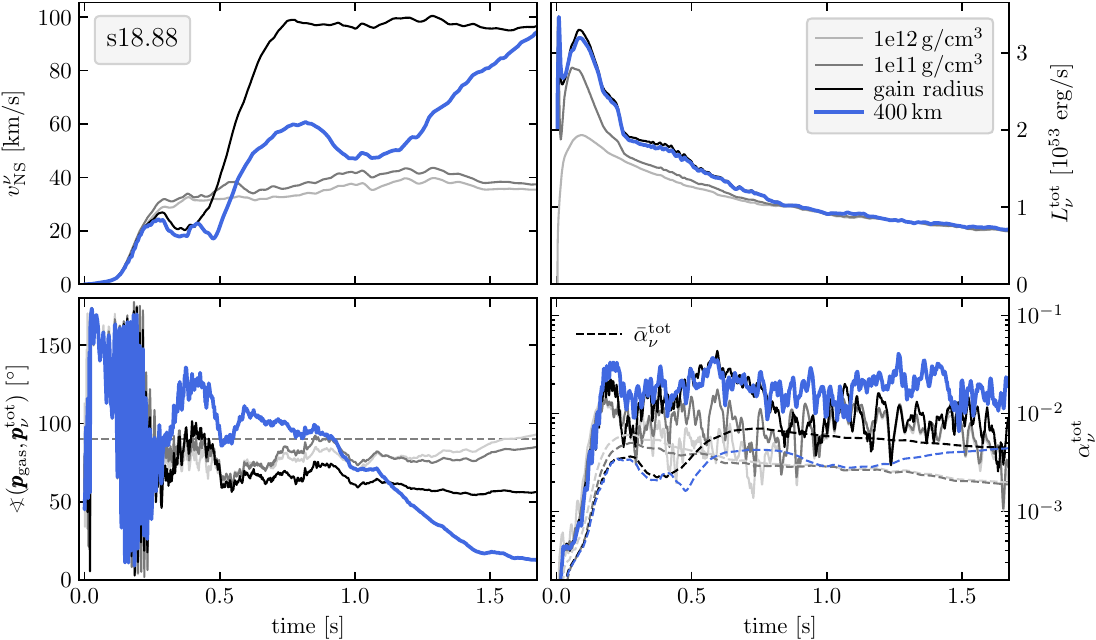}
        \caption{Same as Figure~\ref{fig:vnu_z96}, but for model s18.88. In this case the neutrino-induced NS kick is determined by the neutrino flux asymmetry caused by the LESA dipole before $\sim$0.25\,s after bounce, then it is mainly connected to $\nu_e$ and $\bar\nu_e$ emission and $\nu_x$ scattering in anisotropic accretion flows onto the PNS until $\sim$0.75\,s, and at later times by neutrino absorption and scattering in asymmetric downflows that do not reach the PNS but interact with neutrinos exterior to the gain radius. This can be concluded from the times when the kick velocities evaluated at different locations start to separate from each other or level off at nearly constant values}
        \label{fig:vnu_s1888}
\end{figure*}

\medskip\noindent
{\bf Model m15:}

Model m15 explodes by the neutrino-powered mechanism, because the rapid rotation of the 15\,M$_\odot$ star favors the development of a strong spiral-SASI mode that supports the shock expansion preferentially around the equatorial plane. The shock front possesses a pronounced oblate shape with a sizable dipolar deformation along the equator and accretion downstreams at the poles \citep{Summa+2018}. Due to the rapid rotation of the progenitor, the NS spin vector basically coincides with the progenitor's rotation axis, and the final spin period of the NS is estimated to be about 2\,ms at the end of our hydrodynamic simulations at $t_\mathrm{f}$ (Table~\ref{tab:hydro_kicks}). Here, angular momentum acquired by accretion flows and lost by mass outflow is taken into account according to the evaluation described in Appendix~D of \citet{Stockinger+2020}, but angular momentum carried away by neutrino emission is not included \citep[such an angular momentum drain through neutrinos might increase the final NS spin period by several 10\%;][]{Janka2004}.   

The vector directions of the hydrodynamic kick and of the neutrino kick in model m15 are close to the equatorial plane (see Section~\ref{sec:hydrokicks}), and both have a large angle relative to each other, namely about 50$^\circ$ at $t_\mathrm{f}^\nu$ (Table~\ref{tab:neutrino_kicks} and Figure~\ref{fig:modelm15}) and around 100$^\circ$ at $t_\mathrm{f}$ (Table~\ref{tab:hydro_kicks}). The neutrino kick in this model dominates clearly until the \textsc{Vertex} neutrino transport calculation was stopped at $t_\mathrm{f}^\nu$, because the hydrodynamic kick velocity exhibits large-amplitude and quasi-periodic directional changes until it finds a stable direction only about 150\,ms after the onset of the explosion (Figure~\ref{fig:modelm15}, third panel from top in the right column).  

Since LESA is suppressed in the rapidly rotating PNS \citep[see][]{Walk+2019}, the NS acceleration (momentum derivative) vectors of $\nu_e$ and $\bar\nu_e$ are relatively close to each other (Figure~\ref{fig:modelm15}, bottom right panel). This is expected for an accretion-dominated flux asymmetry, and the importance of accretion emission also explains why the neutrino kick vector stays close to the equatorial plane, because accretion flows from all directions carry large angular momentum, which forces them onto wound trajectories around the PNS's equator. The momentum vector of the $\nu_x$ emission shows only a loose correlation with those of $\nu_e$ and $\bar\nu_e$, because $\nu_x$ mainly originate from the interior of the PNS. Scattering of $\nu_x$ in the accretion flows plays a secondary, but non-negligible role, driving the $\nu_x$ momentum vector also close to the equatorial plane, for which reason the total neutrino kick of the NS is nearly perpendicular to the progenitor's spin axis. Rapid directional variations of the accretion flows lead to a large reduction (by a factor of 10--20) of the time-averaged emission asymmetry parameters $\bar{\alpha}_{\nu_e}$ and $\bar{\alpha}_{\bar\nu_e}$ compared to the instantaneous values of $\alpha_{\nu_e}$ and $\alpha_{\bar\nu_e}$, whereas the reduction is far less extreme (a factor of 2--3 near the end of the \textsc{Vertex} transport calculation) for the more stable $\nu_x$ emission.

\medskip\noindent
{\bf Models s18.88 and s20:}

These two explosion simulations of massive progenitors resemble each other fairly closely in their overall behavior, but model s18.88 was computed with \textsc{Vertex} neutrino transport more than three times longer. Both models are cases where accretion downflows to the PNS clearly determine the neutrino emission asymmetry and the final neutrino-induced kick. Again, the neutrino kick velocities grow initially faster than the hydrodynamic kick velocities, because the latter exhibit rapid, large-amplitude variations of their directions relative to the neutrino kick, which are essentially symmetrically distributed around a relative angle of 90$^\circ$ as expected for randomly varying orientations. When ultimately the explosions take off and develop clear and more stable global asphericities, also the hydrodynamic kick velocities begin to rise (Figures~\ref{fig:models1888} and~\ref{fig:models20}). In the end, the hydrodynamic kicks clearly dominate the total NS kicks in these two models (Tables~\ref{tab:hydro_kicks} and~\ref{tab:neutrino_kicks}; Figures~\ref{fig:vhyd_vtot} and~\ref{fig:scatter_plot}).  

When the LESA dipole in the PNS becomes sizable after $\sim$0.1\,s of postbounce evolution, the LESA emission asymmetry determines the neutrino kick of the PNS, which therefore starts to rise steeply at about this time. The LESA effect is crucial during the first $\sim$0.25\,s after bounce, signaled by the almost perfect agreement of the kick velocities evaluated at all of the different locations in Figure~\ref{fig:vnu_s1888}, indicating that the neutrino kick is defined by anisotropic neutrino transport inside the PNS. Moreover, a detailed inspection shows (the plots are not included in this paper) that the momentum vectors of the radiated $\nu_e$ and $\bar\nu_e$, $\vec{p}_{\nu_e}$ and $\vec{p}_{\bar{\nu}_e}$, are oriented in nearly opposite directions, and $\vec{p}_{\nu_x}$ and $\vec{p}_{\bar{\nu}_e}$ are nearly collinear during this phase, fully consistent with the LESA emission geometry.
For more details on the LESA in model s18.88, the reader is referred to the analysis presented in Appendix~C of \citet{Bollig+2021}.

At $\sim$0.25\,s, the kick velocities computed at the gain radius and in the free-streaming regime at 400\,km begin to deviate considerably from those obtained at densities of $10^{11}$\,g\,cm$^{-3}$ and $10^{12}$\,g\,cm$^{-3}$. At that time, emission of $\nu_e$ and $\bar\nu_e$ from accretion downflows becomes important, mostly taking place in the hemisphere the LESA lepton-number dipole vector points to \citep[as expected if LESA would affect the accretion flows around the PNS;][]{Tamborra+2014}. This accretion emission strengthens the $\nu_e$ flux in this hemisphere and thus increases the $\nu_e$-induced NS kick compared to the LESA-determined kick, whereas it reduces the $\bar\nu_e$-induced NS kick, because the $\bar\nu_e$ LESA dipole vector is in the opposite direction. Since enhanced scattering in the accretion layer damps the $\nu_x$ diffusion out of the PNS on the weaker side of the LESA lepton-number dipole, the $\nu_x$ emission asymmetry and corresponding NS kick are also amplified when measured at the gain radius. This enhancement, however, is compensated to a large degree by $\nu_x$ scattering exterior to the gain radius so that, effectively, the net NS kick by the $\nu_x$ flux is not much altered (only slightly reduced or raised) in the free streaming regime compared to the results obtained in the neutrinospheric region. Until about (0.8--1)\,s after bounce, the accretion-affected contributions of all neutrino species therefore decrease or increase the total neutrino-induced NS kick at 400\,km only moderately relative to the values computed at $10^{11}$\,g\,cm$^{-3}$ and $10^{12}$\,g\,cm$^{-3}$ (Figure~\ref{fig:vnu_s1888}).    

While earlier than (0.8--1)\,s after bounce neutrino emission and scattering in mass flows accreted by the PNS have an influence, accretion becomes insignificant at later times, but absorption and scattering of high-energy $\bar\nu_e$ and $\nu_x$ in asymmetric downflows still play a role outside of the gain radius. This effect mostly concerns $\bar\nu_e$ and $\nu_x$ because of their higher-energy spectra. It is visible in Figure~\ref{fig:vnu_s1888} by the evolution of the relative angle between hydrodynamic NS kick and neutrino-induced NS kick from values larger than 90$^\circ$, which is expected for the accretion phase (Figure~\ref{fig:accretionkick}, left panel), to angles of less than 90$^\circ$, which is a characteristic feature of neutrino interaction in non-accreted downflows (Figure~\ref{fig:accretionkick}, right panel). Moreover, the persistent importance of absorption and scattering of $\bar\nu_e$ and $\nu_x$ in asymmetrically distributed matter in the gain layer at $t \gtrsim 0.8$\,s after bounce can also be concluded from the fact that their NS kicks deduced at the gain radius and at 400\,km exhibit large differences. 

All such effects had also been witnessed in model s12.28 (Figure~\ref{fig:vnu_s1228}), where the interplay of the LESA dipole fluxes with the $\nu_e$, $\bar\nu_e$ accretion emission and $\nu_x$ scattering determined the final PNS kick velocity deduced from the neutrino flux asymmetries measured at large radii. However, in model s18.88, the late-time scattering effects in non-accreted downflows are so strong that the hydrodynamic and neutrino kicks at $t_\mathrm{f}^\nu$ possess a small relative angle of $\sim$10$^\circ$ only (Table~\ref{tab:neutrino_kicks}). During the subsequent long-time evolution, however, the relative angle grows because of changes in the ejecta geometry, and it is large at the end of the simulations in all of the four models of s12.28, s18.88, s20, and s20e (Table~\ref{tab:hydro_kicks}). Remarkably, there is a sizable difference between our two long-term simulations of the 20\,M$_\odot$ model because of their different explosion geometries \citep{Kresse2023}.
Another significant difference between s12.28 on the one hand and s18.88 and s20 on the other hand is the aforementioned fact that in model s12.28 the time-integrated anisotropy parameter $\bar{\alpha}_{\nu_e}$ attains a value close to the (fluctuating) instantaneous values of $\alpha_{\nu_e}$ (Figure~\ref{fig:models1228}), whereas in the s18.88 and s20 models the time-averaged emission asymmetry parameters of all neutrino species become much smaller than their instantaneous values (Figures~\ref{fig:models1888} and~\ref{fig:models20}).

\subsubsection{A note on the evaluation radius}
\label{sec:noteeval}

As we have seen in Figures~\ref{fig:vnu_s1228} and~\ref{fig:vnu_s1888} as well as Figures~\ref{fig:s1228nue}--\ref{fig:s1228nux}, the results for the neutrino kicks can vary drastically depending on the radial position where the neutrino flux anisotropy is determined. Only our ECSN-like model z9.6 is an exception and shows less sensitivity (Figure~\ref{fig:vnu_z96}), because, in this case, asymmetric downflows to the PNS and long-time non-spherical accretion do not play a role. 

There are two alternative approaches to evaluate neutrino-hydrodynamic SN simulations for the resulting NS kicks: Either one considers the balance of the NS's momentum and the total linear momentum carried by the ejected gas and radiated neutrinos as measurable by distant observers (Section~\ref{sec:massdepanalysis}). Or, alternatively, one calculates the total NS kick from the accelerating forces acting at a chosen radius that defines the PNS surface, hydrodynamic forces according to Section~\ref{sec:forces} as well as those connected to neutrinos (Equations~\ref{eq:pdotnui} and~\ref{eq:pdotnu}). This radius has to be the {\em same} for evaluating the hydrodynamic and neutrino kicks. Although the sum of both kick contributions is largely insensitive to the exact radius of evaluation, the individual kicks depend strongly on the selected radius, which implies ambiguity in the interpretation of the values. If the radii for calculating the hydrodynamic and neutrino kicks were chosen to be different, it can imply a double counting of momentum contributions. We mentioned this fact already in Section~\ref{sec:massdepanalysis} and demonstrated by results in Section~\ref{sec:NSneutrinokicks} that such a choice can be problematic, because it leads to an incorrect balancing of hydrodynamic plus neutrino-induced kicks (see also our discussion of neutrino interactions in accretion downdrafts in Section~\ref{sec:neutrino+accretion}).

Therefore, our standard evaluation is based on calculating the linear momentum that can be measured at infinity for both ejecta and radiated neutrinos, using appropriate ways for evaluating the individual contributions, as discussed in Section~\ref{sec:massdepanalysis}. In our opinion, this is the most consistent and unambiguous approach to determine the final NS kicks, because it has a clear interpretation of the underlying physics. Specifically, this approach requires for the hydrodynamic kick the computation of the linear momentum of the ejected gas, i.e., of the gas exterior to the surface of the compact remnant (or the momentary mass cut between ejecta and NS remnant) at any time. In addition, for the neutrino kick it requires a consideration of the linear momentum of the neutrino flux in the free-streaming limit, where further interactions of neutrinos with the stellar matter are no longer relevant. This is ensured by our choice of 400\,km for the radius to calculate the linear momentum carried away by neutrinos that leave the SN interior \citep[following][]{Stockinger+2020}.

\begin{figure*}[t]
        \centering
        \includegraphics[width=\textwidth]{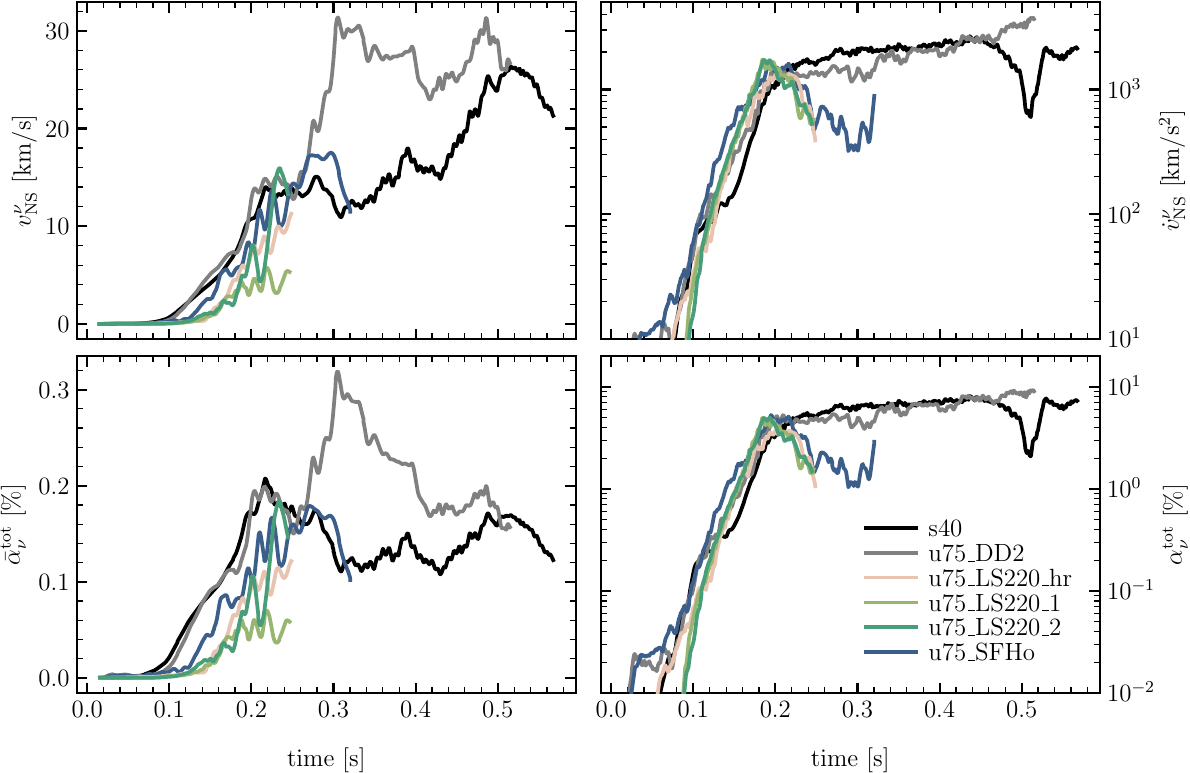}
        \caption{Neutrino-induced kick velocity of the PNS (or proto-BH) ({\em top left}), corresponding acceleration by anisotropic neutrino emission ({\em top right}), time-integrated anisotropy parameter of the total neutrino emission (Equation~\ref{eq:baralpnu}; {\em bottom left}), and corresponding instantaneous neutrino anisotropy parameter (Equation~\ref{eq:alpnu}; {\em bottom right}) for all BH-forming models}
        \label{fig:vnu_anu_alpha_bh}
\end{figure*}

\begin{figure*}[t]
        \centering
        \includegraphics[width=\textwidth]{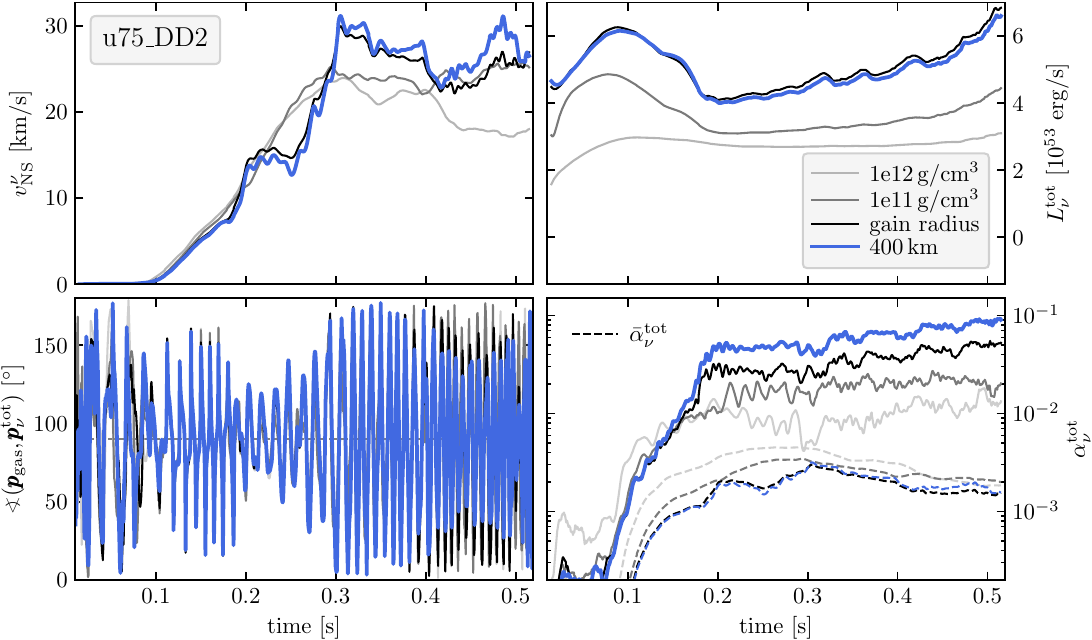}
        \caption{Same as Figure~\ref{fig:vnu_z96}, but for the BH forming model u75\_DD2. Up to roughly one third of the total neutrino luminosity is produced by neutrino emission at densities below $10^{11}$\,g\,cm$^{-3}$ in massive, non-spherical accretion flows onto the transiently existing PNS (or proto-BH). The values of the instantaneous anisotropy parameter of the neutrino emission ({\em bottom right}) are, therefore, very sensitive to the radial position where they are determined. Physics-wise meaningful results require an evaluation of the neutrino-induced kick in the free-streaming regime (400\,km)}
        \label{fig:vnu_u75DD2}
\end{figure*}

On the contrary, \citet{Coleman+2022} and \citet{Burrows+2023b} took the approach of evaluating the total NS kick from the accelerating forces, which has the disadvantage mentioned above that the decomposition of the kick contributions due to hydrodynamic and neutrino forces depends strongly on the choice of the evaluation radius and, therefore, its interpretation is not unambiguous. \citet{Coleman+2022} picked a radius that corresponds to the mass coordinate of the final NS after mass accretion has ended, i.e., they considered a location analogue to the coordinate of the ``mass cut'' often used in 1D models. Although this makes sense in principle, in exploding 3D models there is no Lagrangian mass coordinate of the progenitor that ultimately separates ejecta from NS material. Different from the 1D case, ejecta and NS matter in 3D simulations originate from different radii and thus different enclosed masses of the progenitor profile. In the more recent paper by \citet{Burrows+2023b}, the choice of the ``outer mass surface'' of the final PNS has been replaced by the PNS radius located at a density-isosurface of $10^{11}$\,g\,cm$^{-3}$. This may be the reason why the neutrino-induced kicks are lower in \citet{Burrows+2023b} than in the previous paper of \citet{Coleman+2022}.

Following the approach of \citet{Coleman+2022} for models that form BHs without any explosive mass ejection, one would have to calculate the forces at the edge of the collapsing progenitor, which defines the ultimate mass cut. Evaluated at the surface of the star in the case of complete collapse, or in a spherical progenitor also at any other large radius, the hydrodynamic forces would not transfer any linear momentum to the compact remnant. This is the correct final outcome if no aspherical mass ejection takes place. Such an analysis performed in the free-streaming regime would also yield the correct neutrino recoil kick associated with the linear momentum carried away by the radiated neutrinos. However, this analysis would not permit to monitor the temporary recoil that the PNS receives by anisotropically moving, transiently expanding postshock gas in its surroundings. Evaluating all forces due to hydrodynamic and neutrino effects at a small radius near the neutrinospheres or at the surface of the PNS instead, would lead to incorrect results for the neutrino kick, because it ignores the neutrino interactions in dense, aspherical postshock gas that the neutrinos need to cross before they escape to infinity. Only neutrinos and matter that get unbound from the compact remnant will contribute to its final kick.

We, therefore, apply our standard procedure of computing hydrodynamic and neutrino-induced kicks (according to Section~\ref{sec:massdepanalysis}) also to cases where our 3D simulations lead to BH formation, which means that we determine the (transient) hydro kicks from the gas momentum of (temporary) ejecta, and the neutrino kicks from the momentum of the neutrino flux in the free-streaming regime.

\subsubsection{Neutrino kicks of black holes}

Although some of our 3D simulations with BH formation exhibit a phase of transient shock expansion shortly before the PNS collapses, explosive and asymmetric mass ejection is not expected to happen in any of these models (see Section~\ref{sec:3Dsimulations}). Therefore, the residual BHs receive a natal kick only by the anisotropic loss of neutrinos.

Figure~\ref{fig:vnu_anu_alpha_bh} provides an overview of the evolution of the PNS kicks until the instant when the PNS becomes gravitationally unstable and our simulations were stopped. For the 75\,M$_\odot$ progenitor, this moment is reached between $\sim$250\,ms and 325\,ms after bounce, except with the stiff DD2 EoS, where it takes more than 500\,ms and thus similarly long as in the case of the 40\,M$_\odot$ model, which was simulated with the much softer LS220 EoS. The maximum PNS
kicks are only around 30\,km\,s$^{-1}$. Despite the enormous values of the acceleration of up to several 1000\,km\,s$^{-2}$ at the end of our computed evolution, one cannot count on any relevant further gain in velocity, because the anisotropically distributed postshock matter will be swallowed by the newly formed BH within a few milliseconds at most \citep[see][for similar simulations showing the evolution beyond BH formation]{Rahman+2022}. After this short period, the matter of the non-rotating progenitor stars will fall into the BH basically spherically and the neutrino emission will become very low and effectively isotropic.

Table~\ref{tab:neutrino_kicks} contains a summary of relevant quantities for our BH forming models, too. In the last two columns, we also provide estimates of the final BH kicks, which are smaller than the final PNS kicks by a factor $M_\mathrm{NS,g}(t_\mathrm{f}^\nu)/M_\mathrm{BH}(t\to\infty)$ because of momentum conservation: the final BH will be more massive but will possess the same momentum as the NS at the moment of its gravitational implosion. Here, we distinguish two cases, namely one where the entire pre-collapse mass of the progenitor ends up in the BH, and another case where just the progenitor's helium core is accreted and the hydrogen envelope becomes gravitationally unbound because of the so-called mass-decrement effect, i.e., because of the reduced gravitational potential due to the energy loss in neutrinos \citep{Lovegrove+2013,Nadezhin1980}. The $\sim$15.2\,M$_\odot$ BH formed in the collapse of our 40\,M$_\odot$ progenitor thus receives a natal kick of $\sim$3.3\,km\,s$^{-1}$, whereas the BH left by the 75\,M$_\odot$ progenitor will have a mass in the range between $\sim$54.7\,M$_\odot$ and $\sim$74\,M$_\odot$ and a neutrino-induced kick of less than $\sim$1\,km\,s$^{-1}$. A graphical overview of these results is presented in the right panel of Figure~\ref{fig:scatter_plot}.   

It is important to note that in spite of high values of the instantaneous neutrino-emission asymmetry parameter $\alpha_\nu^\mathrm{tot}$ of up to 10\%, the time-averaged values $\bar{\alpha}_\nu^\mathrm{tot}$ are one to two orders of magnitude lower and only in the range of 0.1\%--0.2\% (Figure~\ref{fig:vnu_anu_alpha_bh}). The enormous reduction is caused by the rapid variations of the kick directions, which force the average effect to much smaller amplitudes. This can be concluded from the lower left panel of Figure~\ref{fig:vnu_u75DD2} and the third panels from top in the right columns of Figures~\ref{fig:models40}--\ref{fig:modelsu75deg5sfho}. The short-timescale variability of the relative angle between hydrodynamic and neutrino kicks seen there is linked to rapid fluctuations of both the gas momentum vector and of the momentum vector of the anisotropically radiated neutrinos. These directional variations are connected to stochastic changes of massive, anisotropic accretion downflows to the PNS, whose neutrino emission fluctuates wildly. 

Figure~\ref{fig:vnu_u75DD2}, which depicts a representative example of our BH forming models, and Figures~\ref{fig:models40}--\ref{fig:modelsu75deg5sfho} reveal the crucial role of accretion emission for the neutrino emission asymmetry. The upper right panel of Figure~\ref{fig:vnu_u75DD2} demonstrates that up to $\sim$30\% of the total neutrino luminosity originate from the accretion mantle exterior to an (angle-averaged) density of $10^{11}$\,g\,cm$^{-3}$, and the lower right panel conveys the information that the emission from the layer between this density and the angle-averaged gain radius as well as neutrino scattering and absorption between the gain radius and 400\,km are mainly responsible for the high values of the instantaneous neutrino asymmetry parameter. Characteristic signatures of accretion-governed neutrino kicks are also visible in Figures~\ref{fig:models40}--\ref{fig:modelsu75deg5sfho}. On the one hand, the bottom right panels there exhibit nearly aligned directions of the force vectors $\vec{\dot{p}}_{\nu_i}$ for $\nu_e$, $\bar\nu_e$, and $\nu_x$ during most of the simulated periods of evolution. On the other hand, they also show that the hydro-kick force vector, $\vec{\dot{p}}_\mathrm{gas}$, and the total neutrino-kick force vector, $\vec{\dot{p}}_{\nu}^\mathrm{tot}$, are mostly in opposite hemispheres (i.e., the relative angle between them is larger than 90$^\circ$ most of the time).  Both of these findings are typical of the situation that is expected when massive accretion flows produce neutrinos of all species (see left panel of Figure~\ref{fig:accretionkick}). 

Only during short phases around $\sim$0.1\,s after bounce in models~s40 and~u75\_DD2, one can witness trends towards the LESA-typical anti-alignment between the $\nu_e$ acceleration direction on the one side and the $\bar\nu_e$ and $\nu_x$ acceleration directions on the other side (Figure~\ref{fig:lesakick}). These indications of LESA effects, however, remain confined to short episodes during the early postbounce evolution of the two models, because the LESA dipole emission is quickly suppressed in BH forming models, where the LESA-driving convective shell in the PNS interior gets buried deep below a thick, convectively stable accretion mantle \citep{Walk+2020}, which grows rapidly due to the extremely high mass-infall rates (see upper left panels of Figures~\ref{fig:models40}--\ref{fig:modelsu75deg5sfho}). 

\section{Discussion and conclusions}
\label{sec:conclusions}

\subsection{Summary}

We have presented a detailed and comprehensive evaluation of hydrodynamic and neutrino-induced NS and BH kicks that we obtained in a set of long-time 3D stellar collapse simulations with the \textsc{Prometheus-Vertex/Nemesis} code of the Garching group. The model set includes successful SN explosions of single-star progenitors with ZAMS masses of 9.0, 9.6, 12.28, 15.0, 18.88, and 20.0\,M$_\odot$ as well as BH forming cases for progenitors with ZAMS masses of 40\,M$_\odot$ and 75\,M$_\odot$. 

The elaborate neutrino transport with the \textsc{Vertex} module was applied until the PNS collapsed to a BH or well beyond the onset of the neutrino-driven SN explosion. Thus, a postbounce evolution of typically $\sim$0.5\,s was covered, in our exploding 12.28\,M$_\odot$ and 18.88\,M$_\odot$ models indeed much longer, namely more than 4\,s and nearly 2\,s after bounce, respectively. All hydrodynamic SN simulations (except the 12.28\,M$_\odot$ model) were continued until the explosion energies were nearly saturated, using our newly developed \textsc{Nemesis} module, which provides an approximate but still sophisticated, computationally efficient treatment of the neutrino effects inside and in the surroundings of the PNS. The \textsc{Nemesis} scheme allows one to obviate the use of time-consuming 3D neutrino transport while employing neutrino results from 1D neutrino-hydrodynamics simulations of PNS cooling with the \textsc{Prometheus-Vertex} code including PNS convection via a mixing-length description.

Our analysis of the kicks is based on the time integration of forces expressed by the time derivatives of the total linear momentum of the ejecta (i.e., gas outside the PNS) and of the linear momentum carried away by escaping neutrinos in the free-streaming limit. In the course of this time integration, we take into account the changes of the gravitational mass of the PNS due to its neutrino losses (Equations~\ref{eq:vnsdot_hyd} and~\ref{eq:vnsdot_nu} in Section~\ref{sec:massdepanalysis}). Thus, our approach is consistent with previous analyses of neutrino and hydrodynamic kicks by the Garching group, for example in \citet{Stockinger+2020} and \citet{Bollig+2021}, but it is slightly more accurate. 

The mapping of our models to the \textsc{Nemesis} treatment for continuing the 3D simulations over longer time scales prevents us from computing the neutrino kicks beyond the end points ($t_\mathrm{f}^\nu$) of the calculations with \textsc{Vertex} neutrino transport. We, therefore, derive estimates for the final values of neutrino-induced kicks by extrapolation, as explained in Section~\ref{sec:asymmetryparameters} (Equation~\ref{eq:vnuinfty}). In contrast, the hydrodynamic kicks should not be affected to a large extent by switching from the \textsc{Vertex} transport to the \textsc{Nemesis} treatment, because the hydrodynamic evolution (e.g., explosion asymmetry and explosion energy) is continuous (without noticeable transients) and in good agreement with calculations continued with transport (as verified by overlapping simulations). Moreover, we tested effects connected to remaining uncertainties of neutrino heating and cooling (leading to slightly different explosion energies) by our model pairs (m15, m15e) and (s20, s20e).

\subsubsection{NS kicks}

Our long-term 3D simulations yield hydrodynamic (total) NS kicks between $\sim$10\,km\,s$^{-1}$ ($\gtrsim$\,60\,km\,s$^{-1}$) and $\sim$1290\,km\,s$^{-1}$ ($\gtrsim$\,1300\,km\,s$^{-1}$), with the highest values still growing when our calculations were stopped after several seconds of postbounce evolution (Table~\ref{tab:hydro_kicks} and Figure~\ref{fig:vhyd_vtot}). The contributions of neutrino-induced kicks to the total values are between $\sim$45\,km\,s$^{-1}$ and $\sim$140\,km\,s$^{-1}$ during the periods computed with full neutrino transport and between $\sim$70\,km\,s$^{-1}$ and $\sim$265\,km\,s$^{-1}$ according to our best-guess estimates (Equation~\ref{eq:vnuinfty}) based on a crude extrapolation of the conditions found at the end of our transport calculations (Table~\ref{tab:neutrino_kicks} and Figure~\ref{fig:scatter_plot}).

We witness a complex interplay of hydrodynamic and neutrino-induced kicks, whose combined effect determines the total natal kick the NS receives at its birth. In line with previous findings \citep{Stockinger+2020,Bollig+2021,Burrows+2023b}, we obtained neutrino-induced NS kicks dominating in the explosions of low-mass progenitors (our 9.0\,M$_\odot$ and 9.6\,M$_\odot$ cases) and hydrodynamic NS kicks clearly dominating in the successful explosion models of our progenitors with 15\,M$_\odot$ and more. The 12.28\,M$_\odot$ model is an intermediate case, where the hydro kick is still higher but the neutrino kick contributes significantly (Figure~\ref{fig:scatter_plot}). Because of a large angle between the hydrodynamic kick and the neutrino kick (which corresponds to the most probable relative orientation of both kicks), the neutrino contribution does not have a great impact on the absolute value of the total NS velocity in this model, but it still has a considerable influence on its direction. We point out that in cases where the NS kicks due to anisotropic neutrino emission and mass ejection have similar magnitudes but opposite directions, these could effectively cancel each other so that the NS would receive a low or vanishing net kick despite the fact that the SN ejecta might possess an observable, global asymmetry. In contrast to morphological or chemical asymmetries in the gaseous SN ejecta and remnants, a nonspherical distribution of radiated neutrinos cannot be measured experimentally.

More generally, however, hydrodynamic recoil should dominate the observed kicks of NSs except for those born in the core collapse of SN progenitors that explode with low energy and low ejecta asymmetry, thus producing only weak hydrodynamic kicks. Since neutrino-induced kicks are limited to fairly low values, they should not have a severe impact on the morphological properties of SN remnants with high-velocity NSs, which are predicted to originate from stronger explosions, where the major fraction of the intermediate-mass and iron-group elements is ejected in the hemisphere pointing opposite to the NS's kick vector \citep{Wongwathanarat+2013,Wongwathanarat+2017}, in agreement with a larger number of closely investigated remnants of relatively recent and near galactic SNe \citep{Holland-Ashford+2017,Katsuda+2018}. 

The magnitude of the hydrodynamic kick velocity is a function of the explosion energy and explosion asymmetry (and of the inverse of the NS mass with less variability, however) as discussed by \citet{Janka2017} (see analytic formulas there and Equation~\ref{eq:vhydscaling} in the present paper). Since the explosion energy scales roughly with the NS mass and progenitor compactness \citep[e.g.,][]{Nakamura+2015,Burrows+2024}, the maximum hydrodynamic NS kick can be expected to do so as well, and the kicks of individual cases will be a fraction of this maximum value, depending on the stochastic variation of the explosion asymmetry \citep[see also][]{Burrows+2024,Burrows+2023b}.

We refrain from making statements requiring or including weighting with the stellar initial mass function (IMF), because our mass grid of models is only coarse and lacks progenitors between 10 and 12\,M$_\odot$, which hampers reliable predictions with respect to dependencies on the IMF.
In addition, our s9.0 and z9.6 models are fairly special cases with extremely small values of their core-compactness parameters, which are not necessarily representative for the majority of low-mass CCSN progenitors in the $\sim$(9--10)\,M$_\odot$ range. Therefore, we stick to our focus on a basic distinction: On the one hand, there are NSs born in low-mass, low-energy explosions with low ejecta asymmetry \citep[for these correlations, see][]{Stockinger+2020,Burrows+2024}, where we expect the neutrino kick to provide the major contribution to the total kick. On the other hand and in contrast, in the higher-energy, more asymmetric explosions of higher-mass (better: higher-compactness) stars, the hydrodynamic kicks (with the gravitational tug as dominant, mediating force) are the clearly stronger effect. There may be intermediate cases, where both kick mechanisms provide contributions of comparable magnitudes, with our s12.28 model being an example. IMF weighting, however, requires information on the fraction of cases with significant explosion asymmetry and a strong hydrodynamic kick below $\sim$12\,M$_\odot$. This is a question of stellar evolution (with still major uncertainties in this mass range; see \citealt{WoosleyHeger2015}, \citealt{Sukhbold+2018}) and is beyond the scope of our study.

In a careful analysis of the individual matter forces \citep[gravity, pressure, momentum flow;][]{Scheck+2006} in all of our models, we demonstrated that usually the gravitational effect as a long-range force between NS and ejecta is the dominant one for determining the magnitude and direction of the final hydrodynamic kick over long evolution times (Figures~\ref{fig:hydro_forces} and~\ref{fig:hydro_forces_integrated}). Since also early fallback (i.e., fallback at the end of PNS cooling and neutrino heating), which can have an impact on the final ejecta and their linear momentum at late times, is governed by the influence of the NS's gravity on the innermost, temporary ejecta, our results justify the use of the term ``gravitational tug-boat effect'' \citep{Wongwathanarat+2013} for the mechanism producing the hydrodynamic NS kicks.

Late-time fallback on time scales of a minute to hours is unlikely to have a significant impact on the NS kicks, because at such late times relatively little mass is expected to fall back (in most cases except fallback SNe) and the ejecta momentum has been transferred already to the fastest-moving outer SN ejecta by pressure forces. Therefore, \citet{Stockinger+2020} and \citet{Janka+2022} found little effect on the NS kicks by the late-time fallback in their 3D simulations, whereas the asymmetric fallback even of little material, if accreted onto the NS, can transfer sizable amounts of angular momentum to the NS.

\subsubsection{BH kicks}

In our BH forming models, the NSs (or proto-BHs) receive neutrino-induced kicks of (10--30)\,km\,s$^{-1}$ until they become gravitationally unstable and collapse. Since shock expansion is weak or absent in our simulations, explosive, asymmetric mass ejection is not expected to happen, but the newly born BHs will swallow the progenitor's helium core or possibly the whole mass of the pre-collapse star. Therefore, applying momentum conservation, the final BH kicks in such failed SNe will be reduced to at most a few km\,s$^{-1}$, exclusively caused by the anisotropic neutrino loss during the transient existence of the NSs (Table~\ref{tab:neutrino_kicks}). These results receive support by a detailed investigation of the massive BH binary system VFTS~243, where a $\sim$10\,M$_\odot$ BH exists in a nearly circular orbit, suggesting that is was born from a stripped progenitor with mass-energy just being lost by neutrinos but no ejected matter \citep{Vigna-Gomez+2024}. The estimated most likely value of about 4\,km\,s$^{-1}$ for the natal kick velocity of the BH and a best-fit value of $\lesssim 0.2$\% for the neutrino emission asymmetry are perfectly well compatible with our results, in particular our s40 model, which is expected to form a $\sim$15.2\,M$_\odot$ BH with a natal kick of 3.3\,km\,s$^{-1}$ and a neutrino emission asymmetry of 0.13\%. 

In fallback SNe, where a powerful, asymmetric SN explosion is launched by the neutrino-driven mechanism before the NS collapses to a low-mass BH, the BH kicks can be much larger, as predicted by \citet{Janka2013} and recently found in 3D simulations by \citet{Chan+2020} and \citet{Burrows+2023,Burrows+2023b}. Here, the NS collapses due to the fallback accretion of initially ejected matter that does not become gravitationally unbound. The fallback mass is sufficiently large to push the NS beyond its upper mass limit, but it is also sufficiently small ---and the SN shock is sufficiently strong--- to allow the initial asymmetry of the blast wave to survive and to expel the far dominant fraction of the star. This combination of conditions requires quite some fine-tuning of the strength of the SN explosion, but the mentioned 3D simulations demonstrate that this possibility, which was pointed out to exist in principle by \citet{Janka2013}, could be realized in some stellar collapse events.   

\subsubsection{Contributions to neutrino kicks}

Our combined analysis of hydrodynamic and neutrino kicks reveals three characteristically different phases for the neutrino kicks in core-collapse simulations of massive progenitors. After the LESA-minted early postbounce phase (Figure~\ref{fig:lesakick}), the LESA effects are overruled by anisotropic neutrino emission (mainly of $\nu_e$ and $\bar\nu_e$) from accretion downflows onto the PNS (Figure~\ref{fig:accretionkick}, left panel). Later, well after SN shock revival, when the mass accretion by the PNS has effectively ended, the non-isotropy of the fluxes of neutrinos escaping to infinity is affected by absorption and scattering (mainly of $\bar\nu_e$ and $\nu_x$) in nonradial downflows that penetrate to the close vicinity of the PNS but are not accreted (Figure~\ref{fig:accretionkick}, right panel).

Therefore, we argued that it is crucial to measure the neutrino flux asymmetry in the free-streaming regime in order to account for the non-negligible interaction of neutrinos in the downflows. The corresponding momentum transfer to the stellar medium changes the linear momentum of both escaping neutrinos and ejected gas. Measuring neutrino emission asymmetries near the neutrinospheres or PNS surface may thus lead to a double-counting of momentum contributions in the sum of neutrino and hydrodynamic kicks, if the hydrodynamic forces are evaluated at a different radius or from the total gas momentum. For the same reason, the neutrino-induced kick velocities of BHs should be evaluated in the free streaming limit, in particular, if the evaluation is stopped at the time of BH formation but the gas momentum is subsequently reduced by fallback of asymmetrically distributed matter to the BH.

\subsection{Discussion}
\label{sec:discussion}

Our results are in overall agreement with the findings reported by \citet{Burrows+2023b} on grounds of an impressively large set of recent long-term 3D simulations that also provide data of neutrino-induced NS and BH kicks, and all together their and our 3D models confirm a variety of insights already discussed in previous literature \citep[e.g.,][]{Podsiadlowski+2004,Janka2013,Janka2017,Tauris+2017,Gessner+2018,Mueller+2019,Stockinger+2020,Chan+2020}. However, our current set of 3D simulations is too small to deduce statistical information and to draw firm conclusions on integral population properties, for example regarding any tendency of NS spin-kick alignment or a bimodal vs.\ single-peaked distribution of NS kicks. Our models compared to those of \citet{Burrows+2023b} exhibit roughly similar magnitudes of the hydrodynamic as well as neutrino kicks and they reveal similar general trends. Nevertheless, our neutrino kicks of BHs tend to be considerably lower, which may be connected either to our earlier BH formation times or to the differences in our analysis of the anisotropic loss of neutrinos as discussed in Sections~\ref{sec:massdepanalysis} and~\ref{sec:noteeval}. Possibly for the same reason (in particular, a different choice of the radius to evaluate the recoil by neutrinos) we also witness values of the instantaneous neutrino emission asymmetry parameter that are significantly lower (by a factor of roughly 2--3) than those plotted in \citet{Coleman+2022}.\footnote{We find better agreement of our instantaneous neutrino anisotropies with the ones plotted in \citet[figure 18]{Vartanyan+2019}, which are evaluated at 500 km (i.e., in the free-streaming regime). However, a detailed, quantitative comparison is hampered by the logarithmic scale of the plots, the rapid time variation of the displayed quantities, the use of different progenitor models, and, possibly, different definitions compared to the formulas provided in our Section~\ref{sec:asymmetryparameters}.}

A specific and quantitative comparison of individual models is impeded by the fact that the only case our model set and the one of \citet{Burrows+2023b} have in common is the 9.0\,M$_\odot$ model, which develops an explosion energy of 0.054\,B in our simulation, whereas \citet{Burrows+2023b} report a value of 0.107\,B \citep[up to 0.111\,B in][]{Wang+2024}, more than twice higher. We generally find explosion energies that are significantly lower, for example also for the 3D model z9.6, for which \citet{Wang+2024} list 0.207\,B but we have 0.086\,B \citep[see also the discussions of 3D models in][]{Stockinger+2020,Bollig+2021}. Moreover, we obtain
self-consistent explosions in all of our 3D simulations only under special circumstances (see summary in Section~\ref{sec:3Dsimulations}), e.g., for low-mass progenitors \citep{Melson+2015a,Melson+2020,Stockinger+2020}, rapid rotation \citep{Summa+2018}, with slight changes in the neutrino opacities \citep{Melson+2015b}, and with 3D perturbations in the convective oxygen-burning layer of the progenitor prior to collapse \citep{Bollig+2021}. Therefore, we expect sizable quantitative differences and probably even qualitatively different outcomes if individual cases could be confronted with each other.

Contrary to what is stated in \citet{Coleman+2022}, we have witnessed clear evidence that LESA dipole emission \citep[i.e., a strong dipole component in the electron-lepton number emission by $\nu_e$ and $\bar\nu_e$;][]{Tamborra+2014} determines the neutrino-induced kick of NSs born in the collapse of progenitors near the low-mass end of stars exploding as SNe. LESA-dominated neutrino kicks manifest themselves by an anti-alignment of the $\nu_e$-induced kick on the one side and $\bar\nu_e$- and $\nu_x$-induced kicks on the other side; the net effect is a total neutrino kick aligned with the LESA dipole vector (Figure~\ref{fig:lesakick}). This concerns our 9.0\,M$_\odot$ model and also the ECSN-like explosion of the 9.6\,M$_\odot$ model \citep[see also][]{Stockinger+2020}. In the latter case, where the explosion develops rapidly, and a phase of massive downflow activity around the PNS after shock revival is absent, the ejecta asymmetry is even influenced by the anisotropic neutrino heating due to the LESA dipole \citep[see][]{Tamborra+2014}, making the blast wave stronger in the anti-LESA direction. Therefore, neutrino kick and hydrodynamic kick of the NS are aligned in this model, both directed collinearly with the LESA dipole vector \citep{Stockinger+2020}.

But we also find that LESA affects the neutrino kick during the first few hundred milliseconds after core bounce in most other models with more massive progenitors, although in these cases LESA is not crucial for the final values of the neutrino kicks. This also holds true for the BH forming cases, where the LESA effects are significantly weakened by the high mass-accretion rates, which lead to the rapid formation of a massive accretion mantle around the PNS, thus burying the LESA-driving convective shell inside the PNS deep below a thick, convectively stable outer layer \citep{Walk+2020}. Our rapidly rotating 15\,M$_\odot$ explosion model is the only case where signatures of neutrino-emission asymmetries connected to LESA are absent, because rapid rotation suppresses PNS convection and thus the LESA dipole emission \citep{Walk+2019}. Nevertheless, this model is expected to develop a high neutrino-induced NS kick because of a large hemispheric asymmetry of massive accretion downflows near the equatorial plane. Both neutrino and hydro kicks have big inclination angles to the rotation axis and a big relative angle, too.

There are two caveats of our calculations of LESA effects and the corresponding neutrino-induced kicks in the low-mass explosion models in particular. First, we had to extrapolate the final values in a crude manner, because our simulations with \textsc{Vertex} neutrino transport followed the PNS evolution only for roughly 0.5\,s (for our lowest-mass models). And second, the RbR+ transport approximation used in the \textsc{Vertex} code might tend to overestimate the growth rate and amplitude of the LESA dipole compared to fully multi-dimensional transport \citep{Glas+2019}. Therefore, we cannot exclude that the LESA-induced neutrino kicks of low-mass NSs might only be around (10--20)\,km\,s$^{-1}$ \citep[as suggested by results of][]{Wang+2024} instead of being between $\sim$(40--50)\,km\,s$^{-1}$ (simulated) and roughly (70--80)\,km\,s$^{-1}$ (extrapolated) as listed in Table~\ref{tab:neutrino_kicks}.

\subsection{Conclusions}

The ubiquitous presence of LESA-induced neutrino kicks of NSs around 50\,km\,s$^{-1}$ (but maybe as low as $\sim$(10--20)\,km\,s$^{-1}$) would define a floor value for the natal kick velocity.\footnote{\citet{BrayEldridge2016,BrayEldridge2018} and \citet{Richards+2023} proposed a floor value for NS kicks in addition to an ejecta-mass dependent second component, but their two components do not add up as vector quantities, different from neutrino-induced and hydrodynamic kicks.}  This floor value is dominant for NSs born in explosions of stars near the low-mass end of SN progenitors. In particular, it also applies to ECSNe of progenitors with O-Ne cores, where the hydrodynamic kicks are predicted to be only on the order of km\,s$^{-1}$ \citep[for a detailed discussion, see][]{Gessner+2018,Stockinger+2020}. Therefore, the LESA dipole emission should determine the kick. In more massive progenitors, this minimal value of the neutrino kick due to the LESA dipole emission is exceeded by the hydrodynamic kicks (and accretion-associated neutrino kicks).

The spatial velocity of the Crab pulsar, which is thought to be the compact remnant of a low-energy explosion of a low-mass progenitor \citep{Smith2013,YangChevalier2015}, is inferred to have a value of around 160\,km\,s$^{-1}$ \citep[with fairly big uncertainties;][]{Hester2008,Kaplan+2008}. This kick magnitude cannot be explained by the hydrodynamic kick in an ECSN, which has been considered as an explanation of SN~1054 because of its low explosion energy and low iron and oxygen abundances \citep{Tominaga2013}. Moreover, the Crab pulsar's velocity is also hardly compatible with a LESA-induced neutrino kick in such an ECSN birth scenario. Therefore, a hydrodynamic kick in a highly asymmetric, low-energy explosion of a low-mass iron-core progenitor seems to provide the most probable and natural explanation, as hypothesized by \citet{Gessner+2018} and \citet{Stockinger+2020} and supported by recent results of \citet{Wang+2024}, who explored the considerable variation of progenitor and explosion properties in the mass range between $\sim$8 and $\sim$10\,M$_\odot$ \citep{WoosleyHeger2015,Sukhbold+2016} by a small set of 3D SN simulations.

The existence of a floor value for neutrino-induced NS kicks without concomitant large hydrodynamic kicks, as suggested by our results as well as those of \citet{Burrows+2023b}, might be consistent with claims of a bimodality of the natal-kick distribution \citep[based on a better representation of the observed velocity distribution of young pulsars by a two-component fit instead of a single-peaked distribution;][]{Arzoumanian+2002,Brisken+2003,Verbunt+2017,Igoshev2020,Igoshev+2021}. However, the floor value should be sufficiently small to be compatible with a low-kick population of NSs required by observations of binary pulsars and pulsars in globular clusters \citep[e.g.,][]{Podsiadlowski+2004,Willcox+2021,Abbate+2023,Wu+2023}, formation paths of low-mass X-ray binaries in globular clusters and NSs that stay bound in these environments~\citep[e.g.,][]{Ivanova+2008}, and the existence of double-NS systems \citep[e.g.,][]{Tauris+2017}. Low-velocity natal NS kicks might thus set severe limits to the magnitude of LESA-induced neutrino kicks or, probably less likely, they might require an efficient (nearly perfect) compensation of hydrodynamic and neutrino-induced kicks in SN explosions where both kicks possess similar magnitudes but can stochastically attain opposite directions. The recent sets of 3D SN models suggest that progenitors in the mass range between $\sim$10\,M$_\odot$ and $\sim$13\,M$_\odot$ might qualify as candidates for such a scenario.        

The work discussed in this paper has led to a better understanding of the interplay of hydrodynamic NS kicks due to the gravitational tug-boat mechanism and neutrino-induced NS and BH kicks due to anisotropic neutrino losses. More work is necessary, in particular a much larger set of models is needed, in order to draw statistical conclusions (e.g., on a trend/lack of NS spin-kick alignment or anti-alignment) and to arrive at information on population-integrated (i.e., IMF dependent) properties (e.g., the measurable distribution of NS kicks). Also longer 3D simulations with our sophisticated \textsc{Vertex} neutrino transport are needed (such as the ones we presented for models s12.28 and s18.88) in order to ameliorate our best-guess estimates of the finally achievable neutrino-induced kick velocities of NSs. 

One may also consider the use of the RbR+ approximation in our \textsc{Vertex} code for neutrino transport in multi-dimensional simulations as a deficiency, although we have shown that the RbR+ neutrino treatment in 3D SN simulations does not lead to any fundamental differences compared to multi-D transport, neither in the hydrodynamical evolution \citep{Glas+2019a} nor in the presence of the LESA dipole asymmetry of the PNS's neutrino emission \citep{Glas+2019}. Nevertheless, RbR+ transport seems to exhibit a tendency to enhance the amplitude of the LESA dipole and to trigger its earlier growth, as suggested by figures~1 and~5 of \citet{Glas+2019}. Therefore, we expect somewhat overestimated LESA kicks with the RbR+ approximation, maybe by up to a factor of 2--3. But quantitative estimates are handicapped by the limited set of models and simulation times in \citet{Glas+2019}.

Another potentially problematic shortcoming in all of the present SN simulations is the fact that the NS is pinned to the center of the polar coordinate grid, and therefore, effects connected to the kick-driven movement of the PNS are neglected. \citet{Janka+2022} discussed possible consequences of the NS's kick motion on late-time fallback accretion and the spin evolution of new-born NSs, because a NS displaced from the center of the explosion can accrete fallback mass from a different volume and with a different angular momentum compared to a NS that remains located at the explosion center.\footnote{These discussed effects concern differences in the physics of the system but not in the numerical treatment of the momentum and angular momentum evolution and conservation. Artificially fixing the PNS at the grid center, which is the case in all existing 3D SN simulations for numerical reasons, does not restrain the evaluation of the NS kick, which is based on assuming linear momentum conservation of the ejecta and neutrinos on the one hand and the NS on the other hand. The assumption of the PNS being pinned to the coordinate center of the grid is equivalent to assuming that it has an infinite inertial mass (like a wall that is hit by a ball bouncing back, which does not cause any noticeable motion of the wall).} Moreover, the PNS movement can affect the values for the predicted NS kick velocities quantitatively on a non-negligible level, because neutrino heating around the propagating NS with a growing off-center displacement can influence the geometry of the SN ejecta and their gravitational back reaction on the PNS. This was demonstrated with a large set of 2D simulations with and without PNS movement by \citet{Scheck+2006}. Nevertheless, such effects, which are currently ignored in 3D models, will not qualitatively change the fundamental physics of PNS acceleration by asymmetric neutrino emission and mass ejection and the general picture emerging from the interplay of such kicks, which was in the focus of our present paper.

Finally, a direct, detailed, and quantitative comparison with other long-term 3D simulations employing sophisticated neutrino transport in the recent literature \citep[e.g.,][]{Burrows+2023b} may be of interest for the wider community.

\bmhead{Acknowledgments}
We are grateful to Bernhard M\"uller and Naveen Yadav for their initial work of performing 3D progenitor simulations and Tobias Melson, Alexander Summa, and Robert Bollig for their 3D core-collapse simulations analyzed in this work. The authors also thank Adam Burrows, Jan Eldridge, Andrei Igoshev, Selma de Mink, Ruggero Valli, Alejandro Vigna-G{\'o}mez, and an anonymous referee for useful comments or discussions. The presented study was supported by the German Research Foundation (DFG) through the Collaborative Research Center ``Neutrinos and Dark Matter in Astro- and Particle Physics (NDM),'' Grant No.\ SFB-1258-283604770, and under Germany's Excellence Strategy through the Cluster of Excellence ORIGINS EXC-2094-390783311. 
The authors are grateful to the Gauss Centre for Supercomputing e.V.\ (GCS; www.gauss-centre.eu) and to the Leibniz Supercomputing Centre (LRZ; www.lrz.de) 
for computing time on the supercomputers SuperMUC and SuperMUC-NG at LRZ under LRZ project IDs pn69ho and pn25me as well as GAUSS Call~13 project ID pr48ra, GAUSS Call~15 project ID pr74de, and GAUSS Call~17 and Call~20 project ID pr53yi.
Computer resources for this project were also provided by the Max Planck Computing and Data Facility (MPCDF) on the HPC systems Cobra, Draco, and Raven.

\section*{Declarations}

\bmhead{Data availability}
Data of the model analysis can be made available upon request.

\bmhead{Code availability}
The codes used for the 3D simulations and post-processing analysis were developed
at the Max Planck Institute for Astrophysics and are not public source.

\bmhead{Conflict of interest}
The authors declare no competing interests.

\bmhead{Ethics approval}
Not applicable.

\bmhead{Consent to participate}
All authors contributed to the work and approved sending the paper for publication.

\bmhead{Authors' contributions}
Both authors contributed equally to this work.
HTJ initiated the project and did most of the writing. DK performed the analysis of the
3D simulations, did most of the figures, and provided inputs for the written text.

\bmhead{Consent of publication}
All authors agree for the work to be published in Astrophysics and Space Science.

\bmhead{Funding note} 
Open Access funding enabled and organized by Projekt DEAL.
Funding through academic grants is listed in the Acknowledgements section.


\begin{figure*}[!]
\vskip-25pt
\centering
\includegraphics[width=\textwidth]{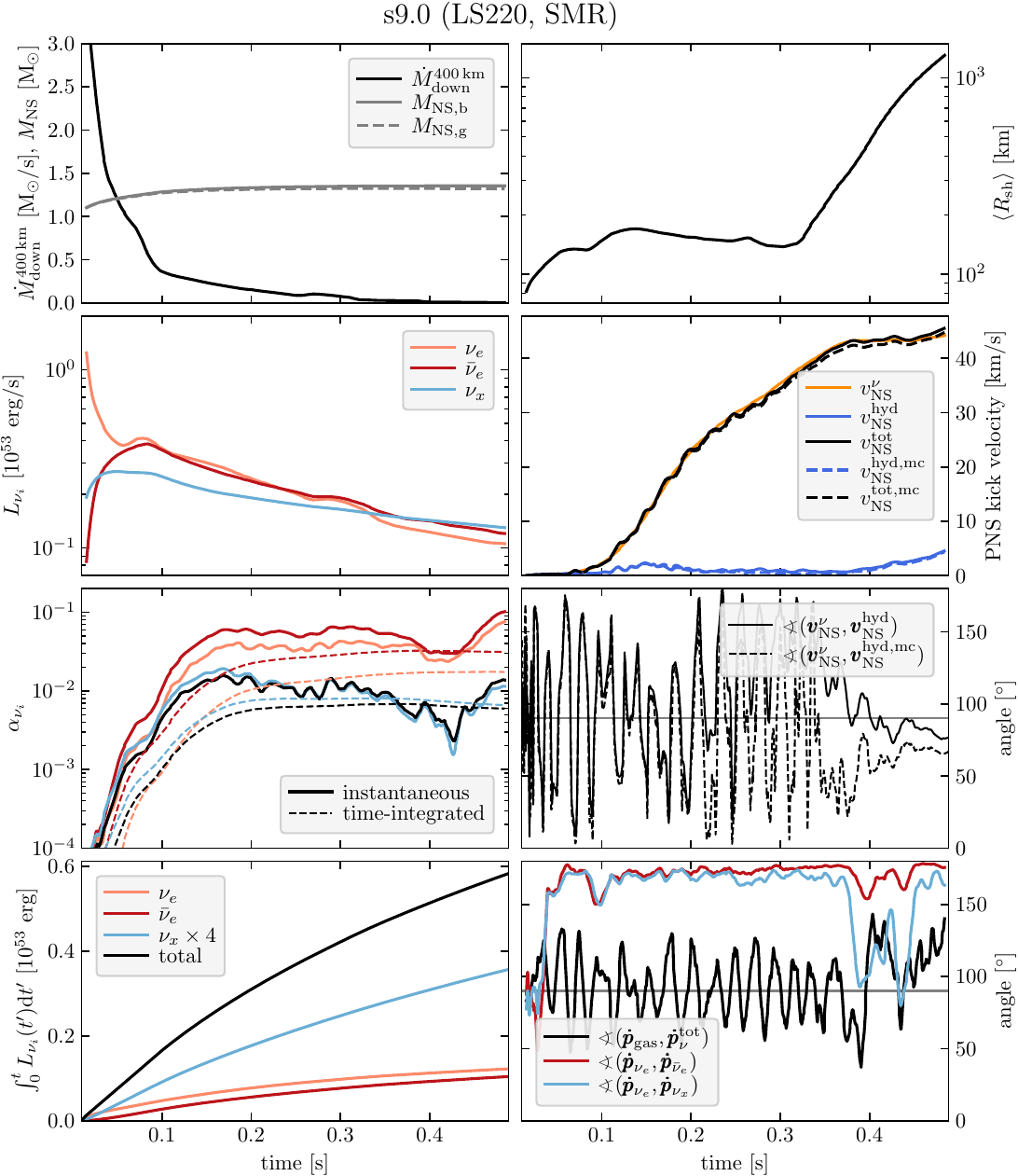}
\caption{Characteristic quantities as functions of postbounce time for the dynamical evolution, neutrino emission, hydrodynamic NS kick ($v_\mathrm{NS}^\mathrm{hyd}$; Equation~\ref{eq:vhyd}), neutrino-induced NS kick ($v_\mathrm{NS}^\nu$; Equation~\ref{eq:vneut}), and total NS kick ($v_\mathrm{NS}^\mathrm{tot}$; Equation~\ref{eq:vtot}) of model s9.0. Results for the kick velocities estimated on grounds of the total gas momentum ($v_\mathrm{NS}^\mathrm{hyd,mc}$ from Equation~\ref{eq:vnsmc1} and $v_\mathrm{NS}^\mathrm{tot,mc}$ from Equation~\ref{eq:vnsmc2}) are also shown for comparison (dashed lines).  
{\em Top left:} Mass accretion rate in downflows at 400\,km and baryonic and gravitational NS masses; {\em top right:} angle-averaged shock radius; {\em second row, left:} luminosities of $\nu_e$, $\bar\nu_e$, and a single kind of $\nu_x$, evaluated at 400\,km, for a lab-frame observer at infinity; {\em second row, right:} hydrodynamic, neutrino-induced, and total NS kicks as labeled; {\em third row, left:} instantaneous neutrino-emission asymmetry parameters (solid) and their time-averaged values (dashed) for all neutrino species and the summed neutrino loss; {\em third row, right:} angles of $\vec{v}_\mathrm{NS}^\mathrm{hyd}$ and $\vec{v}_\mathrm{NS}^\mathrm{hyd,mc}$ relative to $\vec{v}_\mathrm{NS}^\nu$; {\em bottom left:} time-integrated energy losses in $\nu_e$, $\bar\nu_e$, all types of $\nu_x$, and all summed up; {\em bottom right:} relative angle between NS acceleration forces due to asymmetric gas ejection and due to total anisotropic neutrino emission, and angles between momentum force of $\nu_e$ and momentum forces of $\bar\nu_e$ and $\nu_x$, respectively  
}
\label{fig:models9}
\end{figure*}

\begin{figure*}[!]
\centering
\includegraphics[width=\textwidth]{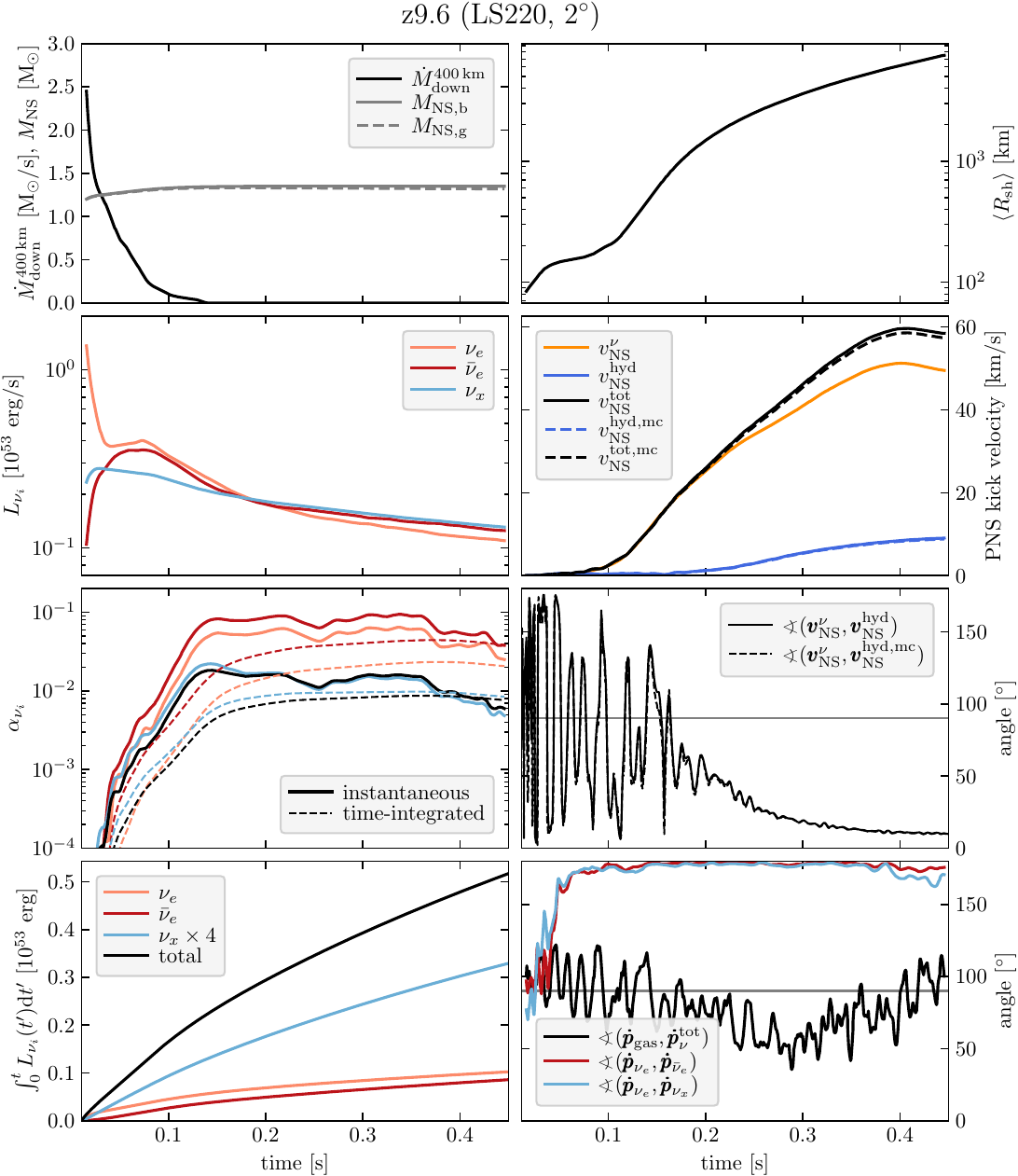}
\caption{Same as Figure~\ref{fig:models9}, but for model z9.6}
\label{fig:modelz96}
\end{figure*}

\begin{figure*}[!]
\centering
\includegraphics[width=\textwidth]{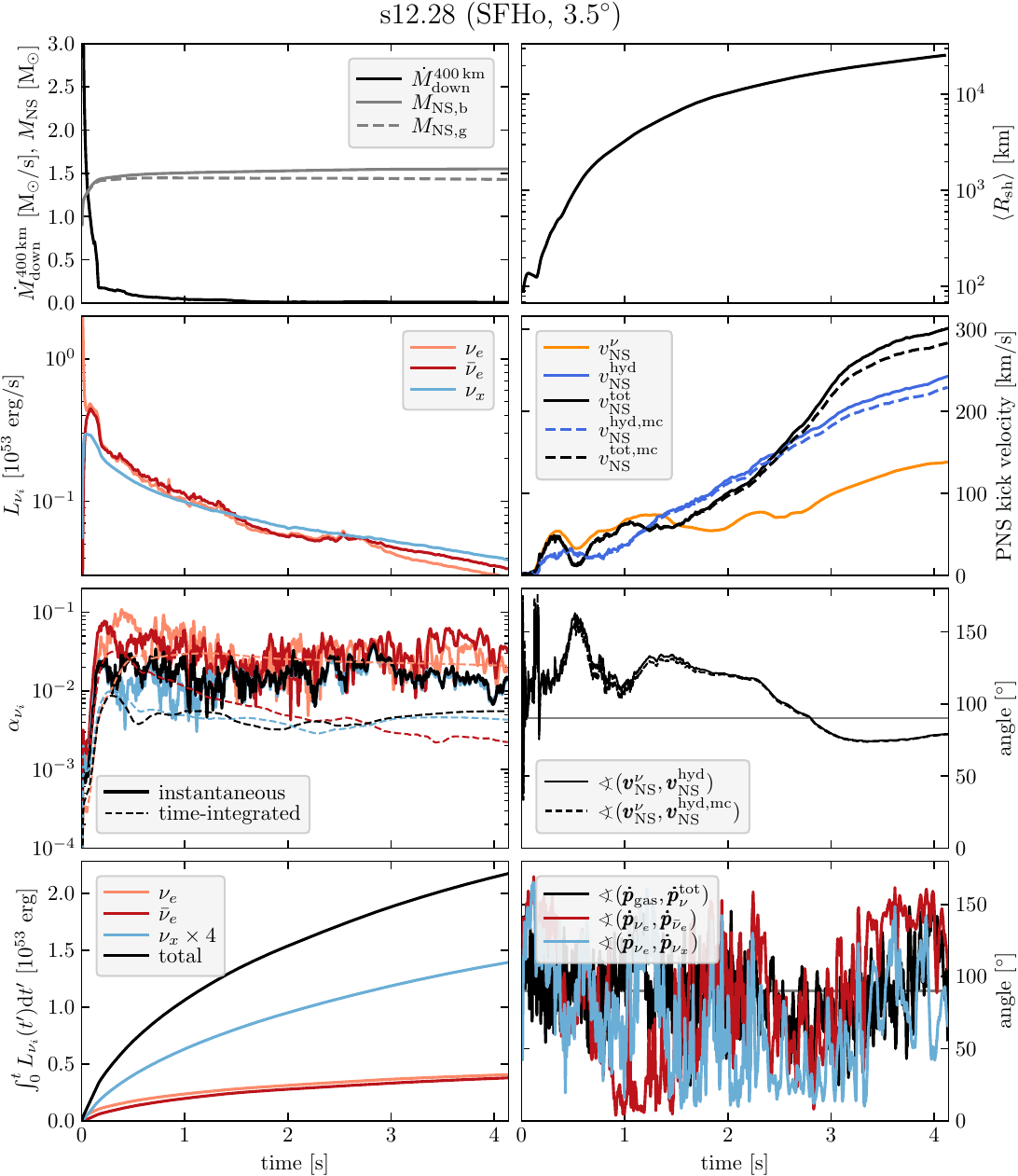}
\caption{Same as Figure~\ref{fig:models9}, but for model s12.28}
\label{fig:models1228}
\end{figure*}

\begin{figure*}[!]
\centering
\includegraphics[width=\textwidth]{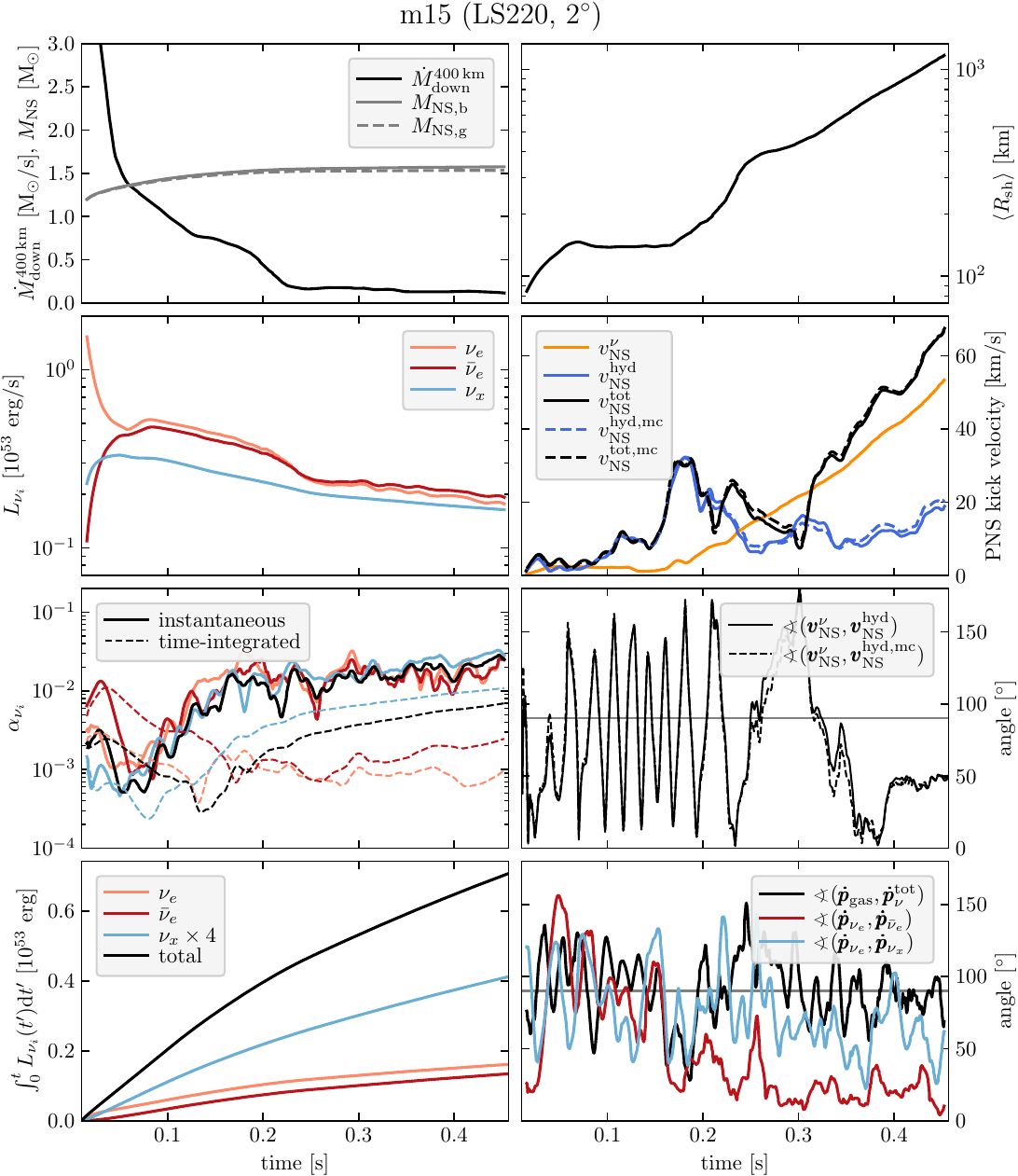}
\caption{Same as Figure~\ref{fig:models9}, but for the rotating model m15}
\label{fig:modelm15}
\end{figure*}

\begin{figure*}[!]
\centering
\includegraphics[width=\textwidth]{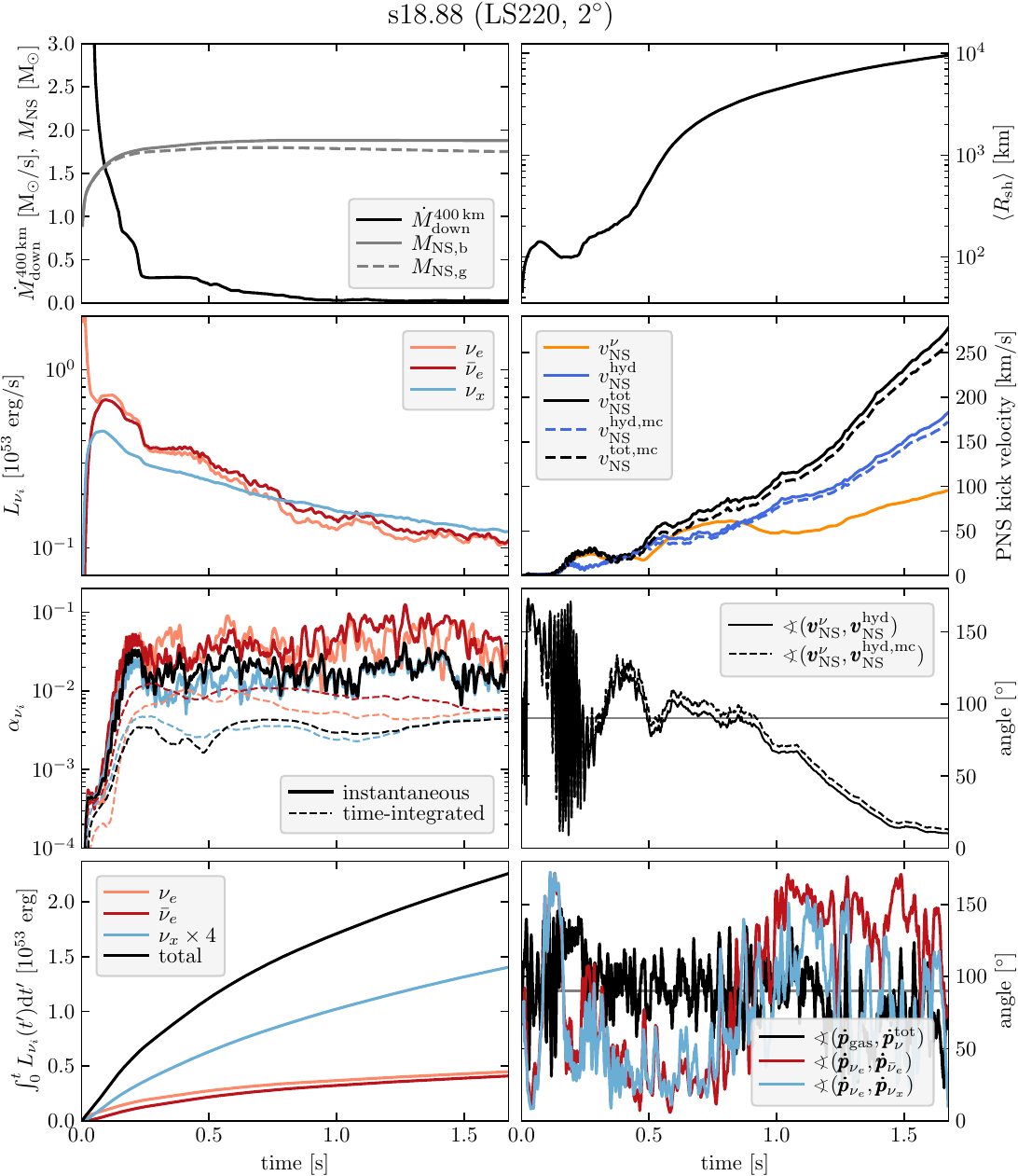}
\caption{Same as Figure~\ref{fig:models9}, but for model s18.88}
\label{fig:models1888}
\end{figure*}

\begin{figure*}[!]
\centering
\includegraphics[width=\textwidth]{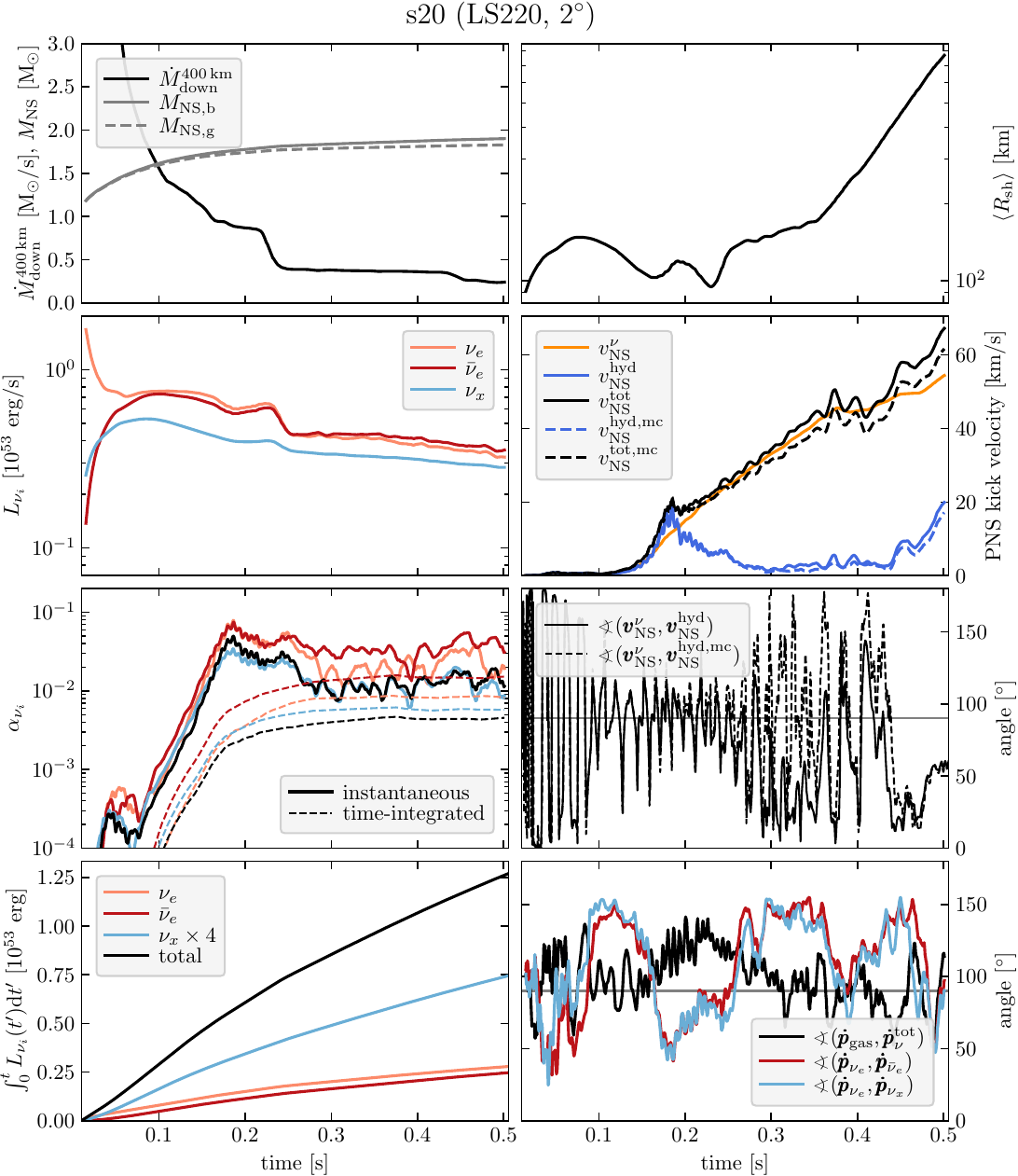}
\caption{Same as Figure~\ref{fig:models9}, but for model s20}
\label{fig:models20}
\end{figure*}

\begin{appendices}

\section{Detailed time evolution of exploding models}
\label{secA1}

Figures~~\ref{fig:models9}--\ref{fig:models20} in this appendix display detailed information about the postbounce evolution of relevant diagnostic quantities for all of our 3D stellar collapse simulations with successful explosions. The displayed time intervals are constrained to the periods that were simulated with full \textsc{Vertex} neutrino transport (ending at times $t_\mathrm{f}^\nu$) in order to relate hydrodynamic and neutrino kicks during the postbounce phase when both are computed, and to compare the characteristic properties specific to the different models against each other.

For the hydrodynamic and total kick velocities as well as their angles relative to the neutrino-induced kick velocity, results for $v_\mathrm{NS}^\mathrm{hyd} = |\vec{v}_\mathrm{NS}^\mathrm{hyd}|$ and $v_\mathrm{NS}^\mathrm{tot} = |\vec{v}_\mathrm{NS}^\mathrm{tot}|$ according to the acceleration integrals of Equations~(\ref{eq:vhyd}) and~(\ref{eq:vtot}) are plotted with solid lines; the time evolution of these kick velocities is also displayed in Figure~\ref{fig:vhyd_vtot}, and their values at $t_\mathrm{f} \ge t_\mathrm{f}^\nu$ are given in Table~\ref{tab:hydro_kicks} and  Figure~\ref{fig:scatter_plot}. We compare them with results for $v_\mathrm{NS}^\mathrm{hyd,mc} = |\vec{v}_\mathrm{NS}^\mathrm{hyd,mc}|$ and $v_\mathrm{NS}^\mathrm{tot,mc} = |\vec{v}_\mathrm{NS}^\mathrm{tot,mc}|$ computed from the total gas momentum and the total momentum of escaping neutrinos via the approximate expressions of Equations~(\ref{eq:vnsmc1}) and~(\ref{eq:vnsmc2}), respectively, which constitute lower bounds for the hydrodynamic and total kicks (dashed lines).

In general, there is excellent agreement obtained with the two complementary methods of evaluation, except for minor systematic differences that are natural when a lower bound is compared with the more accurate values ($v_\mathrm{NS}^\mathrm{hyd,mc} \le v_\mathrm{NS}^\mathrm{hyd}$ and $v_\mathrm{NS}^\mathrm{tot,mc} \le v_\mathrm{NS}^\mathrm{tot}$; see the discussion in Section~\ref{sec:approxestimates}). Clear discrepancies, mainly in the angles of $\vec{v}_\mathrm{NS}^\mathrm{hyd}$ and $\vec{v}_\mathrm{NS}^\mathrm{hyd,mc}$ relative to the neutrino-kick velocity $\vec{v}_\mathrm{NS}^\nu$, can be witnessed only during transient phases with very low absolute values of the hydrodynamic kicks and rapid temporal variations of the corresponding kick directions. The corresponding numerical inaccuracies of the kick evaluation concern mainly models s9.0 and s20, but also lead to episodes when $v_\mathrm{NS}^\mathrm{hyd,mc} > v_\mathrm{NS}^\mathrm{hyd}$ and $v_\mathrm{NS}^\mathrm{tot,mc} > v_\mathrm{NS}^\mathrm{tot}$ holds on a low level in model~m15. 

It should be noted that in Figures~~\ref{fig:models9}--\ref{fig:models20}, the plotted results for all quantities, except the angles of $\vec{v}_\mathrm{NS}^\mathrm{hyd}$ and $\vec{v}_\mathrm{NS}^\mathrm{hyd,mc}$ relative to $\vec{v}_\mathrm{NS}^\nu$ and the time-integrated neutrino energy losses, are smoothed by applying a running average over time intervals of 10\,ms.
The same holds true for all curves displayed in Figure~\ref{fig:vnu_anu_alpha}.
This implies that the first and last 5\,ms of the simulated periods of evolution are not shown, which leads to small discrepancies between the plotted final values and the true final values listed in Table~\ref{tab:neutrino_kicks}.


\begin{figure*}[!]
\centering
\includegraphics[width=\textwidth]{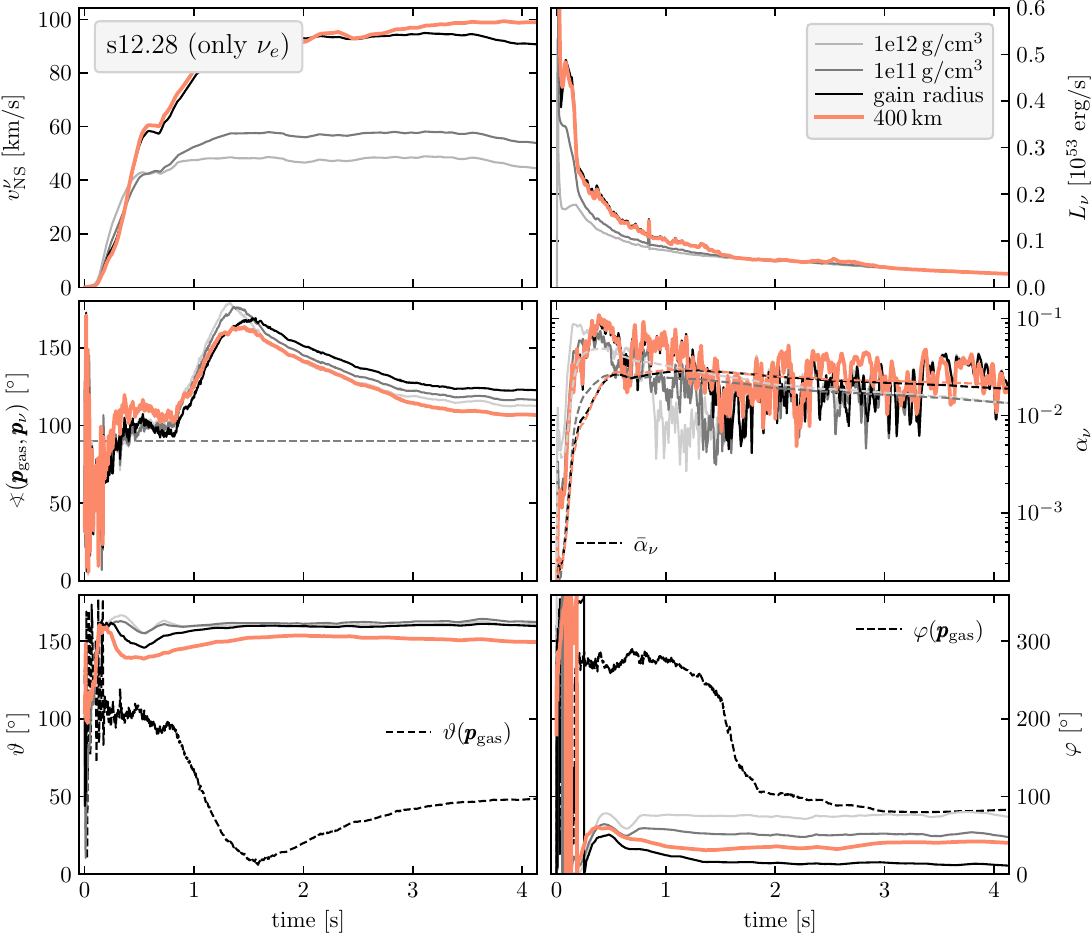}
\caption{Post-bounce evolution of the neutrino-induced NS kick caused by anisotropic emission of electron neutrinos $\nu_e$ in model~s12.28. The displayed quantities are evaluated at radial locations corresponding to angle-averaged densities of $10^{12}$\,g\,cm$^{-3}$ and $10^{11}$\,g\,cm$^{-3}$, at the angle-averaged gain radius, and at a radius of 400\,km, where neutrinos have decoupled from the stellar plasma. {\em Top left:} NS kick velocity; {\em top right:} $\nu_e$ luminosity; {\em middle left:} relative angle between $\nu_e$-kick of the NS and hydrodynamic NS kick; {\em middle right:} instantaneous/time-integrated anisotropy parameter of the $\nu_e$ emission (solid/dashed); {\em bottom left:} latitudinal angle of the $\nu_e$ momentum vector (solid lines) and the ejecta momentum vector (dashed line) in a global polar coordinate system of the simulation output; {\em bottom right:} azimuthal angle of the $\nu_e$ momentum vector (solid lines) and the ejecta momentum vector (dashed line) in a global polar coordinate system of the simulation output. Note that the vectors of the gas momentum and of the neutrino emission point in directions opposite to the corresponding NS kicks}
\label{fig:s1228nue}
\end{figure*}

\begin{figure*}[!]
\centering
\includegraphics[width=\textwidth]{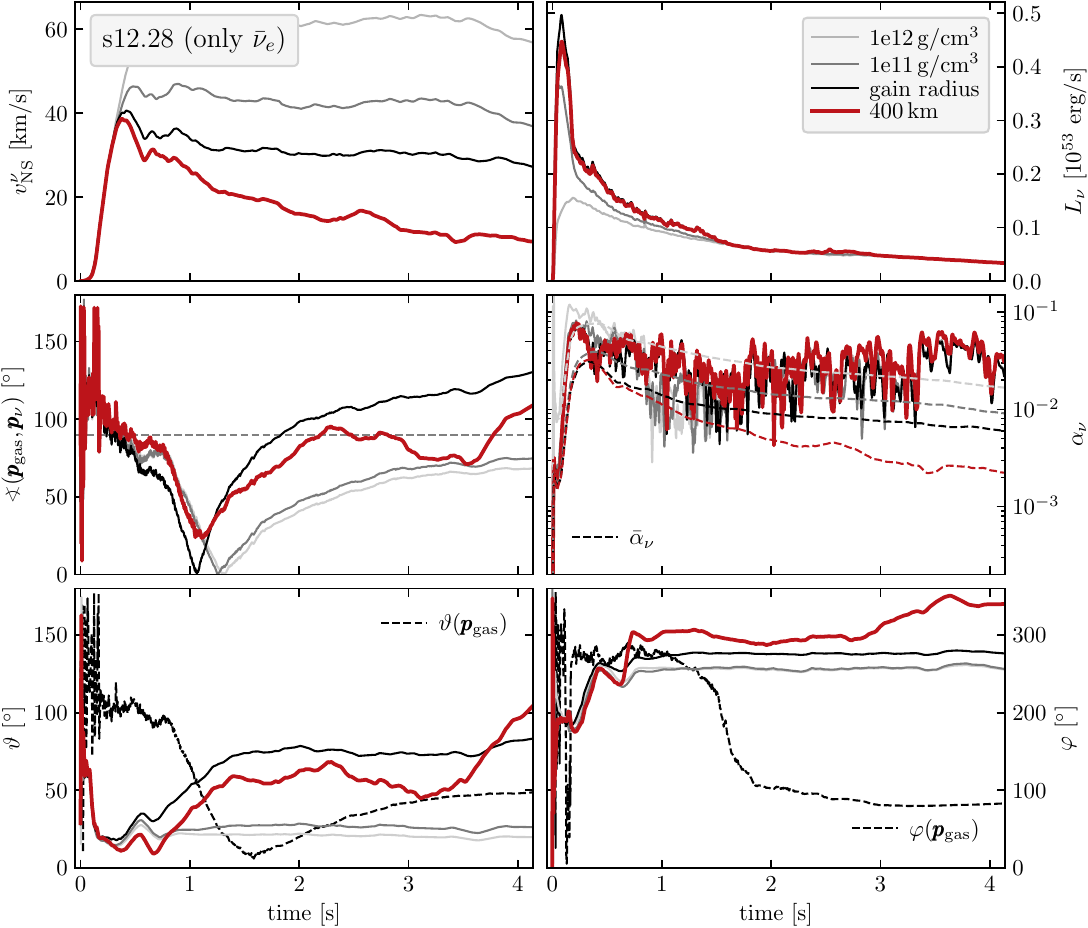}
\caption{Same as Figure~\ref{fig:s1228nue}, but for electron antineutrinos $\bar\nu_e$}
\label{fig:s1228barnue}
\end{figure*}

\begin{figure*}[!]
\centering
\includegraphics[width=\textwidth]{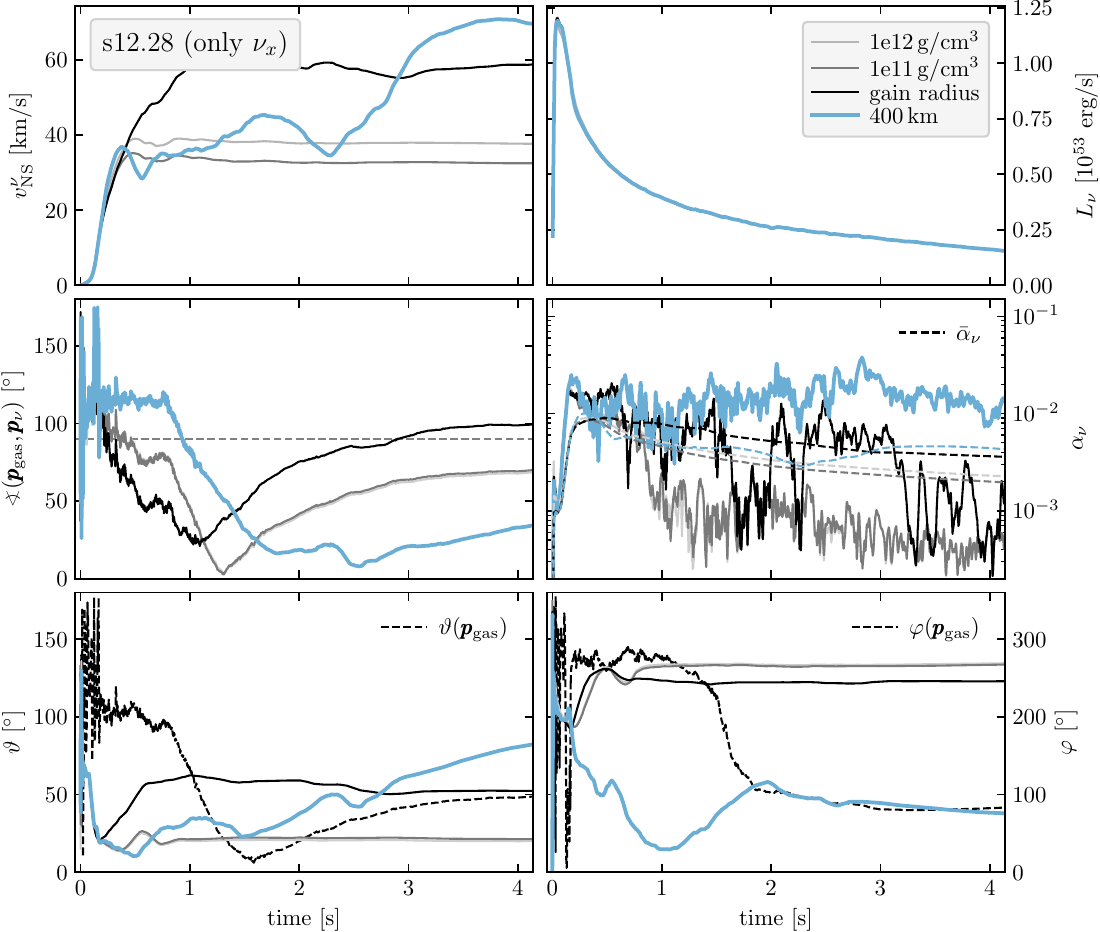}
\caption{Same as Figure~\ref{fig:s1228nue}, but for heavy-lepton neutrinos $\nu_x$. The top panels show results summed up for all kinds of heavy-lepton neutrinos and antineutrinos}
\label{fig:s1228nux}
\end{figure*}

\section{Neutrino-species dependent NS kicks in model s12.28}
\label{secA:s12}

Figures~\ref{fig:s1228nue}, \ref{fig:s1228barnue}, and~\ref{fig:s1228nux}
in this appendix present an in-depth analysis of the neutrino-induced NS kicks for all neutrino species individually and at different radial positions in model~s12.28. The evaluation is performed at radii where the angle-averaged densities are $10^{11}$\,g\,cm$^{-3}$ and $10^{12}$\,g\,cm$^{-3}$, which are locations close to the PNS surface and around the neutrinospheres. In addition, it is also conducted at the angle-averaged gain radius and at a radius of 400\,km. The first two positions at high densities permit to extract information about the neutrino emission originating from inside the PNS and of asymmetry effects associated with the LESA dipole, whereas the additional analysis at the gain radius permits conclusions on contributions by the neutrino radiation produced in accretion flows onto the PNS. The analysis at 400\,km yields the observationally relevant results in the free-streaming regime of neutrinos, which allow for an unambiguous interpretation of the neutrino kicks in their interplay with the hydrodynamic kicks. Moreover, in comparison with the results obtained at the gain layer, it yields insights into the consequences connected to absorption and scattering of neutrinos above the gain radius.

At each of these radii, the neutrino fluxes for computing the neutrino forces according to Equation~(\ref{eq:pdotnu}) are transformed from the comoving frame of the stellar fluid, where the transport is solved, into the laboratory frame of a distant observer, which is coincident with the rest frame of the PNS (because the PNS is fixed at the coordinate center of the polar Yin-Yang grid used for the computations). Note that our \textsc{Vertex} transport solver employs the RbR+ approximation in multi-dimensional simulations, which only provides radial neutrino fluxes. The disregard of nonradial flux components implies approximate results for the neutrino force calculations with Equation~(\ref{eq:pdotnu}). The loss of accuracy should, however, be a minor effect, because near the PNS surface and outside the PNS the radial flux components are the far dominant ones. 

In all of the figures of this appendix (and in Figures~\ref{fig:vnu_z96}--\ref{fig:vnu_s1888} and \ref{fig:vnu_u75DD2}), the curves for the neutrino-induced kick velocities, neutrino luminosities, and neutrino anisotropy parameters have been smoothed by applying running means over time intervals of 10\,ms.


\section{Detailed time evolution of BH forming models}
\label{secA:bh}

\begin{figure*}[!]
\centering
\includegraphics[width=\textwidth]{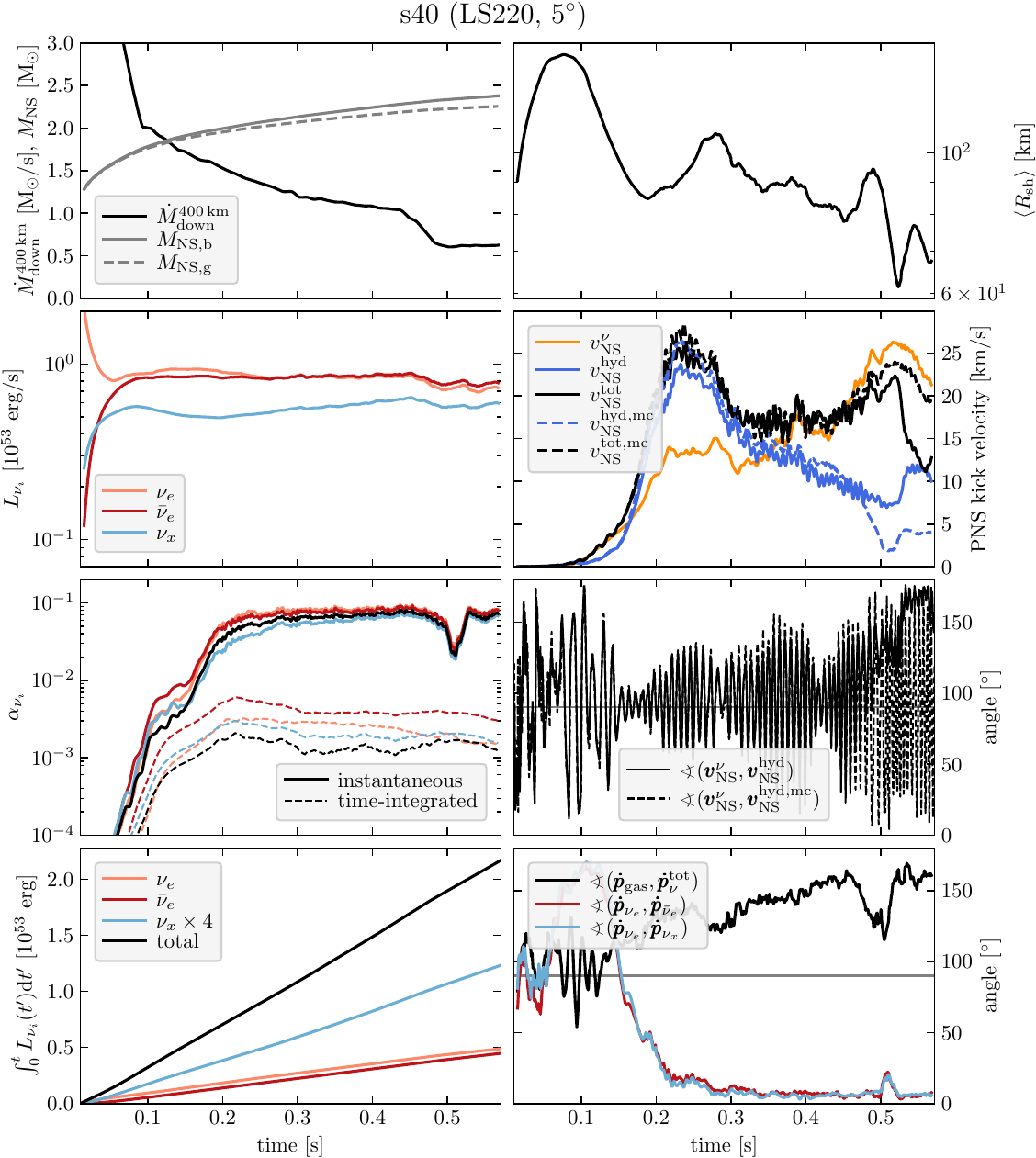}
\caption{Same as Figure~\ref{fig:models9}, but for BH forming model s40}
\label{fig:models40}
\end{figure*}

Figures~\ref{fig:models40}--\ref{fig:modelsu75deg5sfho} display, in analogy to  Figures~\ref{fig:models9}--\ref{fig:models20} in Appendix~\ref{secA1}, the detailed postbounce behavior of the diagnostic quantities that are most relevant for the dynamical evolution, neutrino emission, and proto-BH kicks in all of our 3D stellar collapse simulations that lead to BH formation. The plots cover the evolution until times $t_\mathrm{f}^\nu$, when the PNSs become gravitationally unstable and our calculations with \textsc{Vertex} neutrino transport were stopped.

It is important to keep in mind that the hydrodynamic kicks are only transient, because none of our BH forming models yields an explosion. Therefore, the asymmetric gas around the PNSs constitutes only temporary ejecta, which ultimately fall back and are swallowed by the BHs. Only the anisotropically radiated neutrinos are able to transfer a natal kick to the relic BHs. Our estimates of the neutrino-induced kicks for the final BHs are given in Table~\ref{tab:neutrino_kicks}; the values listed there are calculated with the equation provided in the table notes, employing the PNS kick velocities obtained at the times when the PNSs collapse to BHs.

We repeat that $v_\mathrm{NS}^\mathrm{hyd,mc}$ and $v_\mathrm{NS}^\mathrm{tot,mc}$ (computed via Equations~\ref{eq:vnsmc1} and~\ref{eq:vnsmc2}) are approximate, lower bounds for $v_\mathrm{NS}^\mathrm{hyd}$ and $v_\mathrm{NS}^\mathrm{tot}$ (obtained from Equations~\ref{eq:vhyd} and~\ref{eq:vtot}), respectively. This fact can be disturbed by numerical inaccuracies of the kick evaluation in phases when the hydrodynamic accelerations and kick directions fluctuate wildly on short time scales because of rapid variations of non-spherical accretion flows. This mostly concerns model~s40 and the late phase of model~u75\_DD2. In both cases, the stalled shock retreats and describes violent SASI sloshing and spiral motions, which lead to large-amplitude nonradial variability in the postshock flows at very high frequencies. Discretized calculations of time derivatives by finite differences are particularly prone to inaccuracies under such conditions.   

Finally, we again remark that the displayed curves for all quantities, except the angles of $\vec{v}_\mathrm{NS}^\mathrm{hyd}$ and $\vec{v}_\mathrm{NS}^\mathrm{hyd,mc}$ relative to $\vec{v}_\mathrm{NS}^\nu$ and the time-integrated neutrino energy losses, represent the computed results after smoothing by running means over time intervals of 10\,ms. The same applies for all curves displayed in Figure~\ref{fig:vnu_anu_alpha_bh}. For this reason, the first and last 5\,ms of the simulated periods of evolution are not shown, which leads to small discrepancies between the plotted final values and the true final values listed in Table~\ref{tab:neutrino_kicks}.

\begin{figure*}[!]
\centering
\includegraphics[width=\textwidth]{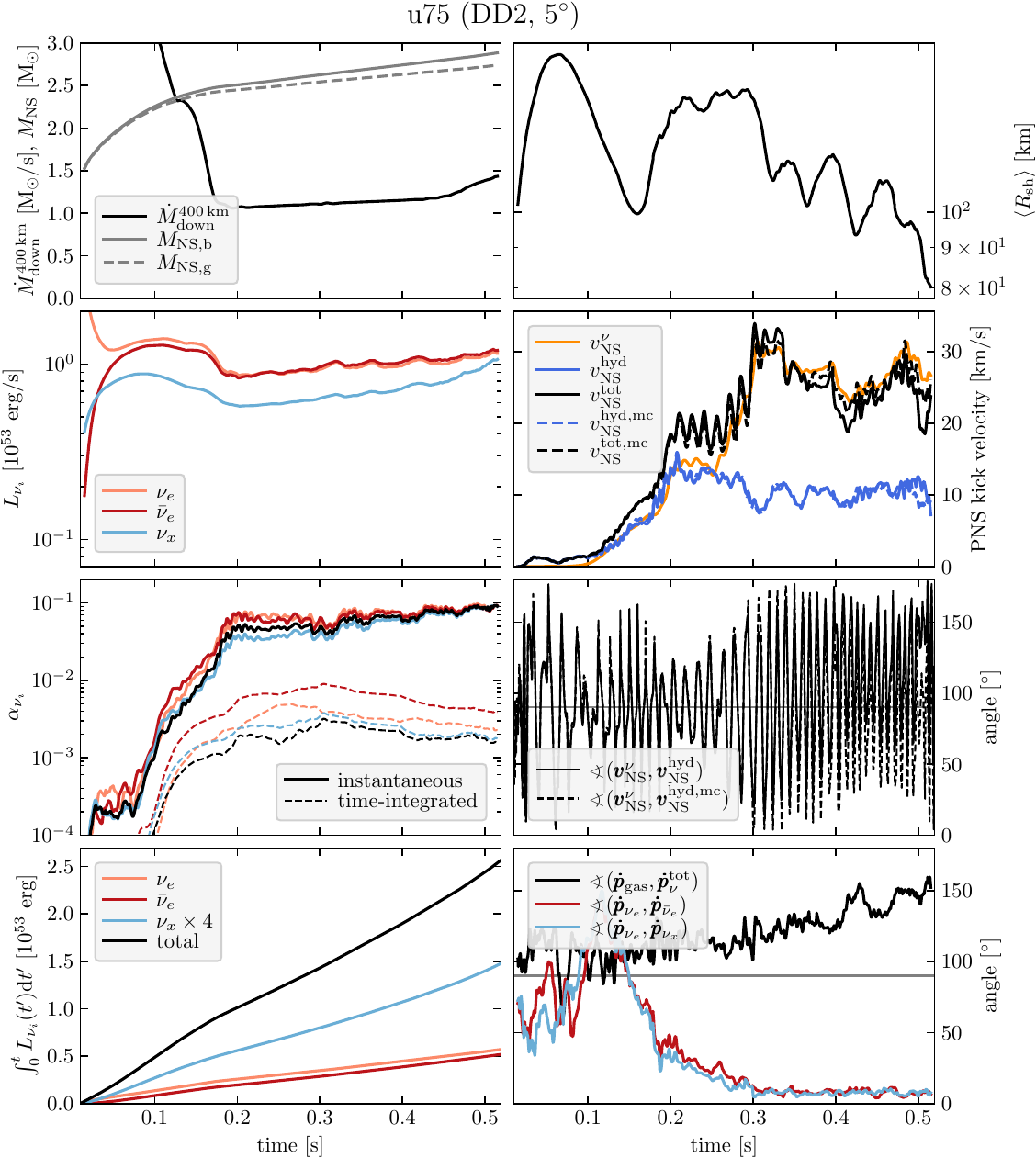}
\caption{Same as Figure~\ref{fig:models9}, but for BH forming model u75\_DD2}
\label{fig:modelsu75deg5dd2}
\end{figure*}

\begin{figure*}[!]
\centering
\includegraphics[width=\textwidth]{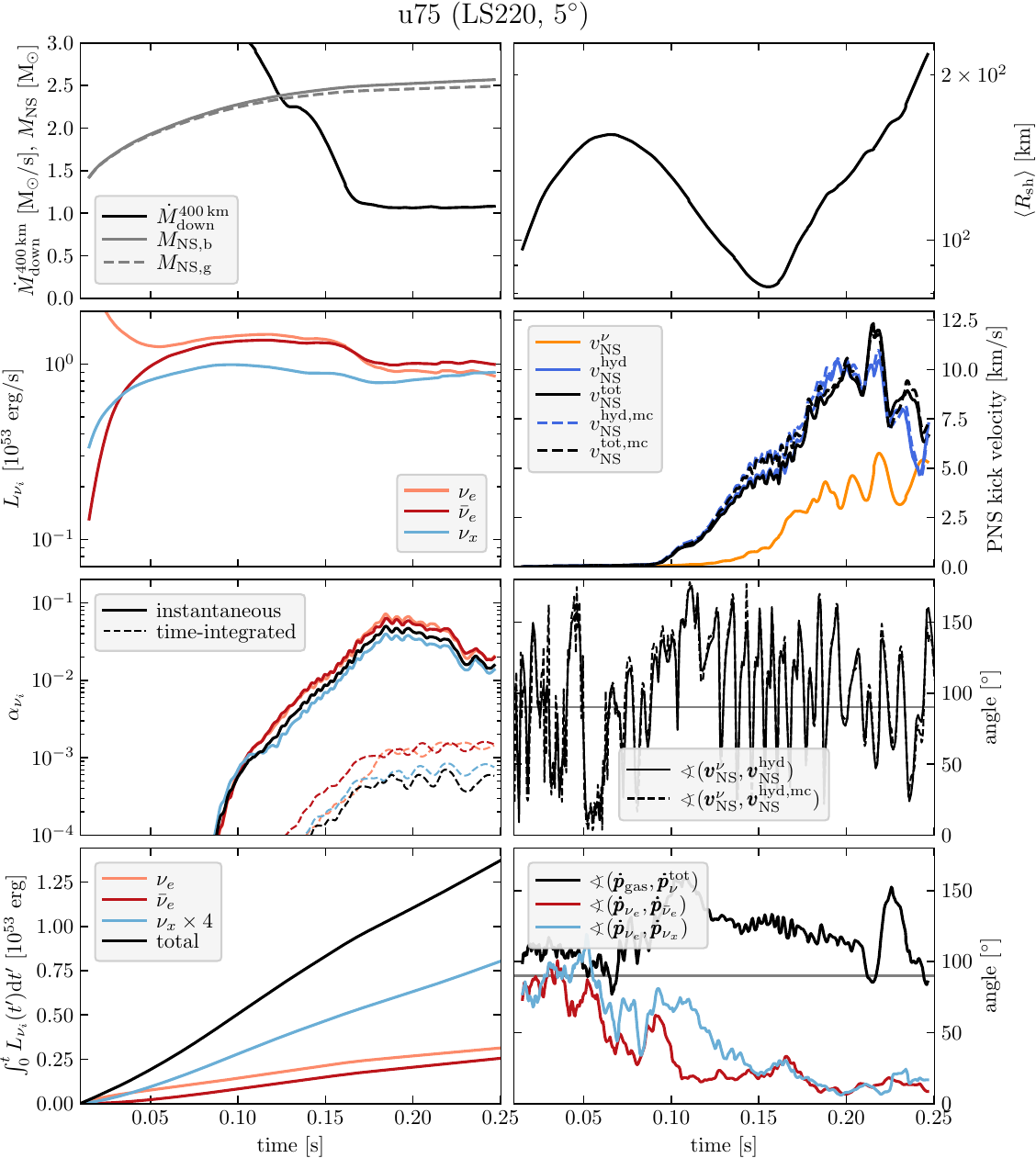}
\caption{Same as Figure~\ref{fig:models9}, but for BH forming model u75\_LS220\_1}
\label{fig:modelsu75deg5}
\end{figure*}

\begin{figure*}[!]
\centering
\includegraphics[width=\textwidth]{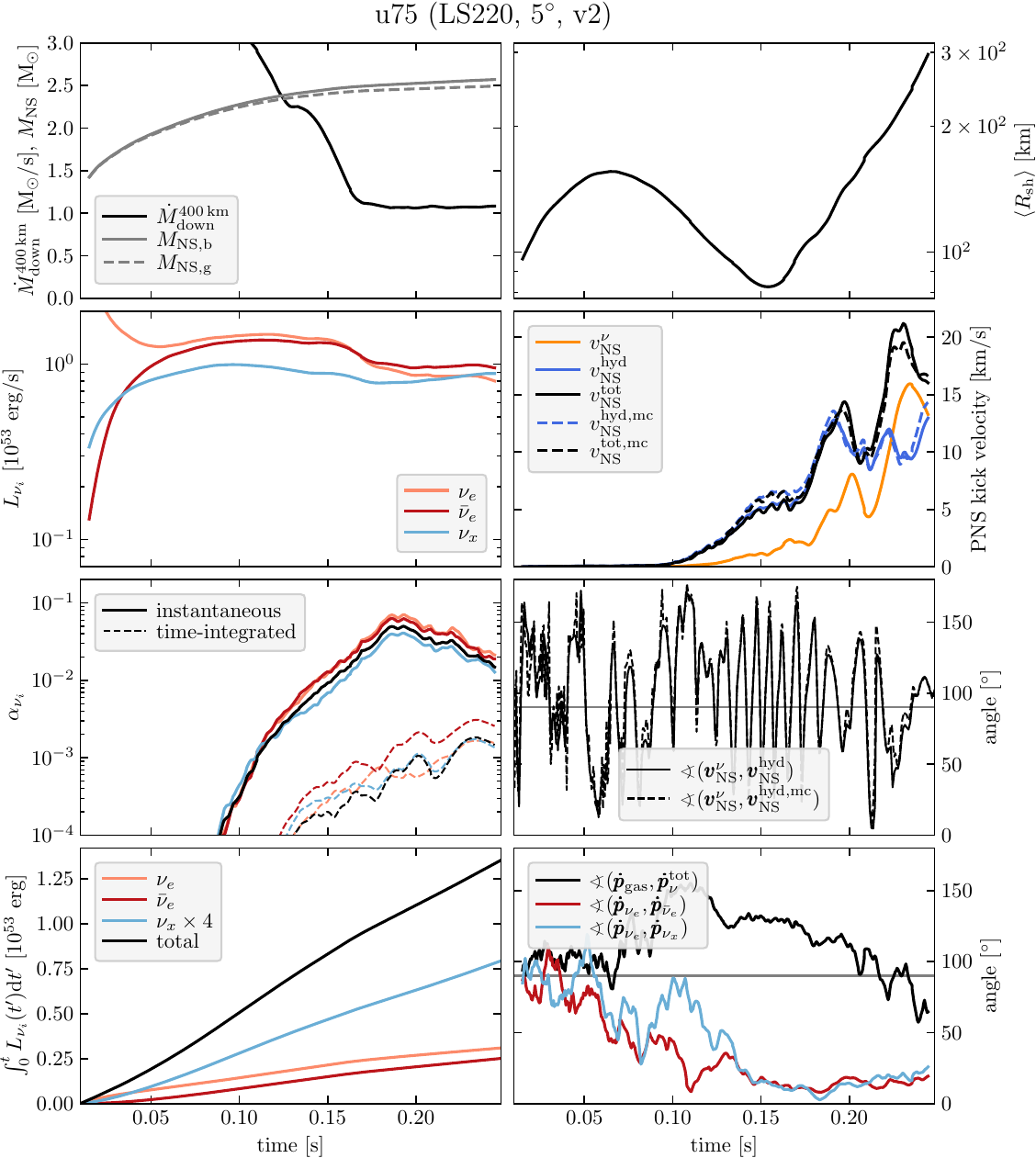}
\caption{Same as Figure~\ref{fig:models9}, but for BH forming model u75\_LS220\_2}
\label{fig:modelsu75deg5v2}
\end{figure*}

\begin{figure*}[!]
\centering
\includegraphics[width=\textwidth]{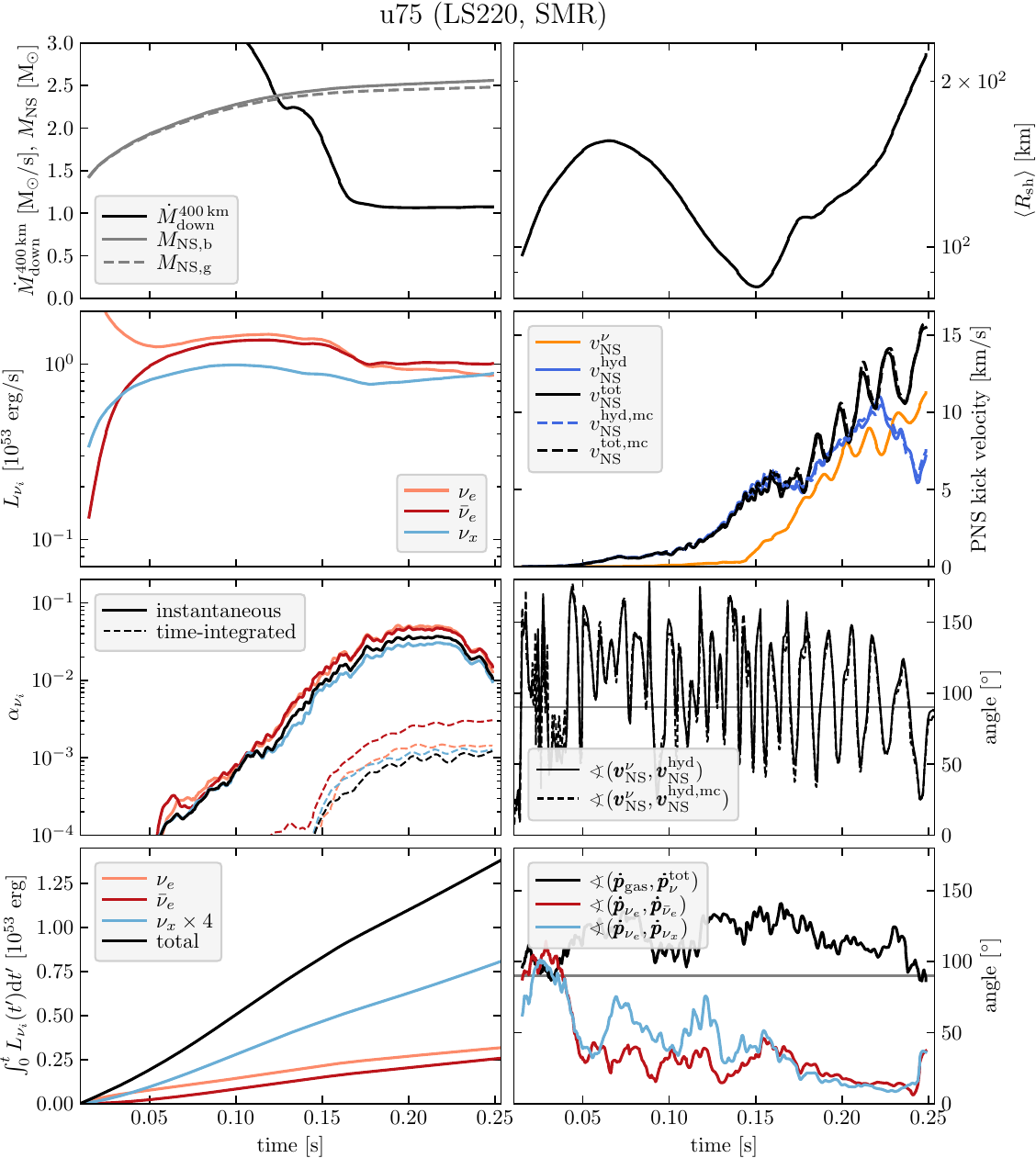}
\caption{Same as Figure~\ref{fig:models9}, but for BH forming model u75\_LS220\_hr}
\label{fig:modelsu75smr}
\end{figure*}

\begin{figure*}[!]
\centering
\includegraphics[width=\textwidth]{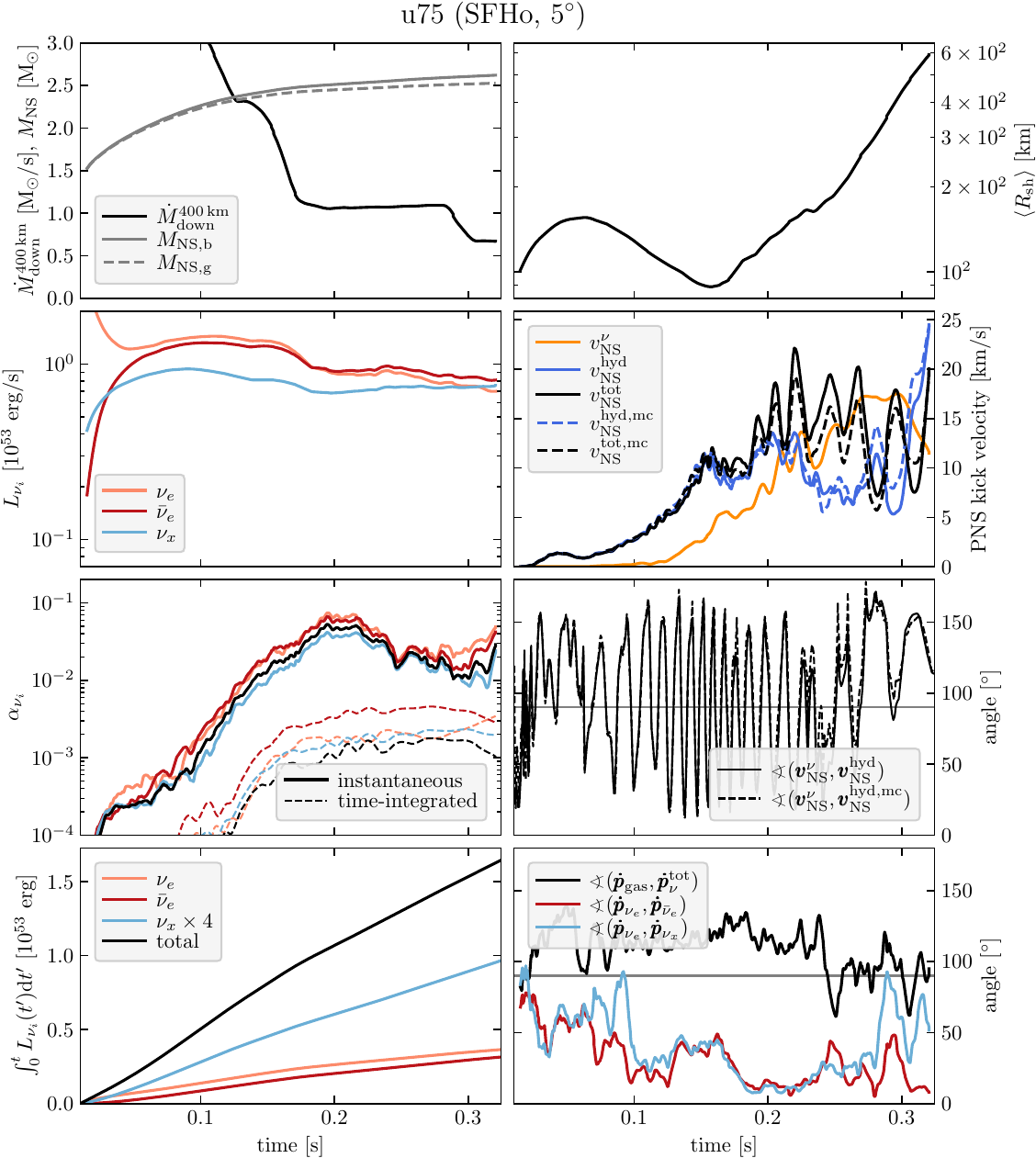}
\caption{Same as Figure~\ref{fig:models9}, but for BH forming model u75\_SFHo}
\label{fig:modelsu75deg5sfho}
\end{figure*}

\end{appendices}

\bibliography{sn-bibliography}

\end{document}